\documentclass[a4paper,11pt]{article}
\usepackage{jheppub}
\usepackage[utf8]{inputenc}
\usepackage{float}
\usepackage{natbib}
\usepackage{graphicx}
\usepackage{comment}
\newcommand{\todo}[1]{{\color{red}{[TODO:~#1]}}}

\begin{document}

\title{Charmonium properties at high temperatures from lattice QCD}

\author[a]{Rasmus Normann Larsen}
\author[b]{Peter Petreczky}
\author[b]{Jorge Luis Dasilva Golan}
\author[c,d]{Johannes~H.~Weber}

\affiliation[a]{Fakult\"at f\"ur Physik, Universit\"at Bielefeld, D-33615 Bielefeld, Germany}
\affiliation[b]{Physics Department, Brookhaven National Laboratory, Upton, New York 11973, USA}
\affiliation[c]{Institut f\"ur Kernphysik, Technische Universit\"at Darmstadt, Schlossgartenstra\ss e 2, 64289 Darmstadt, Germany}
\affiliation[d]{Werner-Heisenberg-Gymnasium Bad D{\"u}rkheim, Kanalstra{\ss}e 19, 67098 Bad D{\"u}rkheim, Germany}
\abstract{
We study charmonium properties at non-zero temperature in 
the temperature range 153 MeV $<T<$ 305 MeV using lattice QCD. We use 
HISQ action for dynamical quarks and Wilson clover action for valence charm quarks
and calculate the correlation function of extended meson operators. 
Our lattice QCD results are consistent with the existence of all charmonium states below the open charm threshold 
in this temperature region. However, charmonium states acquire sizable thermal width, which increases with
increasing temperature.
The size of the thermal width follows the hierarchy of charmonium sizes, i.e.
the smaller ground state charmonium has a smaller thermal width than  the larger excited charmonia. 
}

\maketitle

\section{Introduction}

It was first proposed by Matsui and Satz that quarkonium suppression can serve as a key signature of quark–gluon plasma (QGP) formation in heavy-ion collisions~\cite{Matsui:1986dk}. This idea is based on the expectation that color screening in a deconfined medium reduces the range of the interaction between a heavy quark and antiquark, ultimately leading to the dissolution of bound quarkonium states. Since then, the study of quarkonium properties in the QGP, as well as their production and suppression patterns in heavy-ion collisions, has developed into a major theoretical and experimental research program (see, e.g., Refs.~\cite{Aarts:2016hap,Zhao:2020jqu,Andronic:2024oxz} for recent reviews).
While color screening at high temperature is well established through lattice QCD calculations of the static quark–antiquark free energy~\cite{Bazavov:2018wmo}, its quantitative impact on quarkonium properties and the mechanism of quarkonium melting remain subjects of ongoing investigation. In the potential picture, the binding of a heavy quark–antiquark pair is described in terms of an effective interaction potential. This framework can be systematically justified within the effective field theory of potential non-relativistic QCD (pNRQCD)~\cite{Brambilla:1999xf,Brambilla:2004jw}, which has been extended to finite temperature~\cite{Laine:2006ns,Brambilla:2008cx}. At nonzero temperature, the potential becomes complex: in addition to modifications of its real part, it acquires an imaginary component that encodes medium-induced dissipation and decoherence effects~\cite{Laine:2006ns,Brambilla:2008cx}. Notably, the modification of the real part does not generally correspond to simple exponential (Debye) screening; for distances smaller than the Debye length, thermal effects lead instead to power-law modifications~\cite{Brambilla:2008cx}. In a weakly coupled QGP, the imaginary part of the potential plays a dominant role, leading to finite thermal widths and, ultimately, the dissolution of quarkonium states~\cite{Escobedo:2008sy,Laine:2008cf}.
From first principles, in-medium quarkonium properties are encoded in spectral functions, which are related to Euclidean-time correlation functions accessible in lattice QCD. Extensive efforts have been devoted to studying charmonium~\cite{Umeda:2002vr,Datta:2002ck,Karsch:2002wv,Datta:2003ww,Asakawa:2003re,Jakovac:2006sf,Ohno:2011zc,Ding:2012sp,Ding:2017std,Ali:2025iux} and bottomonium~\cite{Jakovac:2006sf,Aarts:2010ek,Aarts:2011sm,Aarts:2012ka,Aarts:2013kaa,Aarts:2014cda,Kim:2014iga,Kim:2018yhk,Ali:2025iux} in lattice QCD. 
However, most of these studies have relied on point meson operators, where the quark and antiquark fields are located at the same spatial point. Such operators typically have limited overlap with the bound quarkonium states and lead to correlators dominated by the continuum part of the spectral function. As a consequence, their sensitivity to in-medium modifications of quarkonium states is significantly reduced~\cite{Mocsy:2007yj,Petreczky:2010tk,Burnier:2015tda}. Extended meson operators, in which the quark and antiquark fields are spatially separated, offer improved overlap with physical states and have been explored in some early studies of charmonium at finite temperature~\cite{Umeda:2002vr,Ohno:2008cc}.
More recently, substantial progress has been achieved in the study of in-medium bottomonium using extended operators within the framework of lattice non-relativistic QCD (NRQCD)~\cite{Larsen:2019bwy,Larsen:2019zqv,Larsen:2020rjk,Ding:2025fvo}. At zero temperature, such operators are known to have excellent overlap with S- and P-wave bottomonium states~\cite{Davies:1994mp,Meinel:2009rd,Meinel:2010pv,Hammant:2011bt,HPQCD:2011qwj,Daldrop:2011aa}, leading to enhanced sensitivity to their in-medium properties. These studies have enabled the extraction of thermal widths for bottomonium states below the open-flavor threshold~\cite{Larsen:2019bwy,Larsen:2019zqv,Ding:2025fvo}, while the corresponding in-medium mass shifts were found to be consistent with zero. 
This observation suggests that color screening may play a less dominant role than previously anticipated, and that in-medium dissociation is largely driven by dynamical (imaginary potential) effects.
Extending such analyses to charmonium is of considerable importance, both for the phenomenology of heavy-ion collisions and for clarifying the underlying mechanisms of quarkonium modification in the QGP. Charmonium is more sensitive to medium effects due to its smaller binding energy, and its production is a key observable at RHIC and LHC energies. However, unlike from bottomonium, the use of NRQCD is not well justified for charmomium. Therefore, a fully relativistic treatment is required.
In this work, we present a detailed study of charmonium states below the open-charm threshold at finite temperature using correlation functions constructed from extended meson operators. We employ a relativistic formulation for valence charm quarks based on the Wilson clover action, which allows for controlled continuum extrapolation and systematic improvement. Our goal is to assess the in-medium modification of charmonium states with improved operator overlap and sensitivity, thereby providing new insight into  dissociation mechanisms of quarkonia in the QGP.

The rest of the paper is organized as follows: In section 2 we describe our lattice setup. In section 3 we present our lattice QCD results on charmonium correlation functions at zero temperature. The temperature dependence of the charmonium correlations function and the form of the spectral functions used to describe them are discussed in section 4. The 
in-medium charmonium properties are summarized in section 5. Finally, the summary and conclusions are presented in
section 6. Some technical aspects of the calculations and analysis are given in the Appendix.

\section{Lattice QCD setup}
In this study, we use $2+1$-flavor gauge configurations generated by the HotQCD collaboration employing the tree-level Symanzik-improved gauge action and the highly improved staggered quark (HISQ) action for the sea quarks on $64^3 \times N_\tau$ lattices~\cite{Bazavov:2023dci,HotQCD:2025fbd}. The strange quark mass, $m_s$, is tuned close to its physical value, while the light quark mass is set to $m_l/m_s = 1/20$, corresponding to a pion mass of approximately 161~MeV in the continuum limit.

We perform calculations at two lattice spacings corresponding to gauge couplings $\beta = 10/g_0^2 = 7.596$ and $7.825$. The lattice spacing is determined using the $r_1$ scale from the static quark potential, with $r_1 = 0.3106$~fm~\cite{MILC:2010hzw}. A more recent determination of $r_1$ yields consistent results~\cite{Larsen:2025wvg}. This procedure leads to lattice spacings $a = 0.0493$~fm and $a = 0.0404$~fm for $\beta = 7.596$ and $7.825$, respectively. The temperature is varied by changing the temporal extent $N_\tau$, covering the range $152~\text{MeV} < T < 305~\text{MeV}$. For the coarser lattice spacing, we also perform calculations on larger spatial volumes, $96^3 \times N_\tau$, to assess finite-volume effects; these results will be discussed separately.

For the valence charm quark, we employ the clover-improved Wilson fermion action on gauge configurations that have been smeared using one level of hypercubic (HYP) smearing~\cite{Hasenfratz:2001hp}. The use of HYP smearing reduces ultraviolet fluctuations and improves the behavior of the heavy-quark action. The clover coefficient is set to its tadpole-improved tree-level value, $c_{sw} = u_0^{-3/4}$, where $u_0$ is defined from the average plaquette constructed from the HYP-smeared gauge links. This has been implemented into the code SIMULATeQCD \cite{HotQCD:2023ghu}, which was used for measuring the charmonium correlators.
This setup is similar to that used in Refs.~\cite{Izubuchi:2019lyk,Gao:2020ito}; however, in the present case it does not constitute a mixed-action setup, since dynamical charm quarks are not included and their effects are negligible for the observables considered here. Our approach avoids the complications associated with staggered charm quarks in meson correlation functions while maintaining the high quality of the HISQ gauge ensembles.

The parameters of our lattice ensembles on $64^3 \times N_\tau$ lattices are summarized in Tab. \ref{tab:param}, including the temporal extent, temperature, and statistics in terms of the number of gauge configurations and source positions. The use of multiple source positions per configuration improves statistical precision.

The bare charm quark mass
~parameter, $am_c$,
is tuned nonperturbatively by matching the calculated mass of the $J/\psi$ meson to its experimental value. To this end, we perform calculations at several trial values of 
$am_c$
on a subset of configurations and interpolate the resulting $J/\psi$ masses using a functional form given by the square root of a quadratic polynomial. This procedure allows us to determine the value of 
$am_c$
that reproduces the physical $J/\psi$ mass within uncertainties. 
We have verified that the tuning is stable under variations of the interpolation ansatz. Further details of the tuning procedure are provided in the Appendix. The tuned values of the charm quark mass are listed in Table~\ref{tab:param}.

In this work, we focus on correlation functions constructed from extended meson operators, which are described in detail in the following subsection. These operators are designed to enhance overlap with physical quarkonium states and improve sensitivity to their in-medium properties.

\begin{table}[t]
\centering
\begin{tabular}{|c|c|c|c||c|c|c|c|}
\hline
\multicolumn{4}{|c||}{$\beta=7.825,\; a=0.0404$ fm, $N_x=64$ } & \multicolumn{4}{c|}{$\beta=7.596,\; a=0.0493$ fm, $N_x=64$ } \\
\multicolumn{4}{|c||}{$am_c=0.1712,\; c_{sw}=1.0286$} & \multicolumn{4}{c|}{$am_c=0.2285,\; c_{sw}=1.0309$} \\
\hline
$N_\tau$ & $T$ [MeV] & \#conf & \#src & $N_\tau$ & $T$ [MeV] & \#conf & \#src \\
\hline
16 & 305 & 5635 & 32 & 16 & 250 & 3830 & 32 \\
18 & 271 & 5336 & 16 & 18 & 222 & 3722 & 16 \\
20 & 244 & 4448 & 16 & 20 & 200 & 3013 & 16 \\
22 & 222 & 3623 & 16 & 22 & 182 & 4164 & 16 \\
24 & 203 & 3450 & 16 & 24 & 167 & 3285 & 16 \\
26 & 188 & 4255 & 16 &    &     &      &    \\
28 & 174 & 3429 & 16 &    &     &      &    \\
30 & 163 & 2786 & 16 &    &     &      &    \\
32 & 153 & 2139 & 16 &    &     &      &    \\
64 &   0 & 1171 & 80 & 64 &   0 & 1159 & 64 \\
\hline
\end{tabular}
\caption{Parameters of the lattice QCD calculations, including gauge coupling $\beta$, lattice spacing $a$, charm quark mass parameter $m_c$, clover coefficient $c_{sw}$, temporal extent $N_\tau$, temperature $T$, number of gauge configurations, and number of source positions per configuration.}
\label{tab:param}
\end{table}

\subsection{Extended meson operators with Gaussian smearing}
The Gaussian smeared meson operators are defined in the following way
\begin{equation}
    O_{\Gamma}(x,\tau)=\bar \Psi'(x,\tau) \Gamma \Psi'(x,\tau),
\end{equation}
with 
$\Psi'(x,\tau)=\sum_{x^\prime}W(x,x^\prime,\tau) \Psi(x^\prime,\tau)$ is the Gaussian smeared quark field, $\Psi(x,\tau)$ being
the original quark field and $W(x,x',\tau)$ being the Gaussian smearing operator.
The matrix $\Gamma$ fixes
the quantum numbers of the charmonium state. We consider $\Gamma=\gamma_5,~\gamma_i,~1$
and $\gamma_5 \gamma_i$, i.e. pseudo-scalar, vector, scalar and axial-vector channels, corresponding to $\eta_c,~J/\psi,~\chi_{c0}$ and $\chi_{c1}$ states, respectively.
The Gaussian smearing is implemented as follows. We start with a point (delta
function) source $S(x,\tau)$ for the fermion field at space-time point $(x,\tau$ and apply the operator
with discretized Laplacian
\begin{eqnarray}
S'_{\lambda,N}(x_0,\tau) &=& \prod_{n=1}^{N} \left(\delta_{x_{n-1},x_{n}}+\frac{ \lambda ^2\Delta_{3,2}(x_{n-1},x_{n}) }{4N} \right) S(x_N,\tau) \\
\Delta_{3,2}(x,y) &=& \sum _{i=0} ^2 ( U_i(x)\delta_{x+\hat{i},y}+U_i ^\dagger (x-\hat{i})\delta_{x-\hat{i},y} - 2 \delta _{x,y} )
\end{eqnarray}
For sufficiently large $N$ and small $\lambda ^2 /(4N)$ 
this procedure creates a Gaussian source for quark fields of extent $\lambda / \sqrt{2}$ in lattice
units.
The size of the meson operator then is approximately $ \lambda$.
The values
of $N$ and $\lambda$ are given in Tab. 
\ref{tab:param_gauss}.
These values
correspond to $\lambda ^2 /(4N)\simeq 0.08$. We use $\lambda=7$ 
or $10$
for the coarser lattice and $\lambda=7$
, $10$ or $12$ for the finer lattice.
The parameters of the Gaussian smearing used in our analysis are given in Tab. \ref{tab:param_gauss}.
The Gaussian smeared sources are used in the tuning of the charm quark masses. We used 16 sources in
all our calculations with Gaussian smearing.
\begin{table}[H]
\begin{tabular}{ |c|c|c|c|c|c|c|c|}
\hline
\multicolumn{4}{|c|}{$\beta=7.825,~a=0.0404$ fm, $N_x=64$ } & \multicolumn{4}{|c|}{$\beta=7.596,~a=0.0493$ fm, $N_x=64$ } \\
\hline
$N_\tau$ & T[MeV] & $\#$conf &  op. & $N_\tau$ & T[MeV] & $\#$conf & op.\\
 \hline
16 & 305 & 1165  &  (12,450)  & 16 & 250 & 1002  & (7,150),(10,300) \\
 \hline
18 &  271  & 2503  &  (7,150),(10,300),(12,450)   &  &   &  &   \\
 \hline
20 &  244  & 3857  &  (12,450)   &  &    &   &   \\
 \hline
24 &  203  & 3250  &  (12,450)   &  &    &   &   \\
 \hline
28 &  174  & 844  &  (12,450)   &  &    &   &   \\
 \hline
64 &  0  & 1171  &  (7,150),(10,300),(12,450)   & 64 & 0   & 1159  &  (7,150),(10,300) \\
\hline
\end{tabular}
\caption{Parameters of the parameters of Gaussian smearings 
$(\lambda,N)$ with
$N$ being the number of iterations and $\lambda$ being the source size
for different $\beta$ and $N_\tau$, see text. We also give the number of measurements
for each 
$\beta$ and $N_\tau$. 16 sources were used for all the runs.}
\label{tab:param_gauss}
\end{table}

\subsection{Optimized meson operators}
To study charmonium properties in the scalar, pseudo-scalar, vector and axial-vector channels, we consider the following extended meson operators
\begin{eqnarray}
\tilde O_{\alpha} (x,\tau) = \sum _y \phi_{\alpha}(y) \bar{\Psi}(x+y,\tau) \Gamma \Psi(x,\tau),
\Gamma=1,~\gamma_5,~\gamma_i,~\gamma_5 \gamma_i
\label{eq:wf_src}
\end{eqnarray}
In the pseudo-scalar or vector channel
with the proper choice of the trial wave function, $\phi_{\alpha}(y)$ we can optimize
the overlap of the corresponding correlation with the charmonium state $\alpha$ of interest, e.g
$\eta_c(1S)$ or $\psi(2S)$. The choice $\phi_{\alpha}=\delta(y)$ corresponds to a point meson operator. 
Using a set of properly chosen
$\phi_{\alpha}(y)$ we obtain a basis of meson operators, which allows us to extract properties
of several charmonium states.
Since the above extended meson operator is not gauge invariant, we need to fix the Coulomb
gauge. In our study the accuracy of Coulomb gauge fixing is $10^{-8}$.

The correlation function 
\begin{equation}
 \tilde C_{\alpha , \beta}(\tau)=\sum_x \langle  \tilde O_{\alpha}^{\dagger}(x,\tau) \tilde O_{\beta}(0,0)\rangle
\end{equation}
can be expressed in terms of charm quark propagators 
$G_{a,b}(x,x',\tau, \tau ') = \langle \Psi_{a}(x,\tau) \bar{\Psi}_b(x',\tau')\rangle$
as
\begin{eqnarray}
\tilde C_{\alpha,\beta}(\tau) 
        &=& (-)^{L+1} \sum_x \langle\sum _y \phi_{\alpha} ^\dagger(y) \bar{\Psi } _c(x,\tau ) \Gamma^\dagger _{c,d} \Psi_d(x+y,\tau) \bar{\Psi} _{\phi_{\beta},a}(0,0) \Gamma _{a,b} \Psi _{b}(0,0) \rangle \nonumber\\
        &=& (-)^L\sum_x \langle\sum _y \phi_{\alpha}^\dagger (y) G_{b,c}(0,x,0,\tau) \Gamma^\dagger_{c,d} G_{\phi_{\beta},d,a}(x+y,0,\tau,0) \Gamma _{a,b} \rangle \nonumber\\
        &=& (-)^L\sum_x \langle\sum_y 
        \phi_\alpha^\dagger(y)
        {\rm Tr}( \gamma _5 G(x,\tau) ^\dagger \gamma _5 \Gamma ^\dagger G_{\phi_{\beta}}(x+y,\tau) \Gamma)  \rangle. 
\end{eqnarray}
Here we introduced the notation  $\bar{\Psi}_{\phi_{\beta},a} (x,\tau) = \sum _y \bar{\Psi}_a(x+y,\tau) \phi_{\beta}(y)$
for the quark source and $G_{\phi_{\beta},d,a}(x,0,\tau,0)$ for the corresponding propagator. We also
used $G(0,x) = \gamma_5 G(x,0)^\dagger \gamma _5 $. Furthermore, $L$ denotes the orbital angular momentum of the charmonium state: $L=0$ for $\eta_c$ and $J/\psi$, and $L=1$ for $\chi_{c,0}$ and $\chi_{c,1}$.

The correlation function $\tilde C_{\alpha,\beta}$ is non-zero for $\alpha \ne \beta$, i.e. different states can mix. To minimize this, we consider a linear combination of operators 
$O_{\alpha}=\Omega_{\alpha \alpha'} \tilde O_{\alpha'}$, such that 
\begin{equation}
C_{\alpha, \beta}(\tau)=\sum_x \langle  O_{\alpha}^{\dagger}(x,\tau) O_{\beta}(0,0)\rangle = \delta_{\alpha,\beta} C_{\alpha}(\tau).
\end{equation}
The matrix $\Omega_{\alpha \alpha'}$ can be determined by solving the generalized 
eigenvalue problem 
\begin{equation}
\tilde C_{\alpha', \beta'}(\tau) \Omega_{\alpha \alpha'}=\lambda_{\alpha}(\tau,\tau_0) \tilde C_{\alpha',\beta'}(\tau_0)\Omega_{\alpha \alpha'}, 
\end{equation}
see e.g. Ref. \cite{Blossier:2009kd}.
We choose $\tau_0 /a=15$ for $\beta=7.596$ and $\tau_0 /a =20$ for $\beta=7.825$ as the finer lattice spacing at higher $\beta$ takes longer to approach a plateau. We will refer to this correlation function as the wave function optimized 
correlation function or simply as the optimized correlation function. 

\begin{figure}
    \centering
    \includegraphics[width=0.45\textwidth]{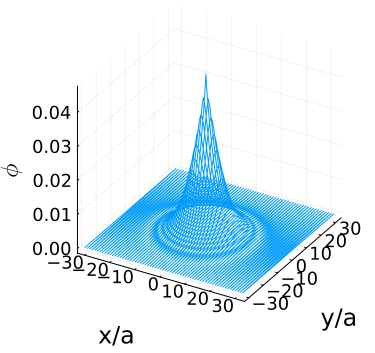}
    \includegraphics[width=0.45\textwidth]{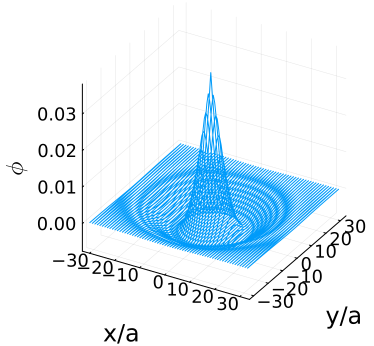}
    \caption{The input wave function for 1S on the left and 2S on the right, generated from a Schroedinger equation with a charm mass of 2GeV, with the potential given in equation (\ref{eq:pot}).}
    \label{fig:wave_sources}
\end{figure}

The larger the basis of operators $\tilde O_{\alpha}$ is, the more states can be extracted from
the calculations. Since in this study we are primarily interested in the charmonium states below
the open charm threshold. To study S-wave charmonia we use a set of 3 operators obtained by using $\phi_{\alpha},~\alpha=1S,~2S,~3S$
that approximate the wave function of $1S,~2S$ and $3S$ charmonium states, plus the point
meson operator that corresponds to $\phi(r)=\delta(r)$, that represents all the states
above the open charm threshold. The trial wave functions $\phi_{\alpha},~\alpha=1S,~2S,~3S$
have been calculated using a potential model.
To fix the parameters of the potential model, we resort to our experience with the calculation 
of the bottomonium Bethe-Salpeter (BS) amplitude \cite{Larsen:2020rjk}. Treating the BS
amplitude as a quarkonium wave function, one can obtain the potential and the constituent heavy 
quark mass by solving the inverse Schr\"odinger problem \cite{Larsen:2020rjk}.
The potential obtained this way can be parameterized as \cite{Larsen:2020rjk}
\begin{eqnarray}
V(r) &=& -0.166522/r+0.1166289 ({\rm GeV})^2 \times r+0.382423\log(r \times{\rm GeV})(GeV). 
\label{eq:pot}
\end{eqnarray}
The constituent bottom quark mass was found to be $5.6$ GeV. We note that this value
is larger than the constituent b-quark mass $m_b=5.18$  used in the Cornell 
potential model \cite{Eichten:1979ms}. To obtain the wave function $\phi_{\alpha},~\alpha=1S,~2S,~3S$ we use the potential given by Eq. (\ref{eq:pot}), and the 
constituent charm quark mass $m_c=2.0$ GeV. This choice is motivated by the fact
that in Cornell potential model $m_c=1.84$ GeV, and for bottomonium the constituent 
mass obtained from the BS amplitude was larger than in Cornell potential model, while giving a good approximation for the splittings between the 1S, 2S and 3S states. The wave functions for 1S and 2S states obtained this way are 
shown in Fig. \ref{fig:wave_sources}.

To study $\chi_{c0}(1P)$ and $\chi_{c1}(1P)$ charmonium states, we use a single extended meson operator with
$\phi_{1P}(r)$ that is equal to the  wave-function of the $1S$ state
obtained from the above potential model. While this choice does not correspond to a realistic 
1P wave function, we still find that the 
corresponding extended meson operators have good  overlap with
$\chi_{c0}$ and $\chi_{c1}$ states.


\section{Zero temperature correlation functions}
\begin{figure}
    \centering
    \includegraphics[width=0.48\textwidth]{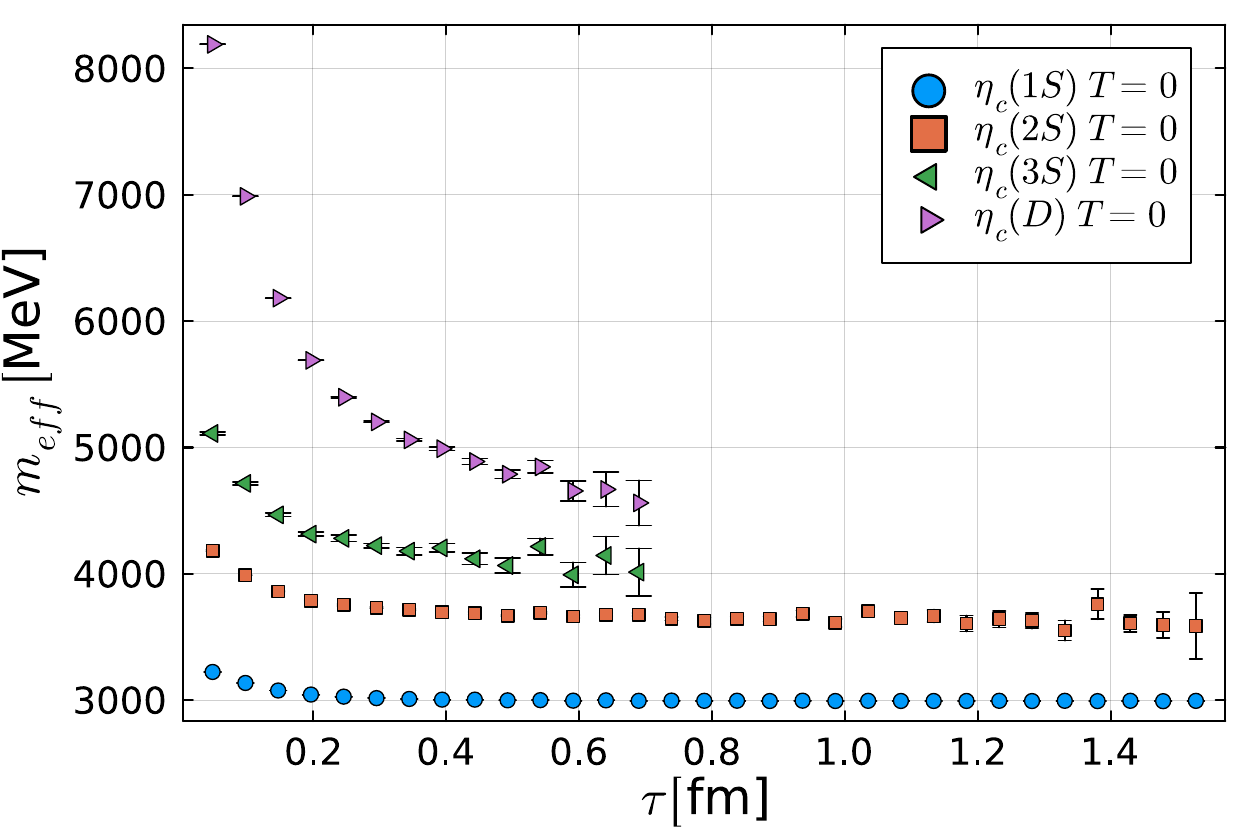}
    \includegraphics[width=0.48\textwidth]{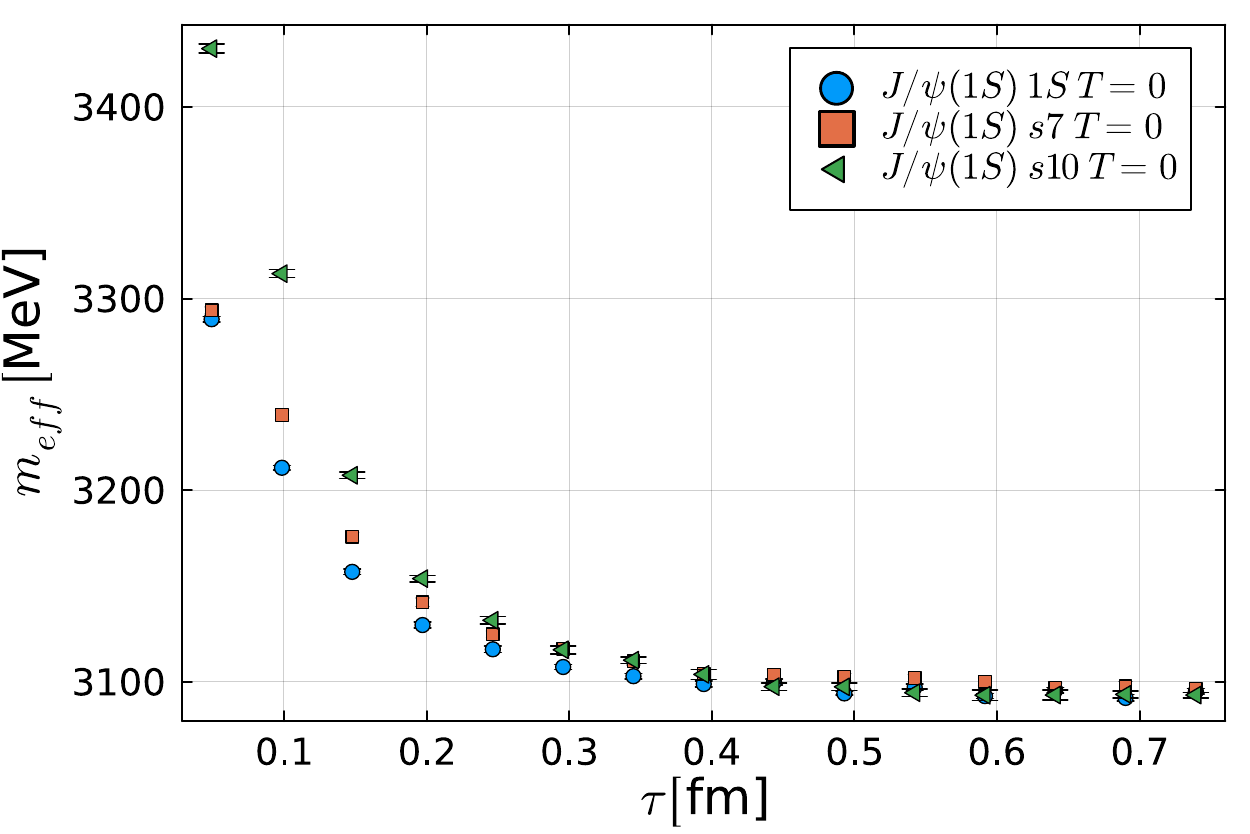}
    \caption{The effective mass from optimized pseudo-scalar correlator for $\beta=7.596$ obtained from the generalized eigenvalue problem around $\tau /a = 15$ (left) and the comparison of the effective masses from optimized correlator for $J/\psi$ and the correlators of
    extended operators with Gaussian smearing for different smearing radiuses.}
    \label{fig:meff_T0}
\end{figure}

In this section we discuss our results on the correlation functions at zero temperature.
The correlation functions are analyzed in terms of the effective masses defined
as 
\begin{eqnarray}
C(\tau)/C(\tau+1) = \cosh(am_{eff}(\tau - N_\tau /2)) / \cosh(am_{eff}(\tau+1 - N_\tau /2)).
\label{eq:meff}
\end{eqnarray}
This definition takes into account the (anti)periodic boundary condition on the lattice and is sometimes referred to as cosh mass.
For $N_\tau\rightarrow \infty$ this definition reduces to $m_{eff} = -\partial _\tau \log (C(\tau))$.
In Fig. \ref{fig:meff_T0} we show the effective masses of optimized correlators as well
as the correlation function of extended meson operators with Gaussian smearing for
S-wave charmonia for $a=0.0493$ fm ($\beta=7.596$). From the optimized correlators we can determine the masses of $1S$
and $2S$ charmonia states. The  two higher energy level states can also be identified, 
but an accurate determination of their masses is challenging. In the right panel of
this figure we compare our results on the effective masses obtained from the correlation functions of 
extended meson operators with Gaussian smearing and the optimized correlation function.
We consider Gaussian smearing with size $\lambda=7$ and $\lambda=10$ labeled as $s7$ and $s10$ in the figure.
While the effective mass corresponding to the optimized correlator approaches the ground state
the fastest, also the effective masses from the correlators of extended operators with Gaussian smearing approach 
the ground state at relatively small $\tau$.  At very small $\tau$, $\tau<0.3$ fm we see significant differences
in the effective masses corresponding to different meson correlators reflecting the fact that these correlators have
different overlap with excited charmonium states.

\begin{figure}
    \centering
    \includegraphics[width=0.5\textwidth]{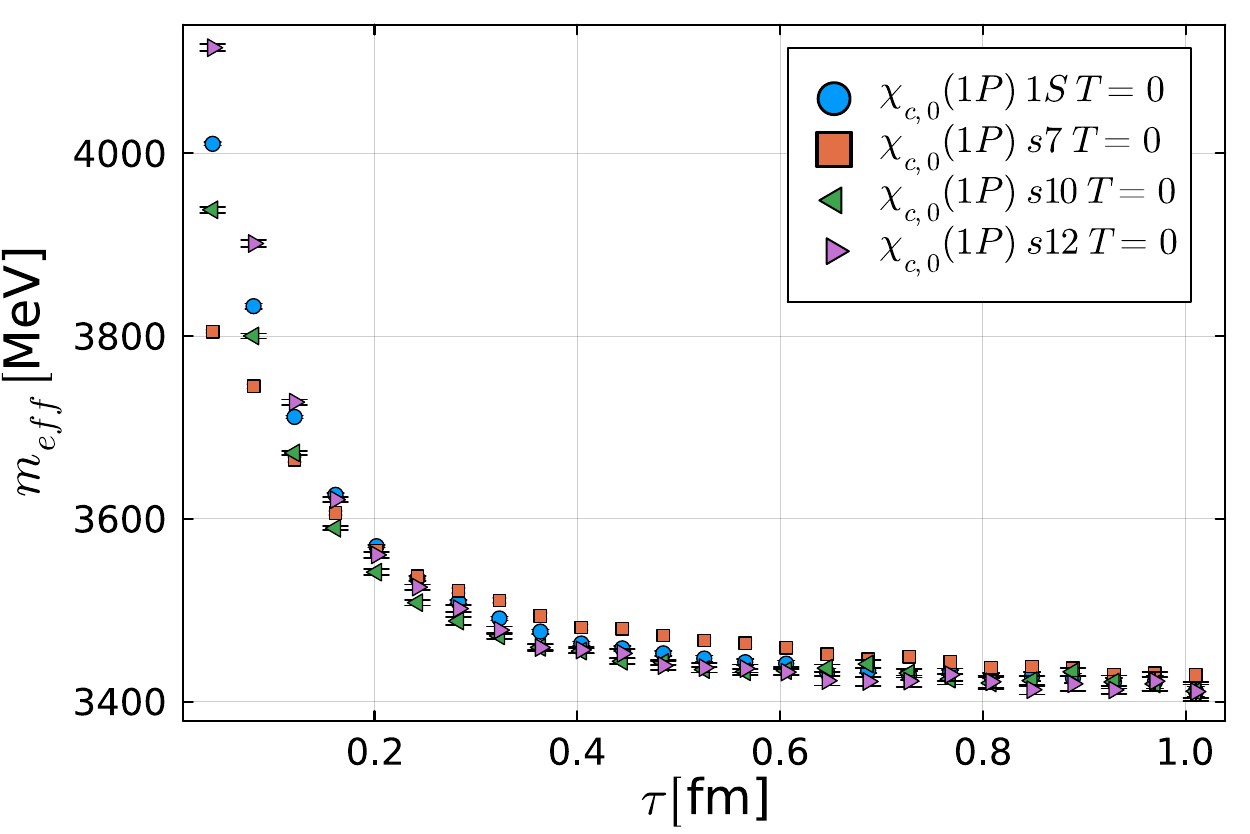}
    \caption{The effective mass for the $\chi_{c,0}$ state at zero temperature for $\beta=7.825$ for
    various meson operators, see text.}
    \label{fig:meff_T0_P}
\end{figure}

To study the $P$-wave charmonia we use the correlation function of extended meson operators
with Gaussian smearing as well as the extended meson operator with 1S wave function as
discussed in the previous section.
The effective masses for $\chi_{c0,1}$ states
are shown in Fig. \ref{fig:meff_T0_P}. Again, the effective masses corresponding to different correlation functions approach
the same ground state plateau for $\tau>0.7$ fm. At smaller $\tau$ excited state contamination is present, but all the extended meson operators are effective in suppressing this contamination at approximately the same level. Only for 
$\tau<0.3$ fm we see significant differences in the effective masses.

By performing 2-exponential fits on optimized correlators, we obtain the masses
of $1S$ and $2S$ charmonia. Similarly, from the 2-exponential fits on
the correlators of meson operators on optimized correlators, we obtain
the masses of the P-wave charmonia. The charmonia masses at zero temperature
are summarized in Tab. \ref{tab:T=0} and compared to the experimental results from
PDG \cite{ParticleDataGroup:2024cfk}. The differences with respect to the experimental results are small, if present, and
are due to the imprecise determination of the input charm quark mass  or lattice artifacts.

\begin{table}[H]
\begin{center}
\begin{tabular}{ |c|c|c| } 
 \hline
State & $\beta$ & Mass[MeV]\\ 
 \hline
$\eta _c (1S)$ & 7.596 & $2993.2 \pm 0.9$ \\
 \hline
$\eta _c (1S)$ & 7.825 & $2995.1 \pm 0.5$ \\
 \hline
$\eta _c (1S)$ & PDG & $2984.1 $ \\
 \hline
$\psi (1S)$ & 7.596 & $3092.1 \pm 0.4$ \\
 \hline
$\psi (1S)$ & 7.825 & $3096.1 \pm 0.8$ \\
 \hline
$\psi (1S)$ & PDG & $3096.9 $ \\
 \hline
$\chi _{c,0} (1P)$ & 7.596 & $3420.1 \pm 2.7$ \\
 \hline
$\chi _{c,0} (1P)$ & 7.825 & $3422.4 \pm 2.8$ \\
 \hline
$\chi _{c,0} (1P)$ & PDG & $3414.71 $ \\
 \hline
 \hline
\end{tabular}
\begin{tabular}{ |c|c|c| } 
 \hline
State & $\beta$ & Mass[MeV]\\ 
 \hline
$\chi _{c,1} (1P)$ & 7.596 & $3505.8 \pm 5.0$ \\
 \hline
$\chi _{c,1} (1P)$ & 7.825 & $3506.0 \pm 4.8$ \\
 \hline
$\chi _{c,1} (1P)$ & PDG & $3510.67 $ \\
 \hline
$\eta _c (2S)$ & 7.596 & $3650.0 \pm 8.8$ \\
 \hline
$\eta _c (2S)$ & 7.825 & $3654.1 \pm 7.1$ \\
 \hline
$\eta _c (2S)$ & PDG & $ 3637.8 $ \\
 \hline
$\psi (2S)$ & 7.596 & $3695.0 \pm 5.7$ \\
 \hline
$\psi (2S)$ & 7.825 & $3703.6 \pm 8.1$ \\
 \hline
$\psi (2S)$ & PDG & $3686.097 $ \\
 \hline
 \hline
\end{tabular}
\end{center}
\caption{Masses of different charmonia states obtained from a 2 exponential fit on the meson correlators at two 
lattice spacings corresponding to $\beta=7.596$ and $\beta=7.825$. We compare our results with experimental values obtained from the Particle Data Group (PDG) \cite{ParticleDataGroup:2024cfk}.}
\label{tab:T=0}
\end{table}

\section{Charmonium correlation functions at non-zero temperature}

Our goal is to gain information on the  charmonium spectral
function in different quantum number channels at non-zero temperature.
The charmonium correlation functions in Euclidean time
that we calculate on the lattice are related to the spectral functions 
as 
\begin{equation}
    C_{\alpha}(\tau,T)=\int_0^{\infty} d \omega \sigma_{\alpha}(\omega,T) K(\tau,\omega,T),~
    K(\tau,\omega,T)=\frac{\cosh(\omega(\tau-1/(2T))}{\sinh(\omega/(2T)}.
\end{equation}
We will consider optimized correlation functions for $1S$ and $2S$ charmonium states
and correlation functions of extended meson operators with wave function for $\chi_{c0}(1P)$
and $\chi_{c1}(1P)$. We will also consider charmonium correlation functions of extended
meson operators with Gaussian smearing. First, we examine the temperature dependence
of the correlation functions to see to what extent this can encode the temperature
dependence of the spectral function and what a suitable Ansatz for the spectral
function can look like. Then we perform fits to the correlation functions using
the Ansatz for the spectral function to determine the charmonium properties at non-zero
temperature. This will be discussed in the following subsections.

\subsection{Effective masses of charmonium correlators at non-zero temperature}

\begin{figure}
    \centering
    \includegraphics[width=0.45\textwidth]{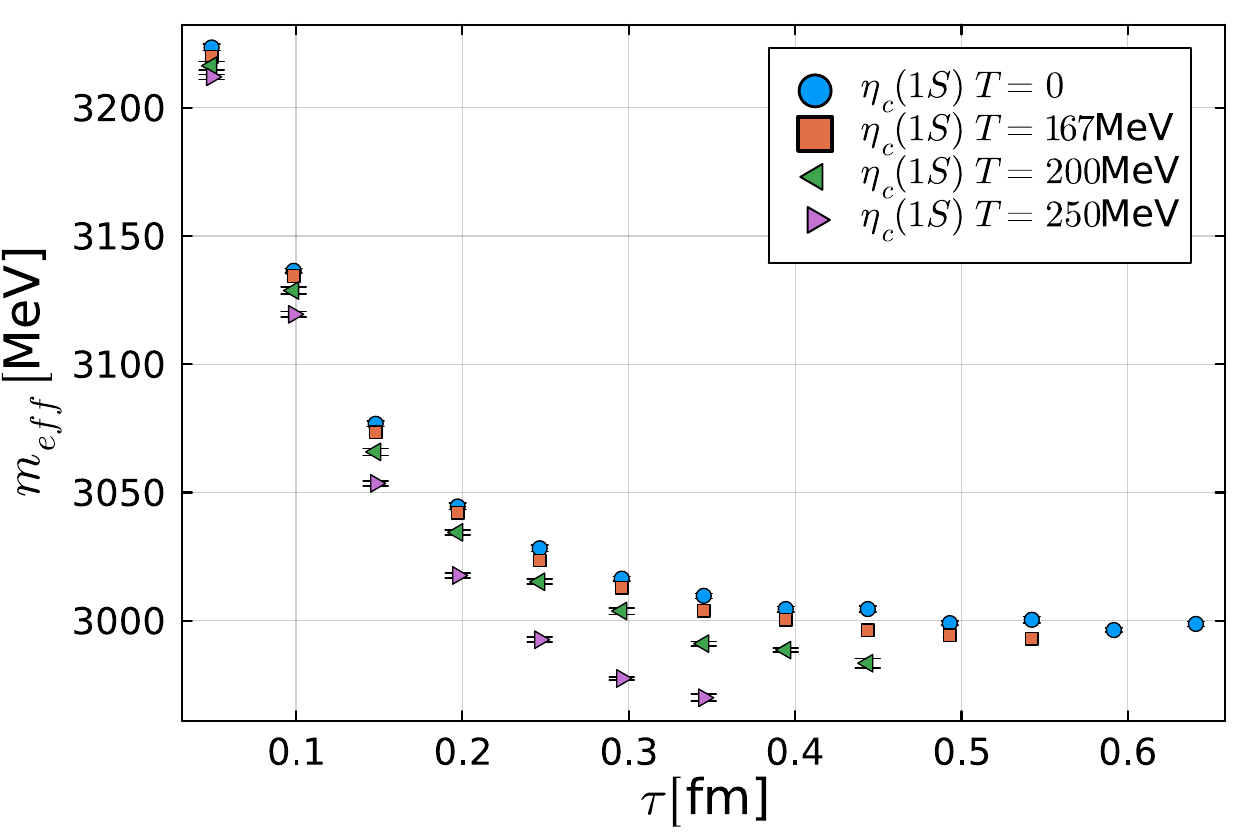}
    \includegraphics[width=0.45\textwidth]{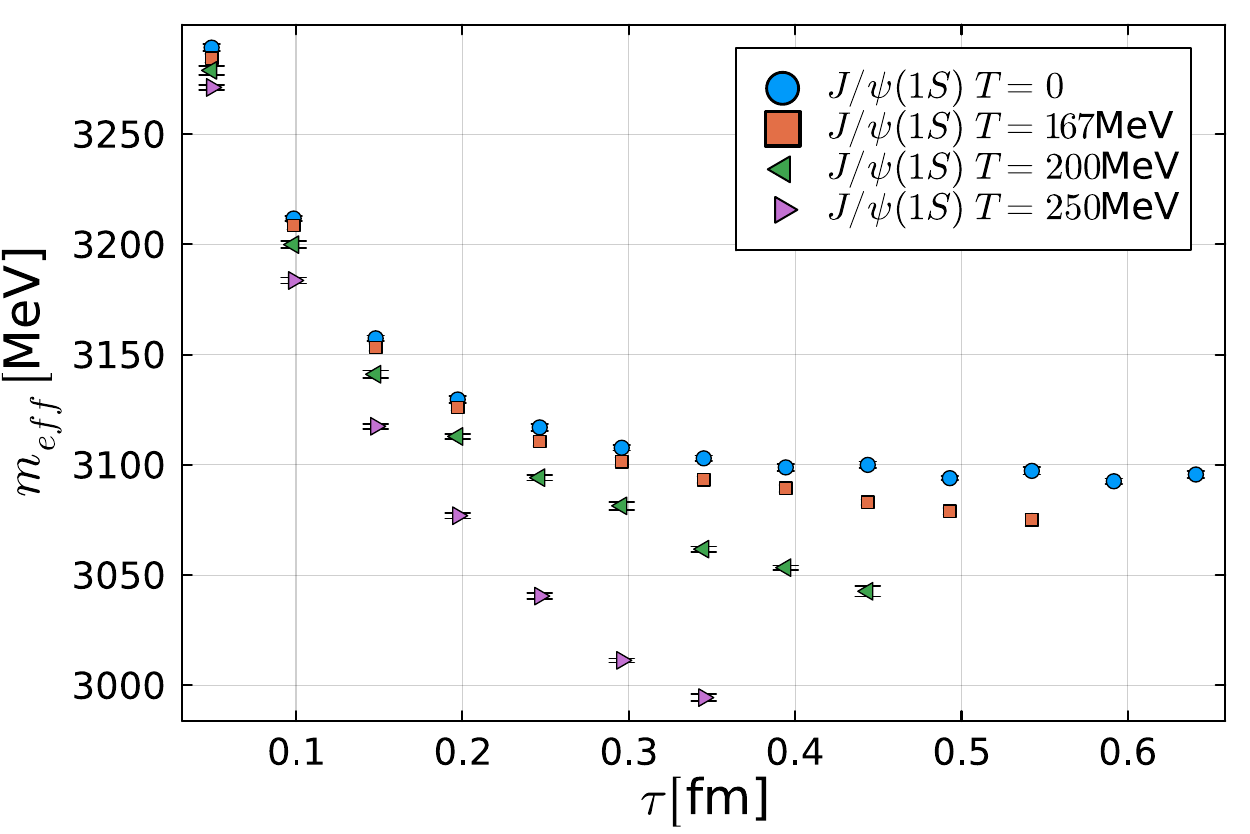}
    \includegraphics[width=0.45\textwidth]{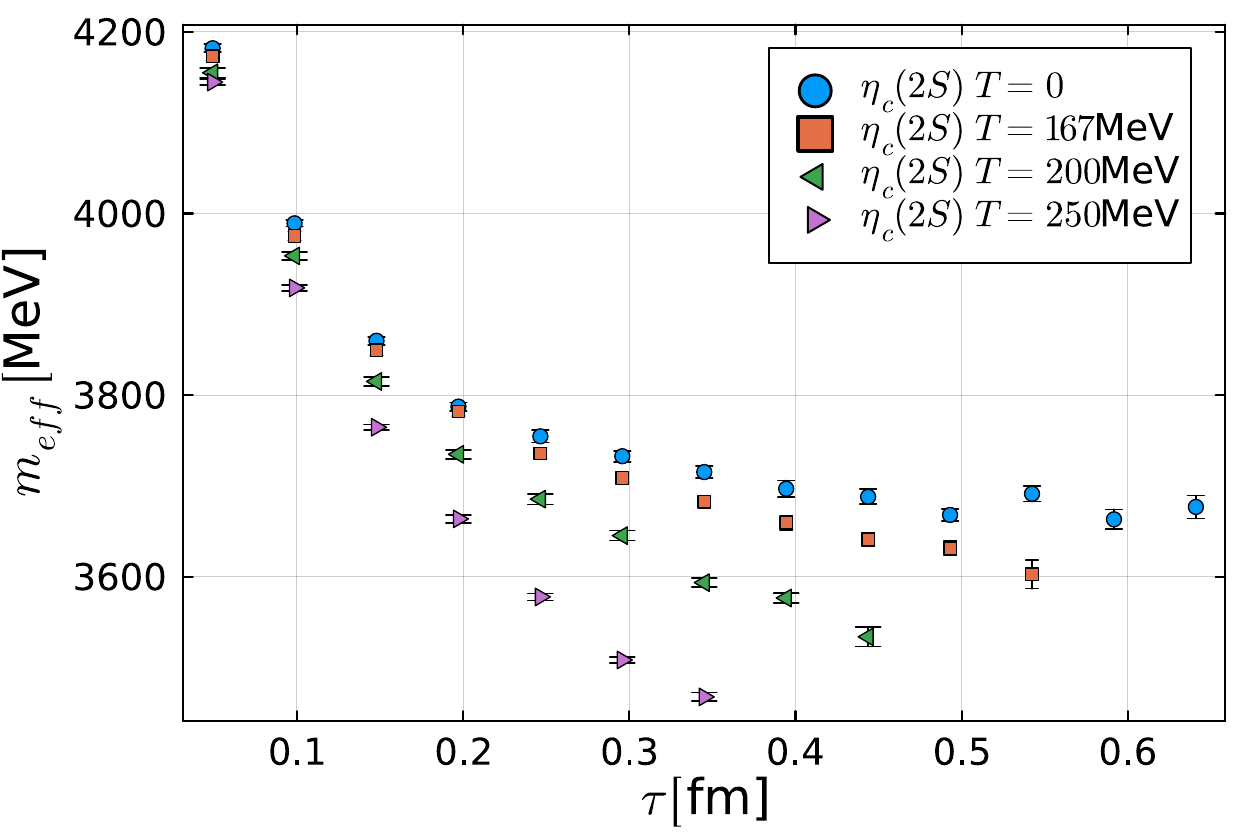}
    \includegraphics[width=0.45\textwidth]{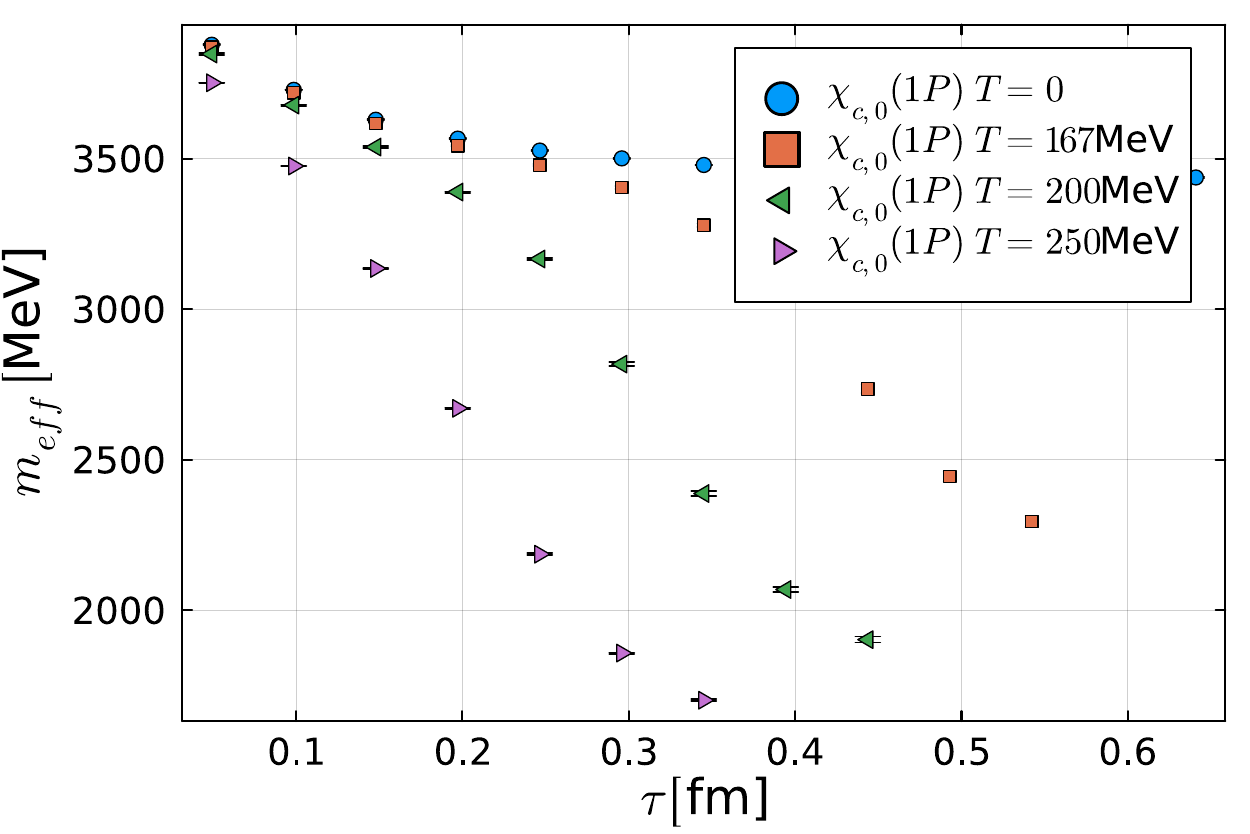}
    \caption{The effective masses for $\eta_c(1S)$ (top left), $J/\psi$ (top right), $\eta_c(2S)$ (bottom left) and $\chi_{c0}(1P)$ (bottom right) at several temperatures. }
    \label{fig:meff_T}
\end{figure}

We study the temperature dependence of the charmonium correlators in terms of
the effective masses defined by Eq. (\ref{eq:meff}). In Fig. \ref{fig:meff_T} we show the effective masses for the optimized correlation function of $\eta_c(1S)$, $J/\psi$, $\eta_c(2S)$ and $\chi_{c0}$
charmonium.
At small $\tau$ the temperature dependence of the effective masses is small, and the
difference with respect to the zero temperature result is also small. The temperature dependence
gradually increases with increasing $\tau$. The temperature dependence is the smallest
for the $\eta_c(1S)$ effective mass, slightly larger for the $J/\psi$ case, and still larger
for $\eta_c(2S)$. For the $\chi_{c0}$ state, we see a much larger temperature dependence at
moderate and large values of $\tau$, and a much larger deviation from the zero temperature 
result. The large medium effects on the $\chi_{c0}$ effective mass cannot be entirely
due to the in-medium modification or melting of the $\chi_{c0}$ state since the effect is much larger
than for $\eta_c(2S)$, which is more loosely bound. While not shown here, we obtain similar
results for the $\chi_{c1}$ state. The temperature dependence of the effective masses of
$\eta_c(1S)$ and $\eta_c(2S)$ are shown in Fig. \ref{fig:meff_T} is quite similar to the temperature
dependence of the effective masses of bottomonium states obtained using NRQCD with extended
operators \cite{Larsen:2019bwy,Larsen:2019zqv,Ding:2025fvo}.

To understand the above behavior of effective masses, we assume the following form of
the spectral function
\begin{equation}
    \sigma_{\alpha}(\omega,T)=\sigma_{\alpha}^{med}(\omega,T)+\sigma_{\alpha}^{high}(\omega),
    \label{eq:spf_gen_decom}
\end{equation}
where $\sigma_{\alpha}^{med}$ encodes the properties of the lowest charmonium state in the medium or in the vacuum, while $\sigma_{\alpha}^{high}(\omega)$ describes the contribution of the excited charmonium states above the open charm threshold and the continuum. This part of the 
spectral function is assumed to be temperature independent. At zero temperature 
$\sigma_{\alpha}^{med}(\omega,T=0)=A_{\alpha} \delta(\omega-M_{\alpha})$ with $M_{\alpha}$
being the mass of the charmonium state $\alpha$ and $A_{\alpha}$ is the corresponding amplitude.
At non-zero temperature $\sigma_{\alpha}^{med}$ contains in-medium charmonium states if the
temperature is not too high. At non-zero temperature there is also a contribution to $\sigma_{\alpha}^{med}(\omega,T)$ at $\omega \simeq 0$ in the scalar, vector and axial vector channels. This contribution to
the spectral function corresponds to an almost constant contribution to the charmonium 
correlation function. For point meson operators such a contribution is well established in both 
numerical lattice QCD calculations \cite{Umeda:2007hy,Petreczky:2008px} and in analytic calculations in free field theory \cite{Aarts:2005hg}. In the vector channel the zero mode contribution corresponds to heavy flavor transport \cite{Petreczky:2005nh}. Furthermore, such contribution
was shown to exist also in the correlators of extended meson operators \cite{Umeda:2007hy}.
Therefore, the difference in the temperature of effective masses of $\eta_c(1S)$ and $J/\psi$
can be attributed to the zero mode. Similarly, the large temperature dependence of the 
$\chi_{c0,1}$ effective masses are due to the zero mode contribution. We will elaborate
on this more in the following subsection. Note, that in NRQCD at non-zero temperature
there is no zero mode contribution.

\subsection{Subtracted correlators and the determination of the properties of
S-wave charmonia} \label{sec:subtracted_correlator}
\begin{figure}
    \centering
    \includegraphics[width=0.45\textwidth]{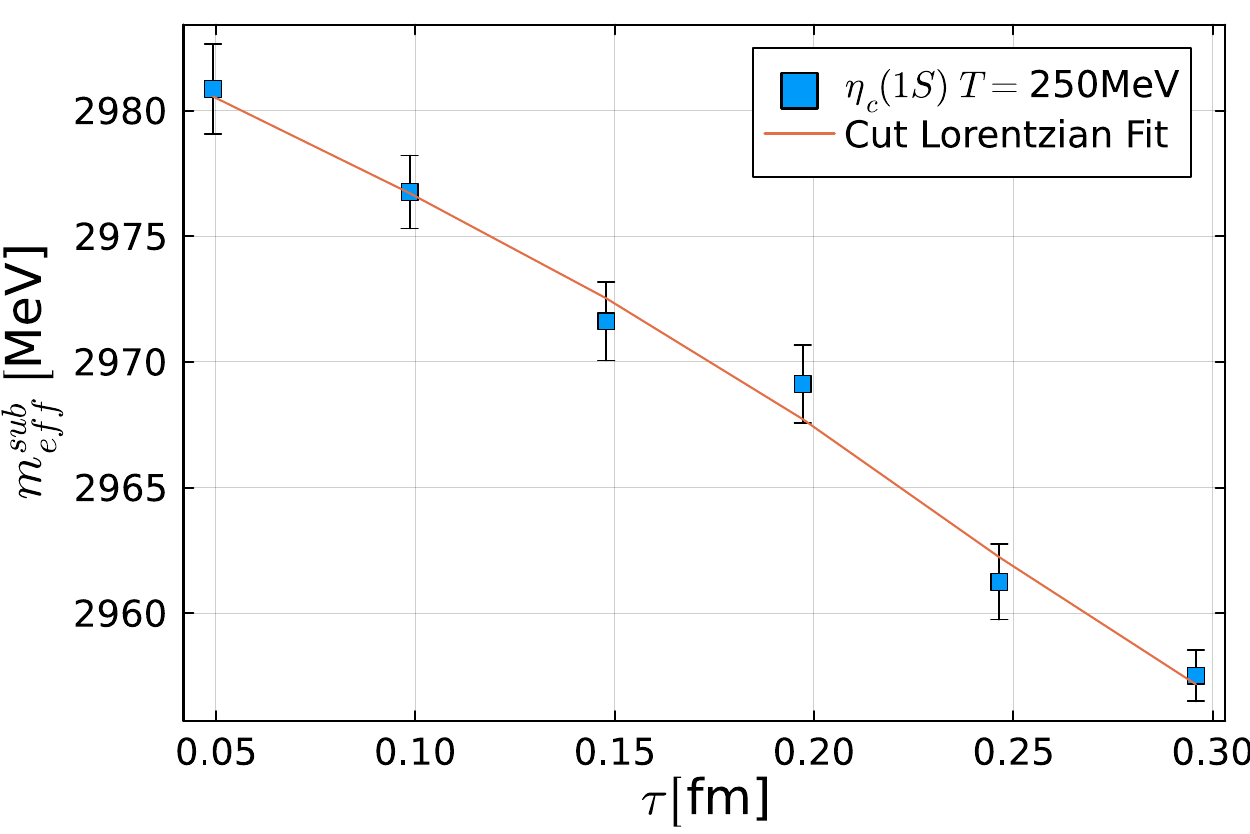}
    \includegraphics[width=0.45\textwidth]{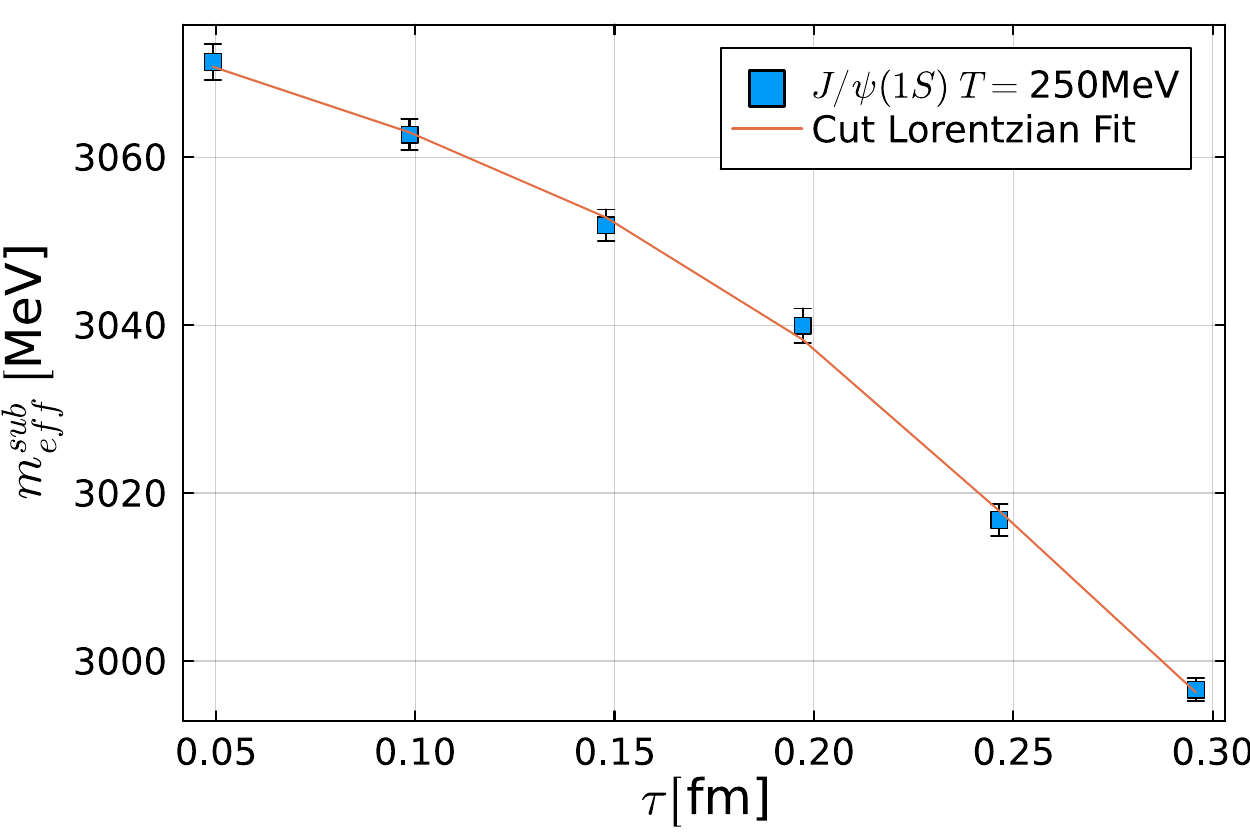}
    \includegraphics[width=0.45\textwidth]{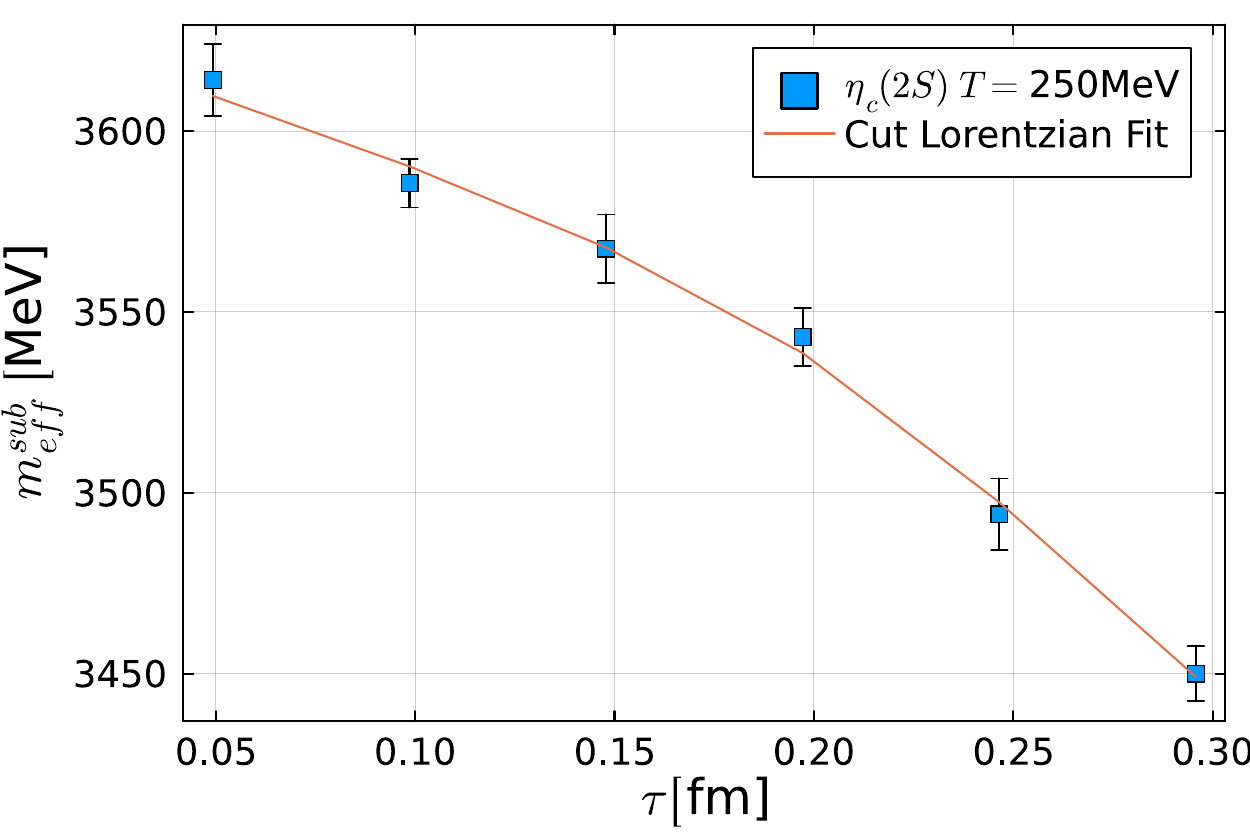}
    \includegraphics[width=0.45\textwidth]{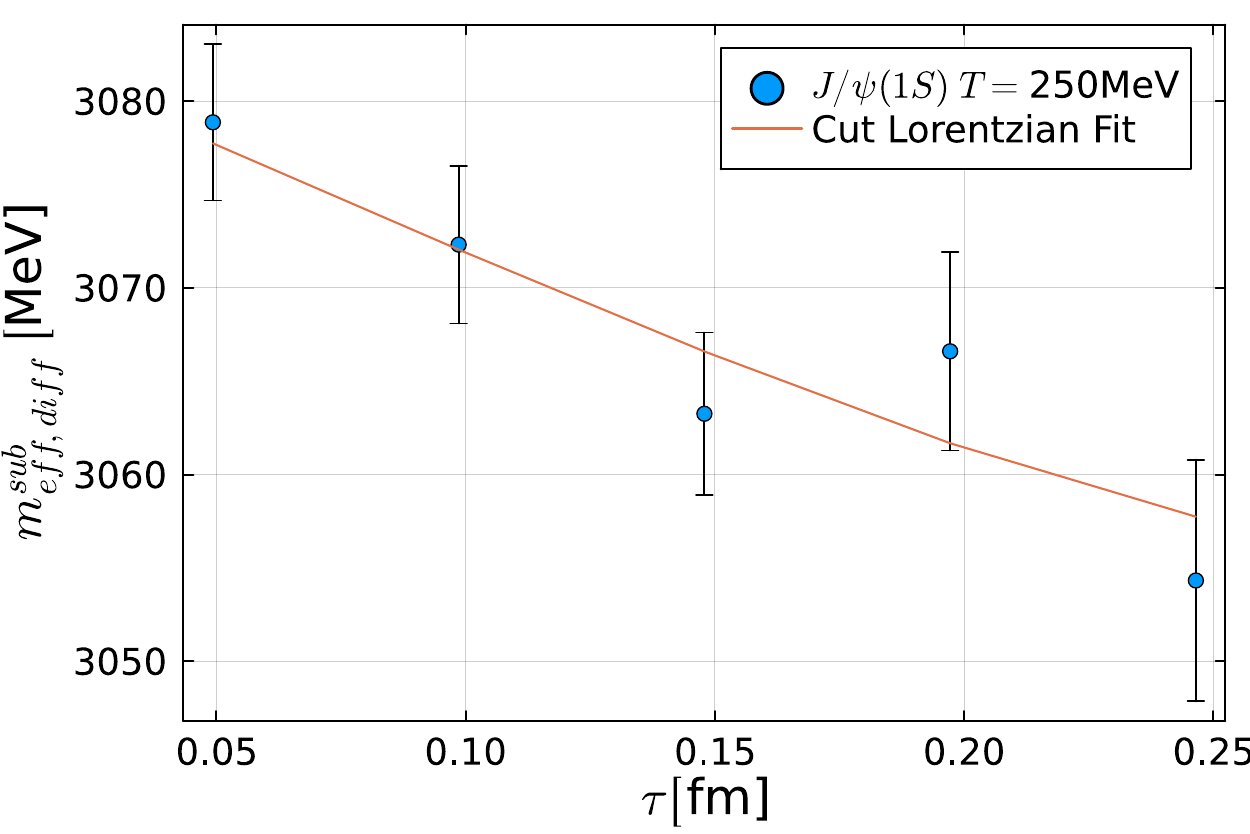}
    \caption{The subtracted effective masses for $\eta_c(1S)$ (left top), $J/\psi$ (right top)
    and $\eta_c(2S)$ (left bottom). The lines show the fits to Eq. (\ref{eq:sigma_med_Ansatz}). (right bottom) Fit to the same data as in Fig. \ref{fig:meff_1S_sub} (right top), but the fits are done by using the difference in the correlator. The effective mass is also reconstructed from the difference in the correlator. Difference fit: Peak position $M_\alpha$ = $3081.4 \pm 3.2$, $\Gamma_{\alpha} ^0 = 107.7 \pm 10.82$. Correlator fit: Peak position $M_\alpha$ = $3080.9 \pm 3.3$, $\Gamma_{\alpha} ^0 = 105.3 \pm 11.7$. }
    \label{fig:meff_1S_sub}
\end{figure}

To determine the properties of the S-wave charmonia, we use the general form of the spectral
function given by Eq. (\ref{eq:spf_gen_decom}). To do so, we first need to determine and subtract
the contribution from $\sigma_{\alpha}^{high}(\omega)$ to the correlation function:
\begin{equation}
C_{\alpha}^{high}(\tau,T)=\int_0^{\infty} d \omega \sigma_{\alpha}^{high}(\omega) K(\tau,\omega,T).
\end{equation}
At zero temperature, calculating this contribution is straightforward 
$C_{\alpha}^{high}(\tau,T=0)=C_{\alpha}(\tau,T=0)-A_{\alpha} \exp (-(M_{\alpha} \tau)$, where $M_{\alpha}$ and $A_{\alpha}$
are the mass and the amplitude of the charmonium state $\alpha$. 
At non-zero temperature corresponding to temporal extent
$N_\tau$ we approximate $C_{\alpha}^{high}(\tau,T)$ as
\begin{equation}
    C_{\alpha}^{high}(\tau,T)=C_{\alpha}^{high}(\tau,T=0) + C_{\alpha}^{high}(1/T-\tau,T=0).
\end{equation}
In principle one needs to add all possible contributions from moving around the periodic time direction of the lattice multiple times, we however find that this does not matter as the charmonium states are heavy, and the additional contributions are suppressed. Then we study the subtracted correlators at non-zero temperature 
\begin{equation}
   C_{\alpha}^{sub}(\tau,T)=C_{\alpha}(\tau,T) - C_{\alpha}^{high}(\tau,T).  
   \label{eq:sub}
\end{equation}
We study the temperature and $\tau$-dependence of the subtracted correlation functions in terms of the subtracted effective masses,
which are calculated by replacing $C_{\alpha}(\tau,T)$ in Eq. (\ref{eq:meff}) by $C_{\alpha}^{sub}(\tau,T)$. 
In Fig. \ref{fig:meff_1S_sub} we show the subtracted effective mass for $\eta_c(1S)$, $\eta_c(2S)$ and $J/\psi$ states
for $a=0.04593$ fm ($\beta=7.596$) and $T=250$ MeV as an example.
The strong $\tau$
dependence of the effective masses is much reduced as the result 
of the subtraction of the UV contribution. At small $\tau$ the subtracted
effective masses show an approximate linear decrease in $\tau$. This feature
of the subtracted effective masses has been observed for NRQCD 
correlation function with extended meson operators \cite{Larsen:2019bwy,Larsen:2019zqv,Ding:2025fvo}. There are differences in the $\tau$-dependence of the effective masses compared to NRQCD case because
of periodic boundary conditions. In Refs. \cite{Larsen:2019bwy,Larsen:2019zqv} it was argued that the linear dependence
of the effective masses is related to the thermal width of bottomonium states
and can be described assuming a Lorentzian form of the spectral function \cite{Ding:2025fvo}.
To see how this linear behavior can arise here, let us consider the spectral decomposition of
the subtracted correlator written as
\begin{equation}
    C_{\alpha}^{sub}(\tau,T)=\int_0^{\infty} d \omega \sigma_{\alpha}^{med}(\omega,T) \frac{\exp(\omega/(2T))}{2\sinh(\omega/(2T)} (e^{-\omega \tau} +e^{\omega \tau-\omega /T}).
\end{equation}
Introducing $ \sigma _{\alpha, eff}(\omega,T)=\sigma_{\alpha}^{med}(\omega,T) \frac{\exp(\omega/(2T))}{2\sinh(\omega/(2T)}$
for small $\tau$ we can write
\begin{eqnarray}
C_{\alpha}^{sub}(\tau,T) &=& \int_0^{\infty} d \omega \sigma_{\alpha,eff}(\omega,T) (e^{-\omega \tau} -e^{\omega \tau-\omega /T})  \nonumber \\
 &\approx &  \int_0^{\infty} d \omega \sigma_{\alpha,eff}(\omega,T) (1-\omega \tau+\frac{(\omega \tau)^2}{2} )\nonumber \\
 &=& N_{\alpha} (1-\langle \omega  \rangle_{\alpha} \tau + \frac{\langle \omega ^2  \rangle_{\alpha} \tau^2}{2}  + \dots).   
\end{eqnarray}
Here we assumed that $\sigma_{\alpha}^{med}(\omega,T)$ and thus also $\sigma _{\alpha, eff}(\omega,T)$ is peaked around the charmonium mass
and for this reason the term proportional to $e^{-\omega/T}$ can be neglected. Furthermore, we introduced the notations:
\begin{equation}
N_{\alpha}=\int_0^{\infty} d\omega  \sigma _{\alpha, eff}(\omega,T), 
~\langle \omega^n \rangle_{\alpha}=\frac{1}{N_{\alpha}}\int_0^{\infty} d \omega \omega^n \sigma_{\alpha}^{eff}(\omega,T),~n=1,2
\end{equation}
It is easy to see that the slope of the effective mass for small $\tau$ is given by the second cumulant of the spectral
function $c_{2,\alpha}=~\langle \omega^2 \rangle_{\alpha}- ~\langle \omega \rangle_{\alpha}^2$. The second cumulant for the spectral function corresponds to
the effective width of $\sigma _{\alpha, eff}(\omega,T)$. One can also define higher order cumulant of the spectral functions. These characterize 
the possible asymmetric nature of $\sigma _{\alpha, eff}(\omega,T)$ around its peak position and will influence the shape
of the effective masses at larger $\tau$. If $\sigma _{\alpha, eff}(\omega,T)$ consists of one approximately symmetric peak, then higher order cumulants will be small.
For example in the nonphysical case 
that $\sigma _{\alpha, eff}(\omega,T)$ were a Gaussian, all higher order cumulants of the spectral function would be exactly zero.
But even for physically motivated Lorentzian form, the higher order cumulants of the spectral function are typically small, see Refs. \cite{Bazavov:2023dci,Ding:2025fvo}.

The effective mass from the subtracted correlator for $J/\psi$ shows more curvature at large $\tau$ and also a stronger decrease. As we argue in the previous section,
this is due to the zero mode contribution to the vector correlator. We can consider the derivative of the subtracted vector correlator instead, 
to which the zero mode
will not contribute and calculate the corresponding effective mass. This is shown in the bottom right panel of Fig. \ref{fig:meff_1S_sub}, where the derivative
of the subtracted vector correlator was estimated using simple finite difference. We see that the $\tau$-dependence of the corresponding effective mass is quite similar
to that of the $\eta_c$, as one would expect. In appendix \ref{sec:grow} we describe an alternative method to get rid of the zero mode. This method turns out to be also useful
in checking our determination of the thermal width as we discuss below.

We will assume that in-medium S-wave charmonia masses and widths are described by a Lorentzian form.
For $J/\psi$ we also need the zero mode contribution to $\sigma_{J/\psi}^{med}(\omega,T)$. Thus,
we write
\begin{equation}
    \sigma_{\alpha}^{med}(\omega,T)=\frac{A_{\alpha}(T)}{\pi} \frac{\Gamma_{\alpha}(\omega,T)}{(\omega-M_{\alpha}(T))^2+\Gamma_{\alpha}^2(\omega,T)}+z_{\alpha} \omega \delta(\omega).
    \label{eq:sigma_med_Ansatz}
\end{equation}
While we do not expect a zero mode contribution to the  $\eta_c$ correlator from the free field theory calculations, we allow such a contribution 
to test for non-linear behavior in the effective mass.
The function $\Gamma_{\alpha}(\omega,T)$ should vanish for large $|\omega-M_{\alpha}|$
and its value or $\omega \simeq M_{\alpha}(T)$ can be interpreted as the thermal width of
S-wave charmonium, while $M_{\alpha}(T)$ is the in-medium charmonium mass.
The simplest choice of this function would be
$\Gamma_{\alpha}(\omega,T)=\Gamma_{\alpha}^0(T) \Theta(|\omega-M_{\alpha}|-cut_{\alpha})$, where
$cut$ is the addition parameter. We call this form of the spectral function 
the cut Lorentzian Ansatz. For this form $\Gamma_{\alpha}^0(T)$ is the nominal
thermal width. In this work we use $cut_{\alpha}=4 \Gamma_{\alpha}^0$ based on
the T-matrix calculations of quarkonium spectral functions \cite{Tang:2024dkz,Wu:2025hlf} as discussed in
Ref. \cite{Ding:2025fvo}. The fits to the effective masses are shown in Fig. \ref{fig:meff_1S_sub}.
As one can see from the figures, the fits to the above form of the spectral function can describe the lattice
results very well.

The above discussion used the wave function optimized sources. For 1S charmonia we also performed a similar analysis using extended meson operators with Gaussian smearing. This analysis
is discussed in the Appendix. The effective masses obtained from the subtracted correlator with Gaussian smearing show similar $\tau$ and temperature dependence as
the 
ones obtained with optimized operators. In particular, we see an approximately linear decrease in the effective masses. The slope of the effective masses 
, however, somewhat depends on the choice of the meson operator. The slope is largest for the largest source size. These differences are barely visible 
for temperatures $T<174$ MeV but in some instances are significant for the largest temperatures. The implications of these findings for the in-medium properties of
1S charmonium will be discussed below.

\subsection{Subtracted correlators and the determination of the properties of
P-wave charmonia}

\begin{figure}
    \centering
    \includegraphics[width=0.45\textwidth]{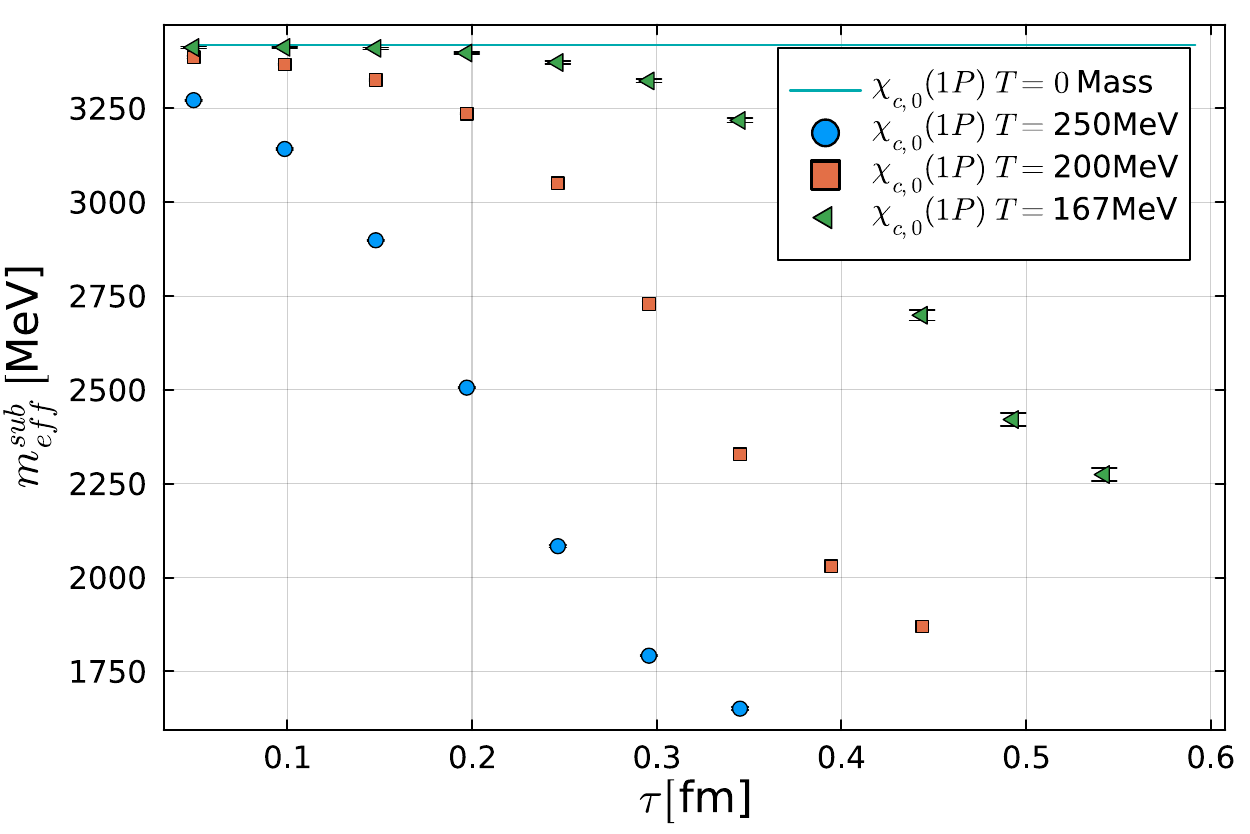}
    \includegraphics[width=0.45\textwidth]{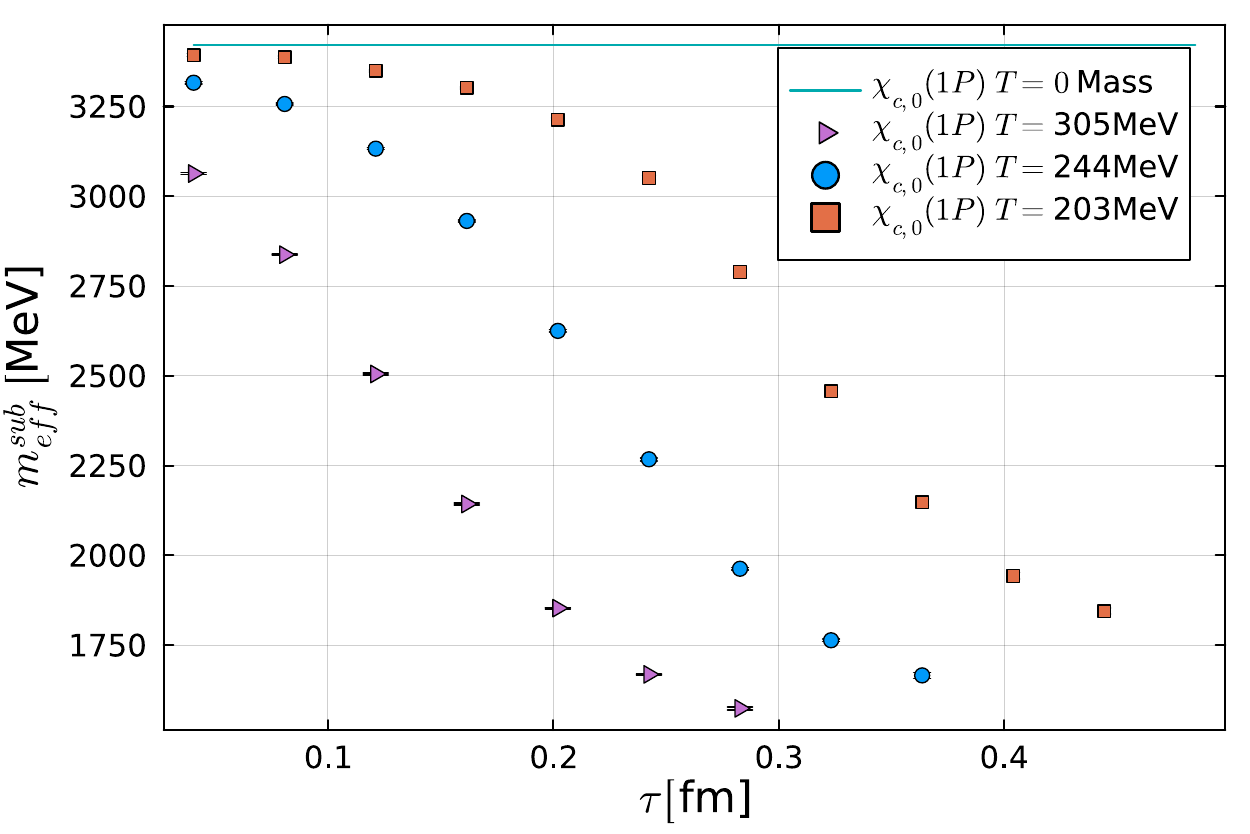}
    \caption{Subtracted effective mass for 1P states for lattice spacing $a=0.0493$ fm (left) and  lattice spacing $a=0.0404$ fm  (right) at several temperatures. }
    \label{fig:1P_meff_sub}
\end{figure}

\begin{figure}
    \centering
    \includegraphics[width=0.45\textwidth]{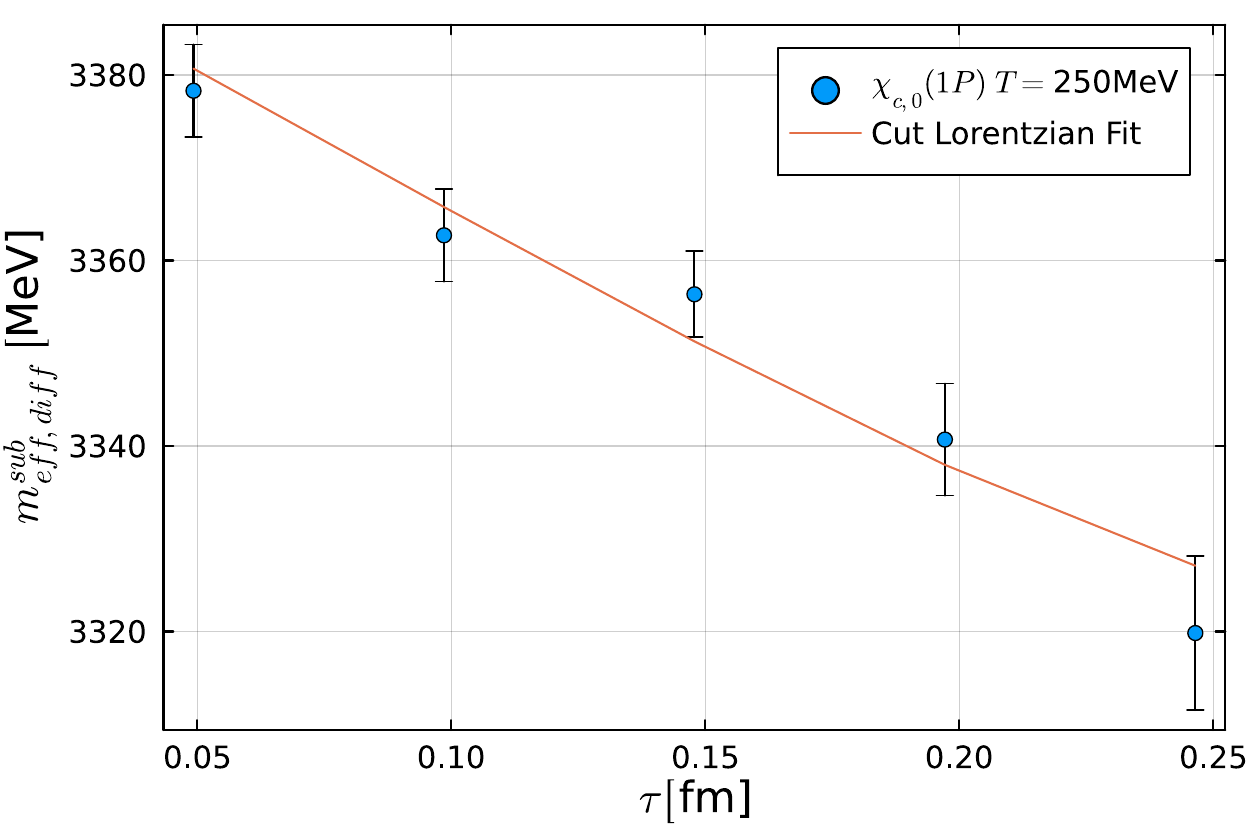}
    \includegraphics[width=0.45\textwidth]{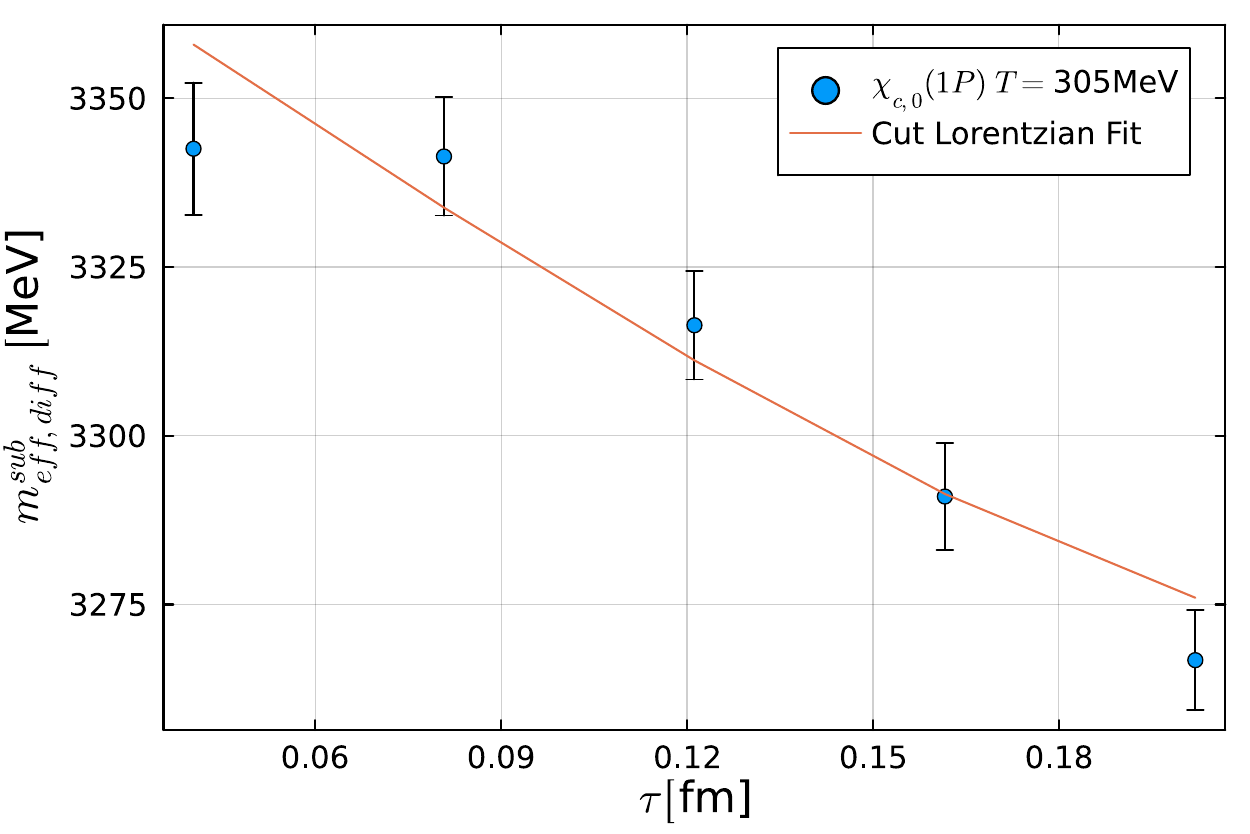}
    \caption{Subtracted effective mass for 1P states, reconstructed from the derivative of the correlator, to remove the zero mode.
    The left panel shows results for $a=0.0493$ fm and $T=250$ MeV. The right panel shows results for 
    $a=0.0404$ fm and $T=305$ MeV.
    Fit is with a cut Lorentzian without a zero mode, done on the difference of the correlator using the Ansatz in eq. (\ref{eq:sigma_med_Ansatz}).}
    \label{fig:1P_meff_sub_diff_fit}
\end{figure}

The analysis of the meson correlation functions described in the previous sub-section has been performed also for the P-wave charmonia, $\chi_{c0}$ and $\chi_{c1}$.
Namely, we first estimate
the subtracted correlation function for P-wave charmonia using Eq. (\ref{eq:sub}), analyze the corresponding effective masses and fit the correlation
using the Ansatz for the spectral function given by Eq. (\ref{eq:sigma_med_Ansatz}). 
In Fig. \ref{fig:1P_meff_sub} we show the effective masses from the subtracted correlator for $\chi_{c0}$ with wave function optimized source for the two lattice spacings used in this study.
The effective masses at small $\tau$ for $T=167$ MeV are close to the vacuum mass of $\chi_{c0}$ state. 
At larger $\tau$ the effective masses drop rapidly with increasing $\tau$. 
As discussed above, this behavior is due to the zero
mode contribution. 
Since the zero mode contribution increases with increasing temperature, this drop also increases.
To get rid of the zero mode we consider the derivative of the $\chi_{c0}$ subtracted correlator. The 
corresponding effective masses are shown in Fig. \ref{fig:1P_meff_sub_diff_fit}. We see from the figure
that these effective masses show qualitatively similar behavior to the effective mass of $\eta_c$, however,
the corresponding decrease is larger, implying the width of the $\chi_{c0}$ state is larger than that of $\eta_c$.
In this figure we also show the fit to the subtracted correlator with cut Lorentzian form of $\sigma_{\chi_{c0}}^{med}$ and $cut=4 \Gamma_{\alpha} ^0$. 
Again, the fit describes the lattice results very well. We obtained very similar results for the subtracted effective masses for the $\chi_{c1}$
correlator. 

We also calculate the correlation function for $\chi_{c0}$ and $\chi_{c1}$ using extended meson operators with Gaussian smearing. The corresponding subtracted masses show
the same $\tau$ and temperature dependence as above at a qualitative level. However, there are quantitative differences between the subtracted effective masses calculated with wave function optimized operators and Gaussian smeared ones, as discussed in the Appendix. The implication of these findings for the in-medium properties of 
P-wave charmonia will be discussed below.

\subsection{Volume dependence}
\begin{figure}
    \centering
    \includegraphics[width=0.45\textwidth]{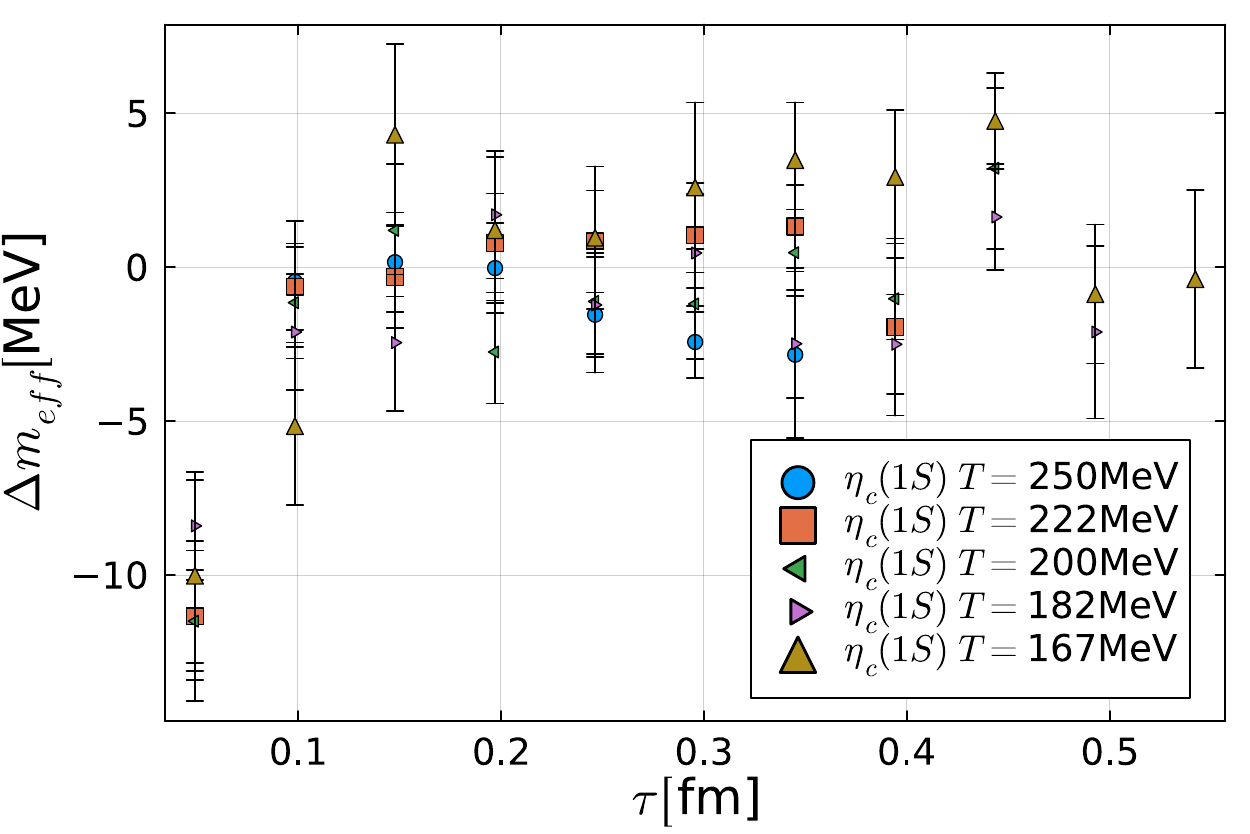}
    \includegraphics[width=0.45\textwidth]{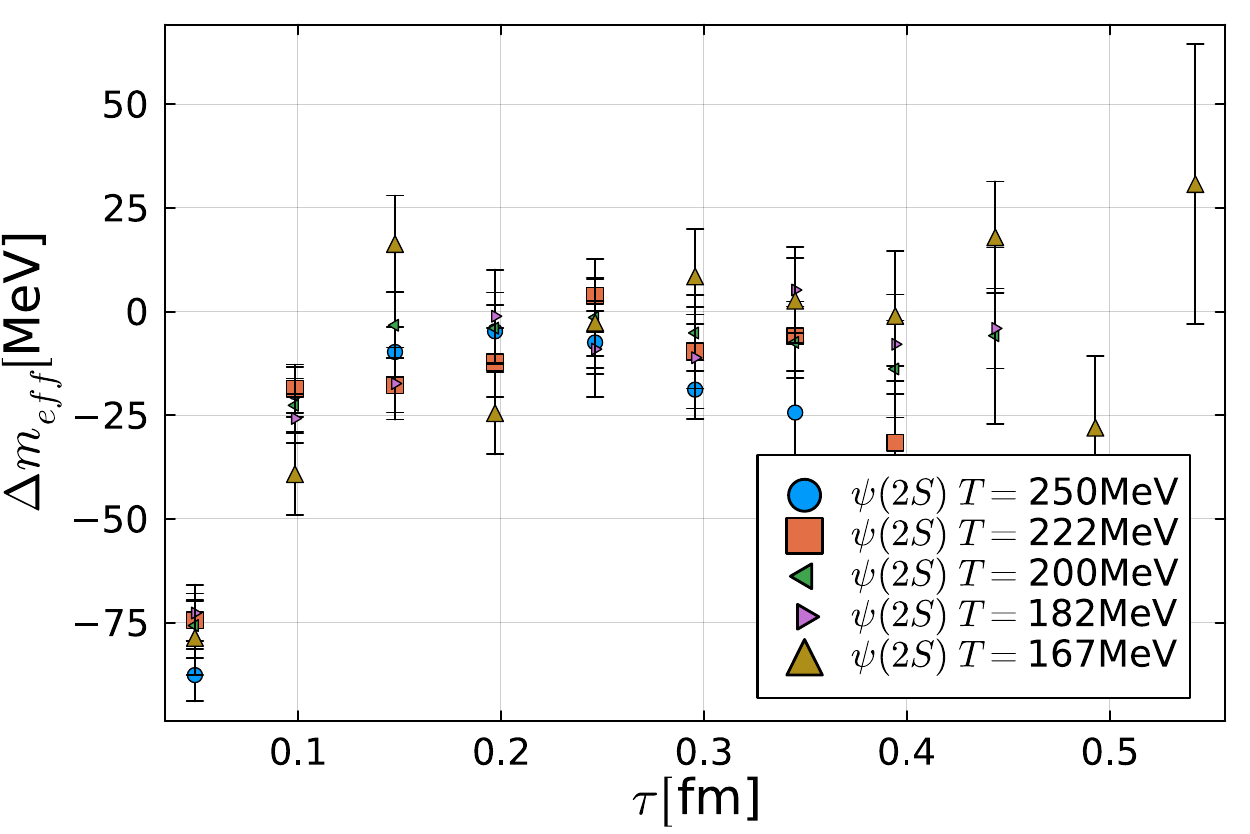}
    \includegraphics[width=0.45\textwidth]{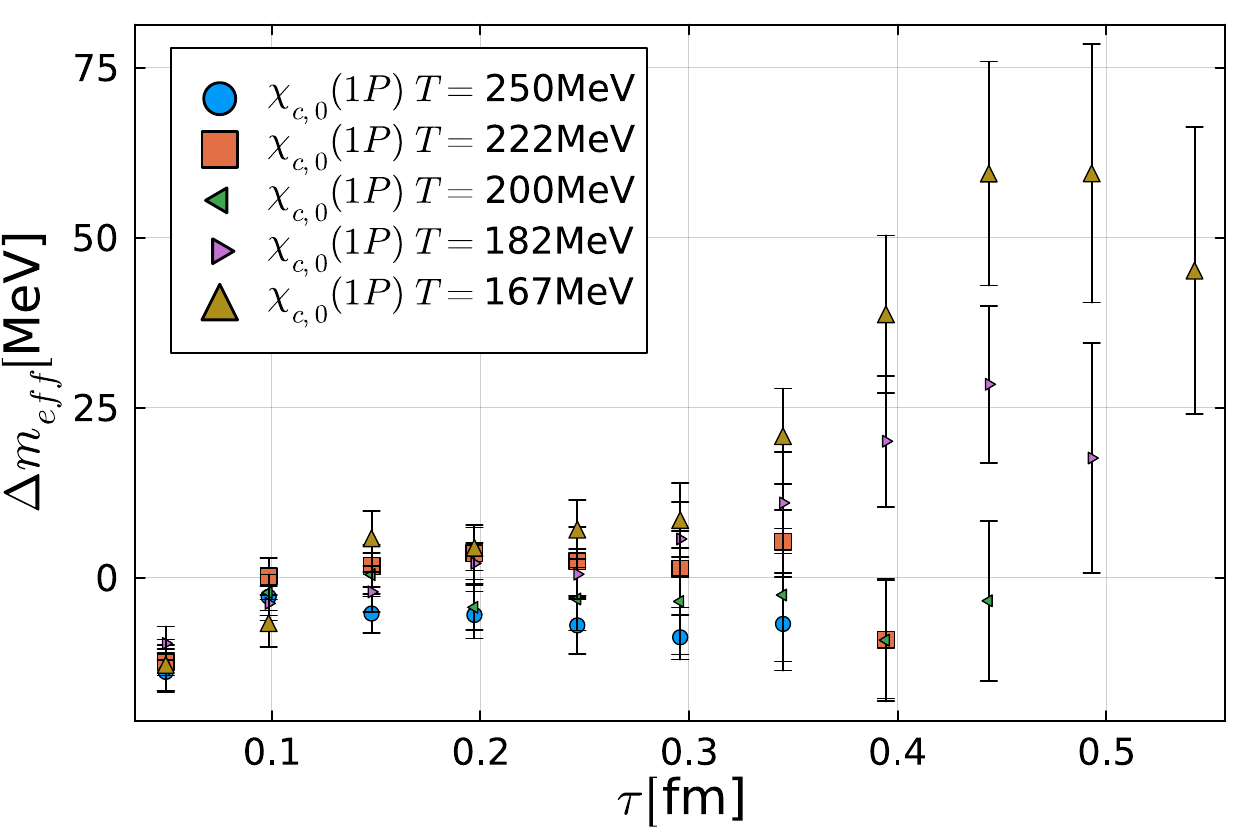}
    \includegraphics[width=0.45\textwidth]{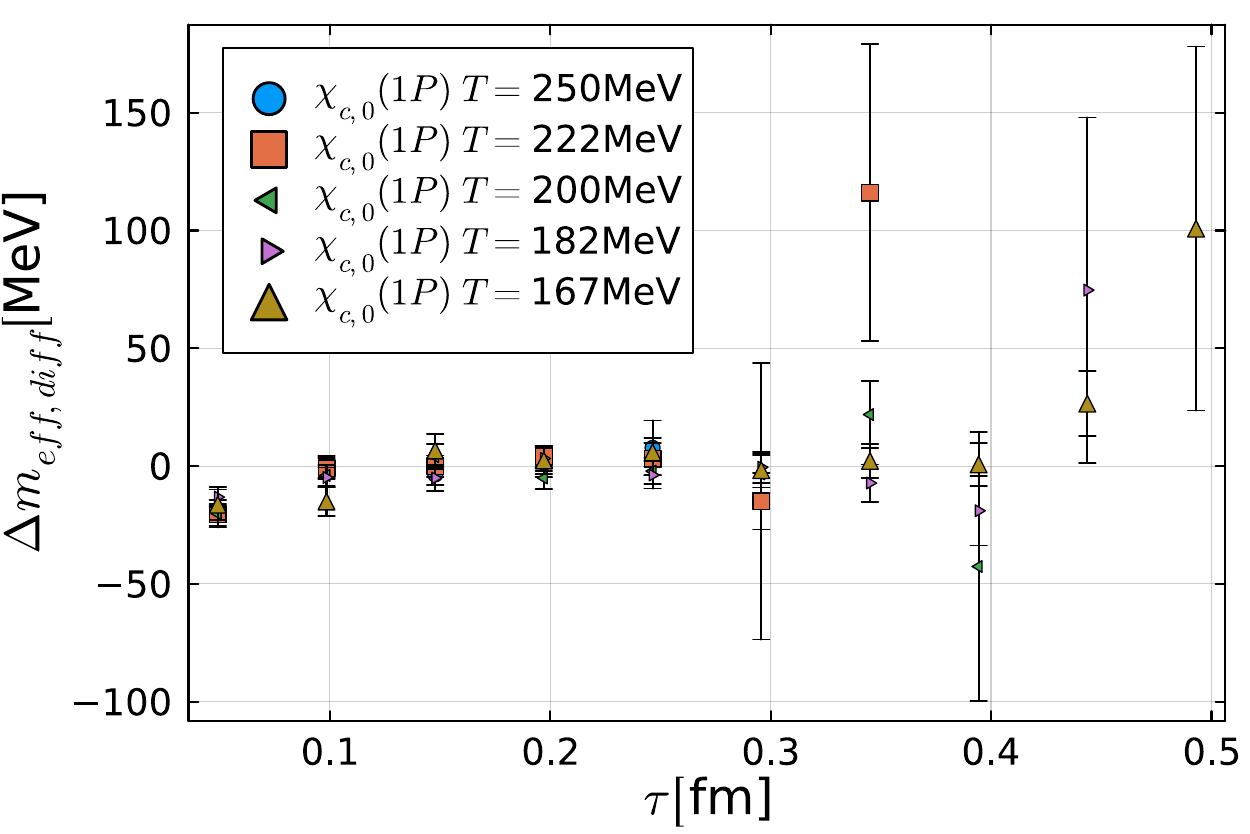}
    \caption{Difference in effective mass between volume $N_x=96$ and $N_x=64$ for $\beta = 7.596$, $a=0.0493fm$. The bottom right plot is done using the difference in the correlator, in order to remove the effect of the zero mode, which for the P-states, is slightly enhanced for $N_x=96$.}
    \label{fig:volume_dependence}
\end{figure}

The spatial lattice sizes used in this paper are quite small and one may wonder to what extent this influences the in-medium properties of various 
charmonia states, especially for the 2S states. As seen in Fig. \ref{fig:wave_sources}, the 2S wave function used as a source, only barely fits inside the used volume.  To test this, we evaluated the finite temperature correlators on a set of configurations for $\beta = 7.596$, which have all the same parameters, but the spatial size of $N_x=96$ instead of $N_x=64$. The parameters of these calculations are shown in Tab. \ref{tab:param_Ns96}. To quantify the volume effects on the correlation functions we consider
the difference in the effective mass $\Delta m_{eff}(\tau,T)=m_{eff}^{N_x=96}(\tau,T)-m_{eff}^{N_x=64}(\tau,T)$. These differences are shown in Fig. \ref{fig:volume_dependence}
for wave function optimized meson correlation functions at different temperatures and for different charmonium states. Except for the smallest two values of $\tau$, 
$\Delta m_{eff}(\tau,T)$ is zero within errors for S-wave charmonia. For P-wave charmonia we find that $\Delta m_{eff}(\tau,T)$ is non-zero for $\tau>0.3$ fm. We think
this is due to the volume dependence of the zero mode contribution. To test this assertion, we calculated the $\Delta m_{eff}(\tau,T)$ using the effective masses
from the $\tau$ derivatives of the P-wave correlator. We find that the corresponding $\Delta m_{eff}(\tau,T)$ is compatible with zero within errors. The largest deviations
from zero are at 1 $\sigma$ level or slightly larger. This is shown in the bottom right panel of Fig. \ref{fig:volume_dependence}. 
Thus the zero mode contribution to the P-wave charmonia correlators may be affected by the finite volume effects within our current precision
we do no see a significant finite volume effect on the width and mass of the P-wave charmomia.

The deviations of $\Delta m_{eff}(\tau,T)$ from zero at the smallest two $\tau$ values appear to be temperature independent and thus are not related to the volume
dependence of in-medium charmonia properties. Instead these reflect the volume dependence of the wave function optimized meson source due to limited precision of
the Coulomb gauge fixing. The precision of gauge fixing has an effect on the shape of the correlation function at small $\tau$ already at zero temperature. We use
the same relative precision for gauge fixing for both $N_x=96$ and $N_x=64$ lattices, but to obtain the same correlation function at small $\tau$ one needs to
increase the precision of gauge fixing on larger lattices. This was observed earlier on in the case of static meson correlation functions in Coulomb gauge \cite{Bazavov:2023dci}. We do not expect to see any significant volume effect in the charmonium correlator of Gaussian smeared meson operators
at small $\tau$. Therefore, 
we calculated $\Delta m_{eff}(\tau,T)$ for the charmonium correlator of Gaussian smeared meson operators for $\lambda=7$ for $T=250$ MeV.
We find that $\Delta m_{eff}(\tau,T)$ is zero within errors for all $\tau$ in the case of charmonium correlator of Gaussian smeared meson operators.
\begin{table}[H]
\begin{tabular}{ |c|c|c|c|c|}
\hline
\multicolumn{5}{|c|}{$\beta=7.596,~a=0.0493$ fm, $N_x=96$ } \\
 \multicolumn{5}{|c|}{$am_c=0.2285,c_{sw}=1.030944$}\\
\hline
 $N_\tau$ & T[MeV] & $\#$conf & $\#$sources & op.\\
 \hline
 16 & 250 & 4184 & 16 & wf. (7,150),(10,300) \\
 \hline
18 & 222 & 3423 & 16 & wf.\\
 \hline
20 & 200 &  2783 & 16& wf.\\
 \hline
22 & 182 & 2085  & 16 & wf.\\
 \hline
 24 & 167 & 1630 & 16 & wf.\\
\hline
\end{tabular}
\caption{
The type and amount of measurements done for the meson operators at a larger volume.
We use wave function optimized meson operators denoted as wf., and Gaussian meson operators. The latter are labeled by numbers $(\lambda,N)$ with
$N$ being the number of iterations and $\lambda$ being the source size
in lattice units, see text.}
\label{tab:param_Ns96}
\end{table}
Furthermore, to verify that the deviations from zero in $\Delta m_{eff}(\tau,T)$ at the smallest two $\tau$ values seen in Fig. \ref{fig:volume_dependence} for
correlators of wave function optimized correlators are present already at zero temperature, we also calculated
the corresponding effective masses of different charmonium states on $80^4$ lattices with $a=0.0493$ fm. We have found that the differences in the effective masses obtained on $80^4$ and
$64^4$  lattices for $\tau<0.15$ fm 
are similar to those seen in Fig. \ref{fig:volume_dependence} for $\tau<0.15$ fm and consistent with zero for larger $\tau$.

\section{In-medium charmonium properties }

In the previous section we have seen that the subtracted charmonium correlation functions can be described well using a cut Lorentzian spectral
function  plus a zero mode contribution given by Eq.  (\ref{eq:sigma_med_Ansatz}) for 153 MeV $\le T \le$ 305 MeV. We interpret the peak position $M_{\alpha}$ and the width parameter $\Gamma_{\alpha}^0$ obtained from
the corresponding fits to the subtracted charmonium correlators as in-medium charmonium  masses and width. This interpretation tacitly assumes that all charmonium states
can exist in this temperature region.

\begin{figure}[H]
  \centering
 \includegraphics[scale=0.34]{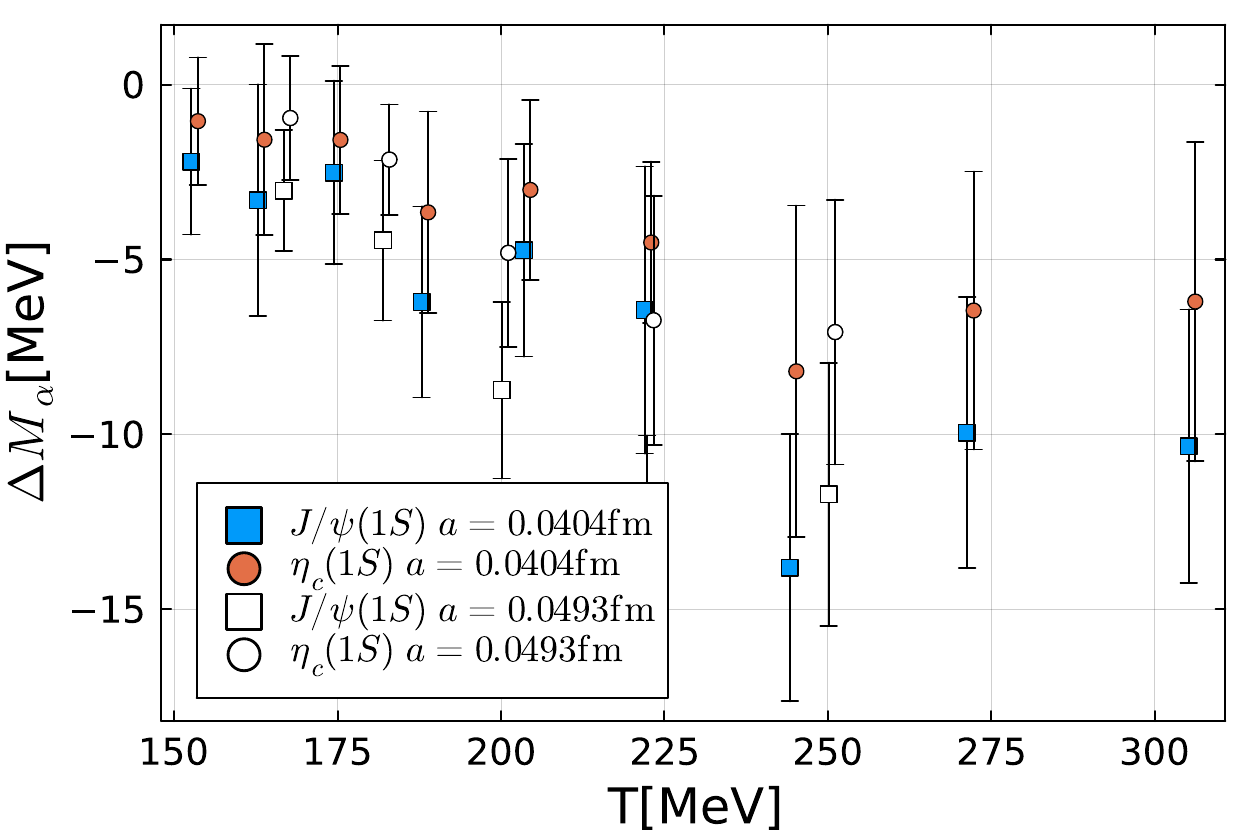}
  \includegraphics[scale=0.34]{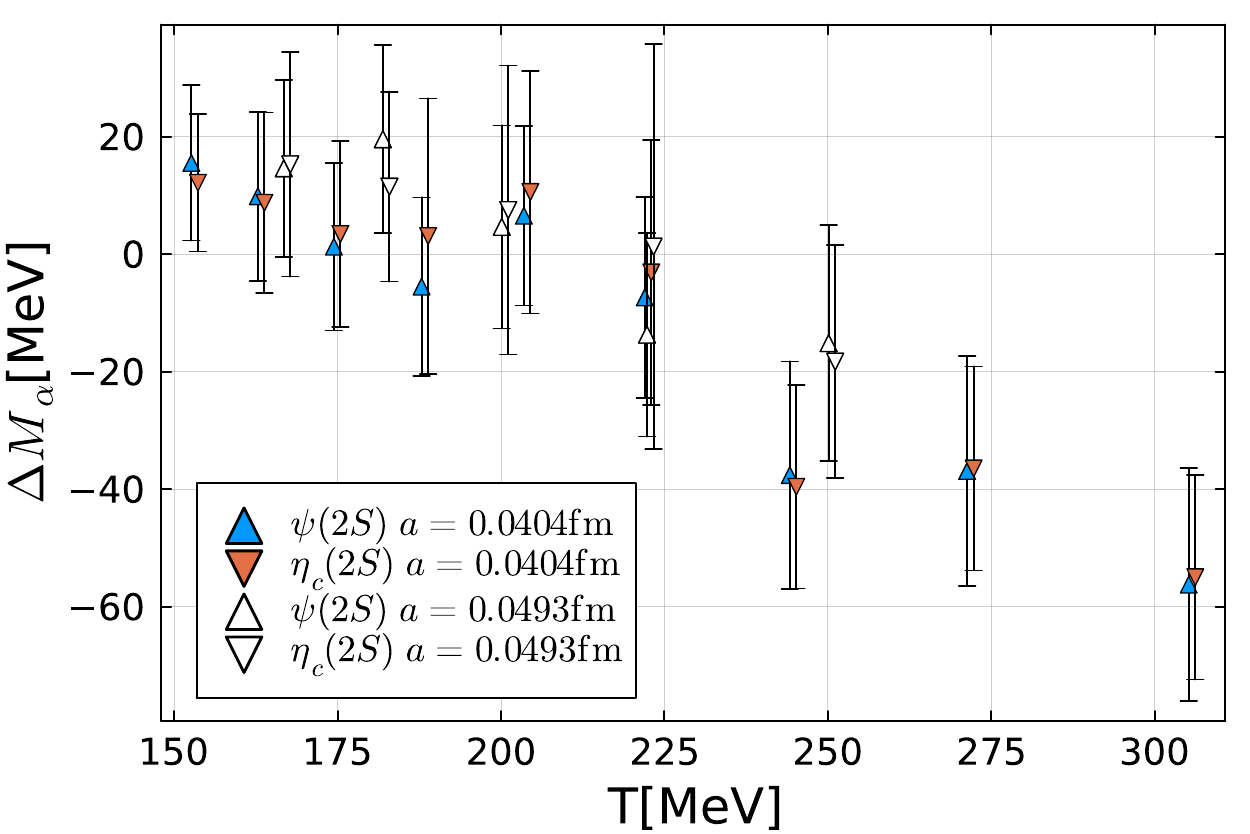}
  \label{fig:sub1}
\caption{The in-medium mass shift of 1S (left) and 2S (right) charmonium states as a function of the temperature at two different lattice spacings. Pseudo-scalar results shifted by 1 MeV to the right for better visualization. }
\label{fig:Deltam_S}
\end{figure}

\begin{figure}[H]
  \centering
  \includegraphics[scale=0.5]{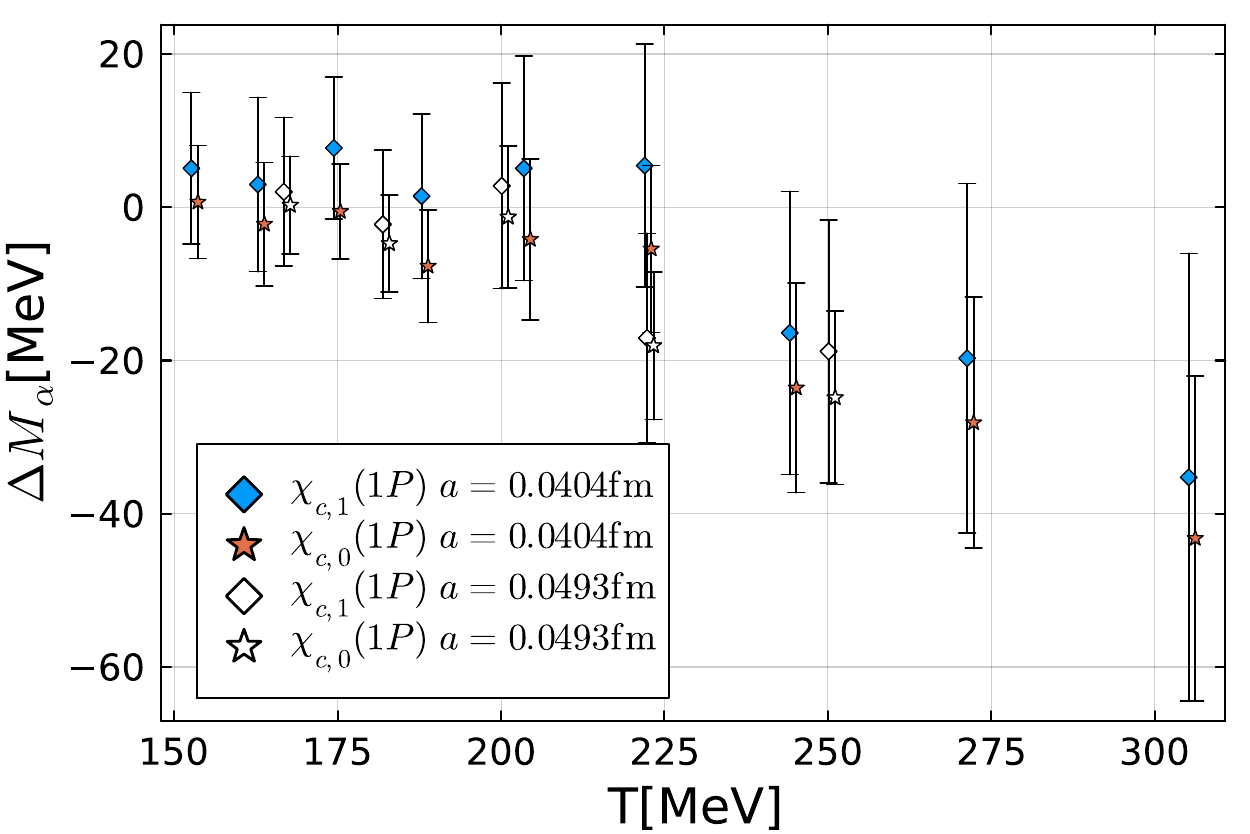}
\caption{The in-medium mass shift of P-wave charmonium states as a function of the temperature at two different lattice spacings. $\chi_{c,0}$ results shifted by 1 MeV to the right for better visualization. }
\label{fig:Deltam_P}
\end{figure}

The in-medium charmonium masses can be studied in terms of the thermal mass shift $\Delta M_{\alpha}(T)=M_{\alpha}(T)-M_{\alpha}(T=0)$ with index $\alpha$ labeling different
charmonium states. Our results for $\Delta M_{\alpha}(T)$ of 1S and 2S charmonia states obtained using wave function optimized meson operators and $cut_{\alpha}=4 \Gamma_{\alpha}^0$
are shown in Fig. \ref{fig:Deltam_S}. 
The errors shown in the figure combine the statistical errors and the systematic errors, which are 
discussed in the Appendix.
For $T<200$ MeV the in-medium mass shift is small and, in many cases, compatible with zero. But for higher temperatures it is significant, especially for the 2S states.
The overall magnitude of the mass shift is smaller than 15 MeV for 1S state but could be as large as $60$ MeV for 2S charmonia. Our results of the $\eta_c(1S)$ mass shift are
compatible with the ones obtained in the lattice study of Ref. \cite{Ali:2025iux}. From Fig. \ref{fig:Deltam_S} we also see that within errors
the mass shift does not depend on the lattice spacing.
For 1S charmonia we also estimated  $\Delta M_{\alpha}(T)$ for extended operators with Gaussian smearing for different smearing radii as discussed in the Appendix.
We find that in most cases
the peak positions are the same as the ones obtained using wave function optimized correlators within estimated errors. In two instances we see differences in the peak
positions at 0.3\% level with statistical significance of $3 \sigma$. 
Thus, the peak positions are to a large extent 
independent of the choice
of the meson operators. We also performed fits using the cut Lorentzian form with $cut=6 \Gamma_{\alpha} ^0$ discussed in the Appendix and found that the peak position is the same within errors.
Therefore, the peak positions are robust and these 
peaks correspond to the 1S and 2S charmonia state in the deconfined medium.
In Fig. \ref{fig:Deltam_P} we show the thermal mass shift for P-wave charmonia obtained using wave function optimized meson sources and $cut_{\alpha}=4 \Gamma_{\alpha}^0$. Again the errors are a combination
of the statistical and systematic errors.
There is no significant lattice dependence of the mass shift within the estimated errors.
Again, the mass shift is small for $T<200$ MeV but is significant at larger temperatures, reaching up to 40 MeV at the highest temperature.
We also 
obtained the peak position from the P-wave charmonia from the correlators of extended meson operators with Gaussian smearing as discussed
in the Appendix. We find that the peak position obtained 
through fits to these operators agrees with the above results. Furthermore, we also performed
fits of the wave function optimized operator using $cut_{\alpha}=6 \Gamma_{\alpha}^0$. These fits result in peak positions that are consistent with the ones for
$cut_{\alpha}=4 \Gamma_{\alpha}^0$. Therefore, also the in-medium masses of the P-charmonia are robustly determined in our analysis. 

\begin{figure}[H]
  \centering
 \includegraphics[scale=0.5]{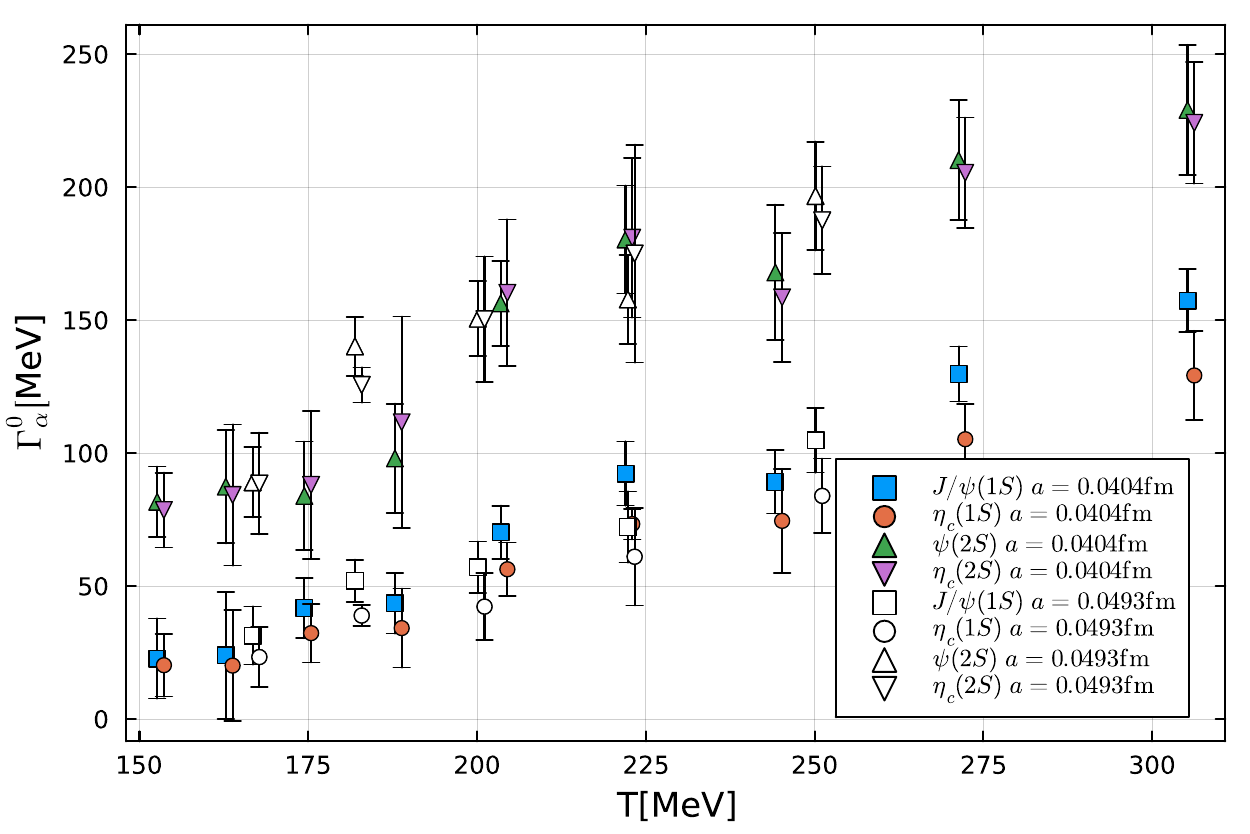}
\caption{The extracted width $\Gamma_{\alpha} ^0$ from Lorentzian fits cut off at 4 times the width on the subtracted correlator for 1S and 2S wave function sources. Errors are a combination of statistical and systematics from 2 fits for 2 different subtraction parameters, and for pseudoscalar, with or without including a zero mode, for a total of 4 in that channel. Pseudo-scalar results shifted by 1 MeV to the right for better visualization.}
\label{fig:width_S}
\end{figure}

\begin{figure}[H]
  \centering
 \includegraphics[scale=0.5]{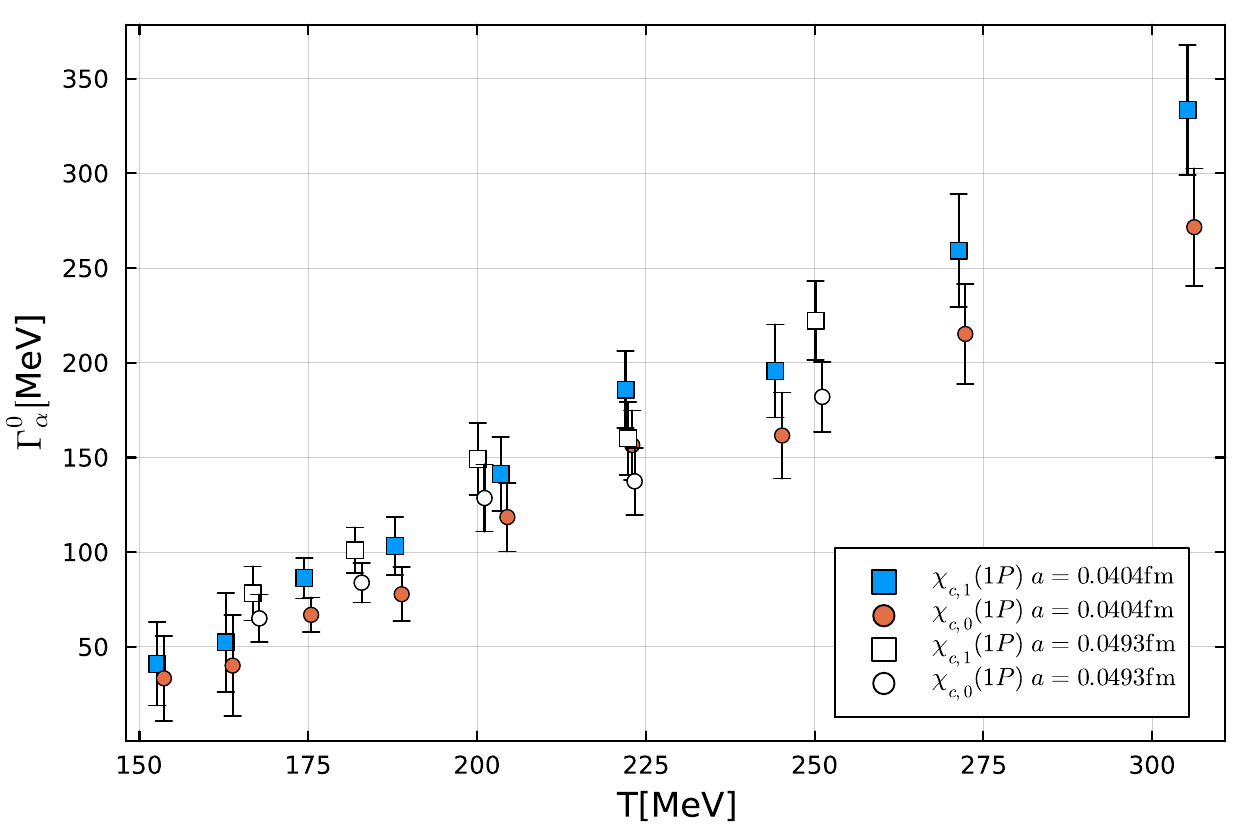}
\caption{The extracted width $\Gamma_{\alpha} ^0$ from Lorentzian fits cut off at 4 times the width on the difference of the subtracted correlator for 1S wave function source for the P-states. Errors are a combination of statistical and systematics from 2 fits for 2 different subtraction parameters. $\chi_{c,0}$ results shifted by 1 MeV to the right for better visualization. }
\label{fig:width_P}
\end{figure}

The thermal width of S-wave charmonia is shown in Fig. \ref{fig:width_S} as a function of temperature obtained from the wave function optimized operators and 
$cut_{\alpha}=4 \Gamma_{\alpha}^0$. We see a clear increase in the thermal width with the increasing temperature. 
The results for the thermal width obtained with two different lattice spacings more or less agree with each other within the estimated errors. The thermal width of the 2S states
is larger than for the 1S states in accordance with general expectations: the larger 2S states are more affected by the hot medium than the more tightly bound 1S state.
We also see that the different spin states, $J/\psi$ and $\eta_c(1S)$, and $\psi(2S)$ and $\eta_c(2S)$ have widths that agree within errors. This is expected for
bound states of heavy quarks, where spin-dependent interactions are suppressed. Our estimate of the width shown in Fig. \ref{fig:width_S} for $\eta_c(1S)$ is about factor two smaller
than the one obtained in Ref. \cite{Ali:2025iux} at comparable temperatures. Note that our definition of the width parameter is twice smaller than used in Refs. \cite{Ali:2025iux}.
Recently the thermal width of charmonium states has been estimated in the T-matrix approach \cite{Wu:2025hlf}. Our results for 1S charmonia width in the temperature interval 200 MeV $<T<$ 305 MeV 
agree with the findings of the T-matrix approach \cite{Wu:2025hlf}.  However, we find smaller width for the 2S charmonia in the same temperature interval compared to the T-matrix approach.
Interestingly, the in-medium width of 1S charmonia is not very different from the in-medium width of 1P bottomonia obtained in NRQCD \cite{Ding:2025fvo}. The 1P bottomonia
and 1S charmonia have similar sizes suggesting that the magnitude of the in-medium quarkonium width correlates with the quarakonium size, as intuitively expected.

Our results for the thermal width of the 1P charmonia obtained from the correlation functions of wave function optimized operators and $cut_{\alpha}=4 \Gamma_{\alpha}^0$
are shown in Fig. \ref{fig:width_P}. As in the case of S-wave charmonia we see a clear increase of the width with increasing temperature and the results obtained at two
different lattice spacings agree with each other. Furthermore, there is no statistically significant difference in the thermal width of the $\chi_{c1}$ state 
$\chi_{c0}$ state, which is what we expect for heavy quark bound states. Our results are in rough agreement with the results obtained for 1P charmonia in the T-matrix approach \cite{Wu:2025hlf}.
We also see that the width of 1P charmonia is similar in magnitude to the width of 3S bottomonia obtained in NRQCD \cite{Ding:2025fvo}. This again supports the idea
that the thermal quarkonium width is correlated with its size.

So far we have discussed the thermal width of charmonium states obtained from wave function optimized operators. As mentioned in the previous section, the effective masses
shows some dependence on the type of meson operators used in the analysis. For extended meson operators with Gaussian smearing, the slope of the effective mass is larger
for operators of large size, as discussed in the Appendix. This translates to larger charmonia width if the same value of $cut_{\alpha}$ is used.
We find that for 1S charmonia this effect could be as large as 25\%, while for 1P charmonia this effect is about 40-60\% depending on the temperature. 
We present the corresponding analysis in the Appendix. 
We also tested the sensitivity of the thermal width to the choice of the parameter $cut_{\alpha}$ by performing fits with $cut_{\alpha}=6 \Gamma_{\alpha}^0$, which are
discussed in the Appendix. We find that these fits give thermal widths that are 22\% smaller than the ones obtained for $cut_{\alpha}=4 \Gamma_{\alpha}^0$.

\section{Conclusion}
In this paper we studied charmonium correlation functions of extended meson correlators with the aim to constrain the properties of charmonia states
in the temperature interval 153 MeV $\le T \le $ 305 MeV. We performed calculations at two different lattice spacings, $a=0.0493$ fm and $a=0.0404$ fm and found
no significant lattice spacing dependence. Furthermore, we also checked that the finite volume effects have no significant impact on the in-medium properties of charmonium states
within present errors.

We have found that the behavior of these charmonia correlators is consistent with a spectral function
that has a dominant peak corresponding to 1S, 2S or 1P charmonia. The corresponding peak positions and widths can be interpreted as the charmonium masses and widths
in the hot QCD medium. Using different meson operators and different shapes of the spectral function we showed that the determination of the in-medium
masses is robust. For $T<200$ MeV the in-medium charmonium masses are not very different from the corresponding vacuum masses. At higher temperatures 
we see a downward shift of the charmonium masses, which is at most $10$ MeV for 1S charmonia but could be as large as $40-60$ MeV for 1P and 2S charmonia.

Our analysis shows that the thermal width of different charmonium states increases with increasing temperature. The thermal width of 1P and 2S charmonia is larger than for 1S charmonia as expected. Qualitatively the temperature
dependence of the thermal width is similar to that of bottomonia states 
\cite{Ding:2025fvo,Larsen:2019zqv,Larsen:2019bwy}.
While the temperature dependence of the width is quite robust, the size of the width depends somewhat
on the precise form of the spectral peak and the type of extended operator. 
For the largest temperature of $305$ MeV the width can be as large as 300 MeV.
This is comparable to the  level splitting, i.e. difference in the masses of  charmonia states. At this temperature the interpretation of the dominant peak
as the in-medium charmonium state may be questionable and further studies using correlators of point meson operators may be needed. In fact, the study of
the charmonium spatial meson correlation function indicates that all charmonium states may be melted for $T>300$ MeV \cite{Karsch:2012na, Bazavov:2014cta}.
For bottomonium a similar study of the spatial meson correlation function \cite{Petreczky:2021zmz} indicates melting of the bottomonium state for $T>500$ MeV.
Thus our interpretation of the peaks as charmonium states is only justified for temperatures of about 300 MeV or lower.

\section*{Acknowledgments}

This material is based on work supported by the U.S. Department of Energy, Office of Science, Office of Nuclear Physics under Contract No. DE-SC0012704  and the Topical Collaboration in Nuclear Theory "Heavy-Flavor Theory (HEFTY) for QCD Matter".

R.N.L. was supported by the Ministry of Culture and Science of the State of Northrhine Westphalia (MKW NRW) under the funding code NW21-024-A (NRW-FAIR). R.N.L. acknowledge support from the Deutsche Forschungsgemeinschaft (DFG)
through the CRC-TR 211 “Strong-interaction matter under extreme conditions” (Project No.
315477589 - TRR 211).

This research used computing time provided by the ALCC and INCITE programs at the Oak Ridge Leadership Computing Facility, a DOE Office of Science User Facility supported under contract no. DE-AC05-00OR22725; the National Energy Research Scientific Computing Center (NERSC), a DOE Office of Science User Facility at Lawrence Berkeley National Laboratory under Contract No. DE-AC02-05CH11231; PRACE awards on JUWELS at GCS@FZJ, Germany; Marconi100 at CINECA, Italy; and EuroHPC JU award for the project ID EHPC-EXT-2024E01-039 access to the LUMI-G supercomputer at the Finnish IT Center for Science (CSC). Part of the computations in this work were performed on the GPU cluster at Bielefeld University. We thank the Bielefeld HPC.NRW team for their support.

\section*{Appendix}
\appendix
In this appendix we discuss some technical 
aspects of the calculations, including the tuning
of the charm quark mass parameter, details of the fits
of the spectral functions and the determination of the
in-medium charmonium masses and widths.

\section{Determination of the charm quark mass parameter}
\begin{figure}[H]
 \includegraphics[scale=0.35]{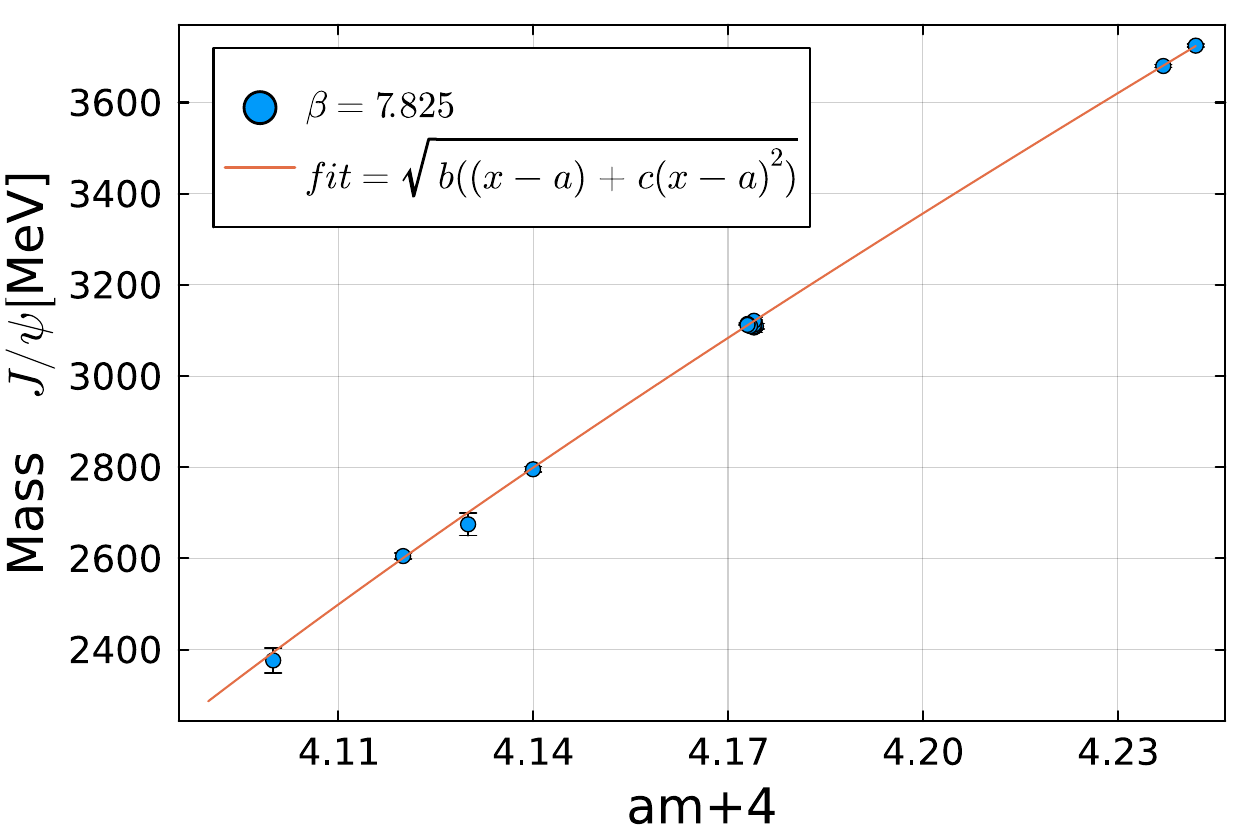}
 \includegraphics[scale=0.35]{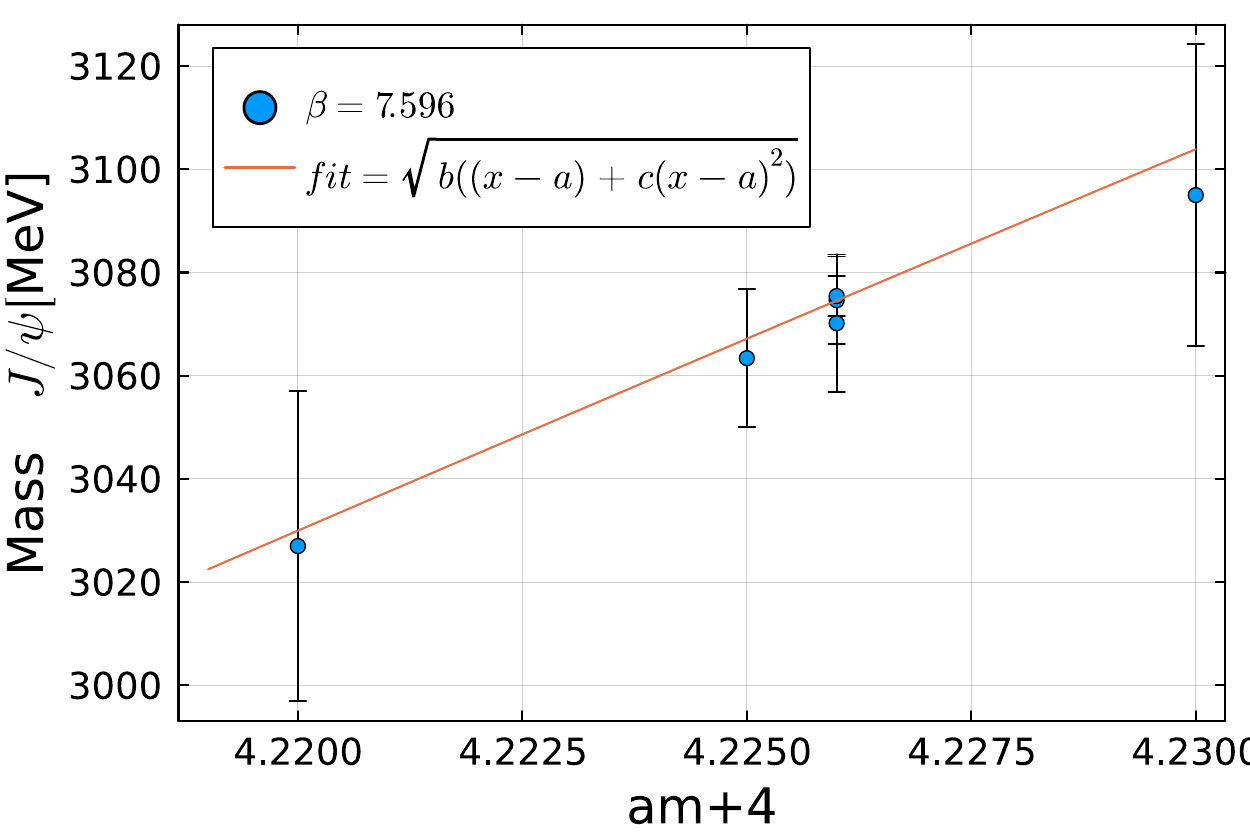}
\caption{The dependence of the $J/\psi$ mass on $4+a m_c$ for $\beta=7.825$ (left)
and $\beta=7.596$ (right).}
\label{fig:mass_tune}
\end{figure}

In the lattice QCD calculations we need to determine
the bare
charm quark mass parameter $am_c$ that corresponds
to the physical value of the charm quark mass.
The parameter $am_c$ is
related to the usual hopping parameter in the 
Wilson fermion action $\kappa=1/(2(4+am_c))$.
To do this, we calculate the zero temperature mass of $J /\psi$ at different input masses to find which mass parameter corresponds to the physical mass of $3096.8 MeV$. The results were fitted with the square root of a second order polynomial and then used to set the mass parameter for each beta value. The fits are shown in 
Fig. \ref{fig:mass_tune}. The final values of the $J/\psi$ have slightly shifted due to an increase in statistics compared to the initially used values. From the fits we obtain $m_{J/\psi}=3096.0 \pm 1.3$ MeV at $4+am_c=4.1712 \pm 0.0001$ for $\beta=7.825$
and $=3092.4 \pm 4.8$ MeV at $4+am_c=4.2285 \pm 0.0007$ for $\beta=7.596$.

\section{Fit details and estimating systematic effects on the in-medium charmonium properties}

\subsection{Systematic effects from UV subtraction}

Our analysis method of the spectral functions is based upon the subtracted
correlation functions as explained in the main text. The high energy part of the
correlation functions used for subtraction is obtained from the zero temperature correlation function. To determined it we subtract the contribution of the lowest
energy state from the zero temperature correlation function. This determination
is clearly very sensitive to the amplitude and the mass of the lowest lying state
obtained from exponential fits.
We have performed all fits on 2 different sets of subtracted correlator, where the range used to extract the plateau for zero temperature is different. The ranges
were chosen to give a good fit for a plateau for zero temperature, with one being as large a possible range with a good fit, while the other was a small range instead.

\begin{figure}
    \centering
    \includegraphics[width=0.45\textwidth]{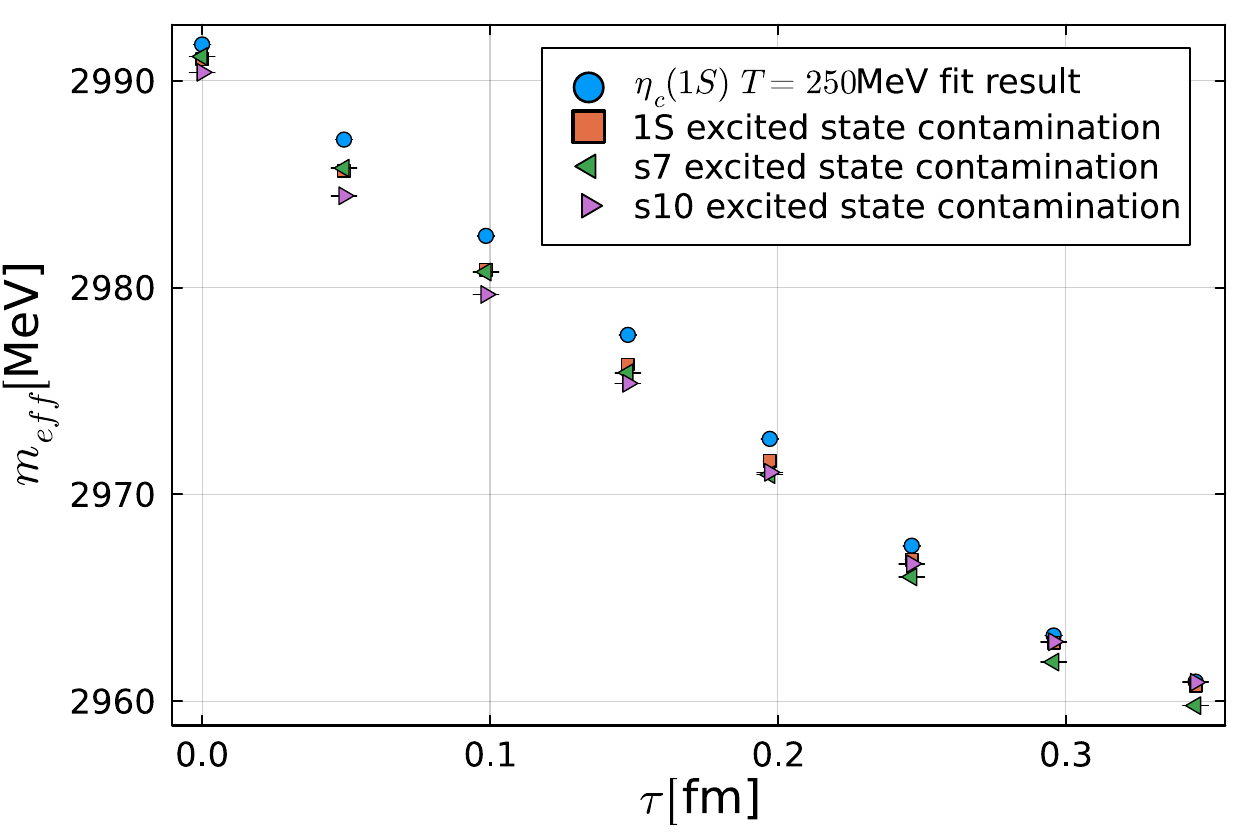}
    \includegraphics[width=0.45\textwidth]{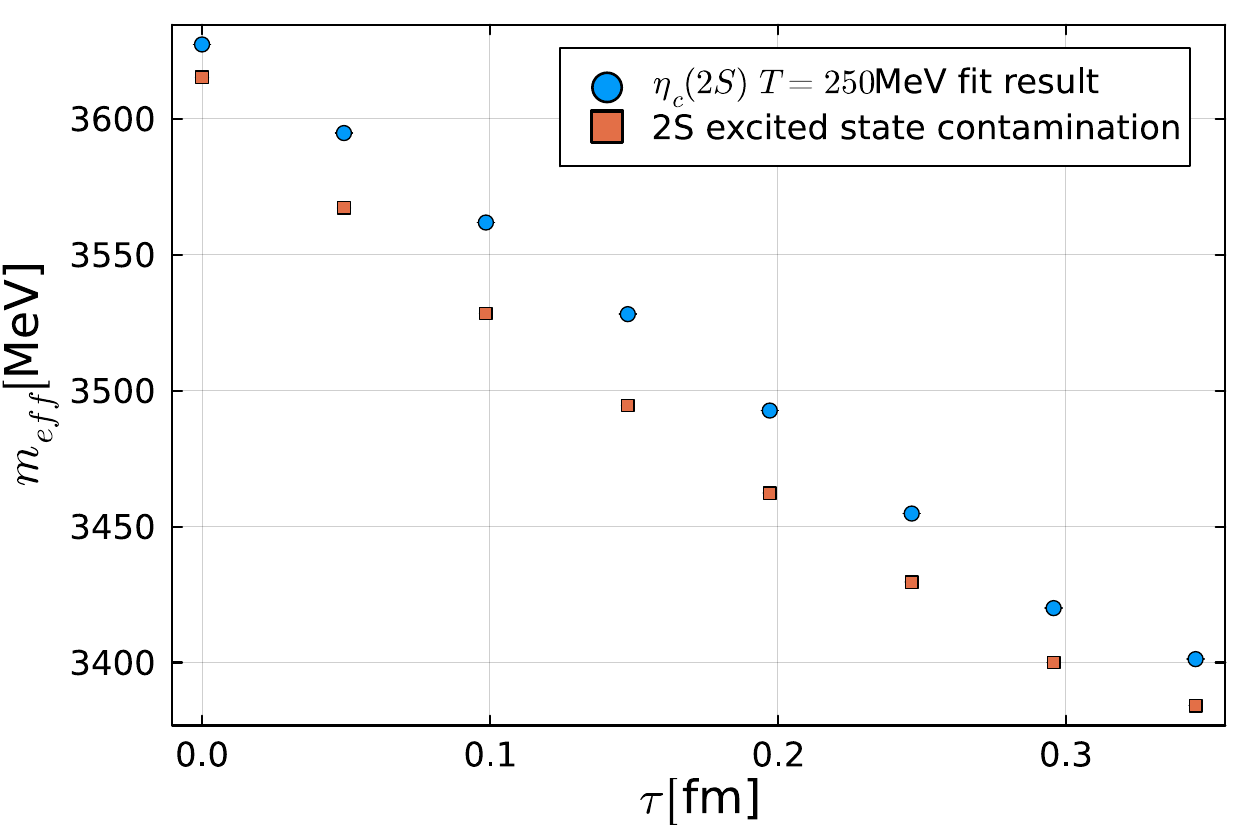}
\caption{
The effective mass of a cut Lorentzian for $\eta_c(1S)$ (left) and $\eta_c(2S)$ (right) for $a=0.0493$ fm and $T=250$ MeV obtained from fits to the subtracted correlator described in the main text and shown as blue circles compared to the predicted effect of contamination of excited states in the zero temperature subtraction method from using the 1S wave function as source (1S) or smeared Gaussian sources of size $7a$ or $10a$. The contamination from the excited states is assumed to be (left) 2 times and (right) 1.5 times larger than the lowest states' width $\Gamma_{\alpha} ^0$.} 
    \label{fig:subtraction_systematics}
\end{figure}

The subtraction of the high energy part of the correlation function from the lattice results on the finite temperature correlation functions assumes that
$\sigma_{\alpha}^{high}$ is strictly temperature independent. While largely correct, this is an oversimplification. In reality  $\sigma_{\alpha}^{high}$
will have some temperature dependence. To model this temperature dependence
we assume that $\sigma_{\alpha}^{high}$ contains a second state peak, which broadens at high temperatures, and many other higher excited states for which in-medium
modification is ignored. While there is no physically motivated reason for ignoring the temperature modifications of the higher excited states, it is reasonable to
assume that the cumulative effect of these modifications on the charmonium correlation function is small. At least this is the case in the potential model calculations
of the spectral functions \cite{Mocsy:2007yj}. In this picture the proper subtraction of the finite temperature charmonium correlation function is not just the subtraction
of $G_{\alpha}^{high}$ but also a contribution arising from the difference of the spectral function corresponding to the in-medium  broadened second state peak and the 
delta function corresponding to the vacuum second state. The in-medium peak corresponding to the second state is modeled by a cut Lorentzian with peak position equal to
the mass of this state at $T=0$ and a width which is twice the estimated width of the lowest state for 1S charmonia or one and a half time the width of the lowest state for 2S charmomonia.
Here we use $cut_{\alpha}=4 \Gamma_{\alpha}^0$. The result of such an analysis is shown in Fig. \ref{fig:subtraction_systematics} for $a=0.0493$ fm and $T=250$ MeV. In the case of 1S charmonia the mass and amplitude
of the second state at $T=0$ were determined from wave function optimized correlator as well as from correlators of Gaussian extended sources of size $\lambda=7$ and $\lambda=10$.
We see from the figure that performing the subtraction this way results in an effective mass of 1S charmonia that is shifted down by about 3 MeV with no significant change in the slope. For the 2S charmonia
we see a larger shift in the effective mass of about 20 MeV, but again no significant change in the slope.

\subsection{Quantifying the zero mode contribution}
\begin{figure}[H]
 \includegraphics[scale=0.36]{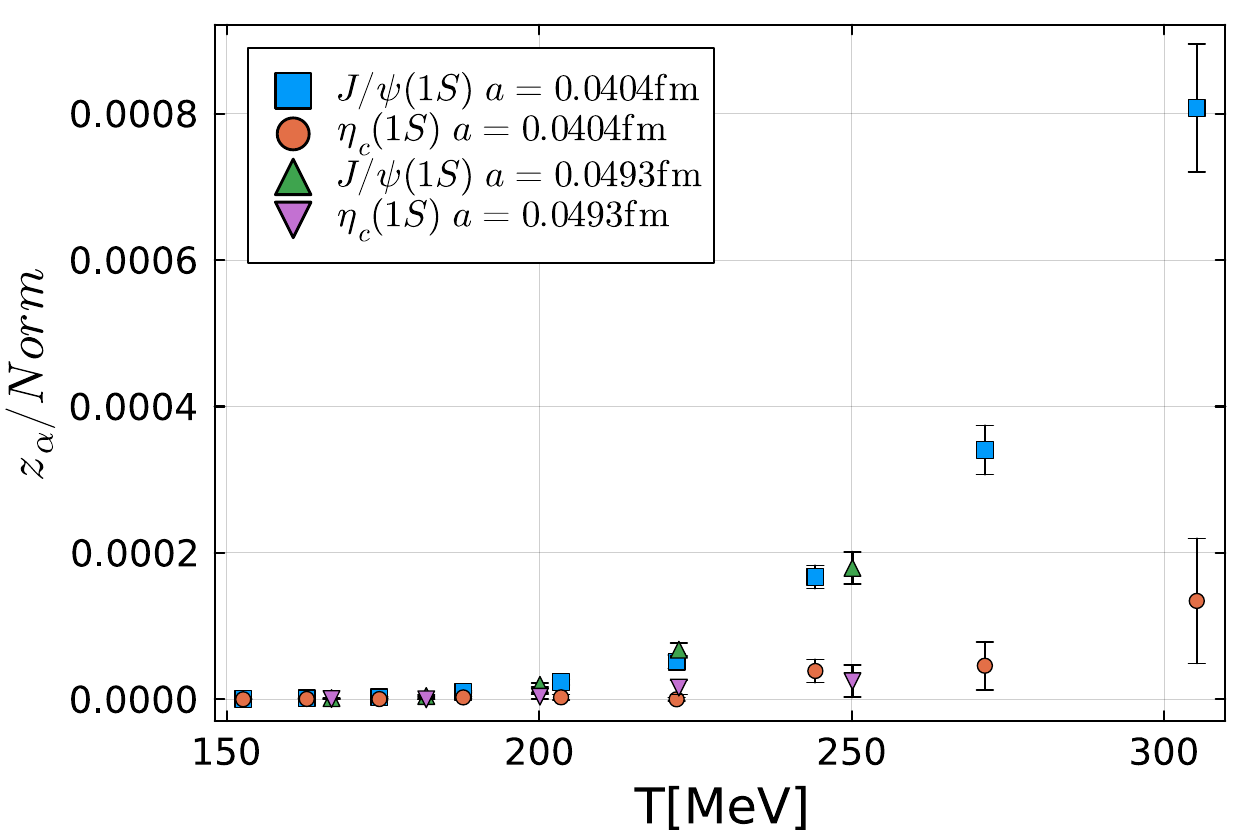}
  \includegraphics[scale=0.36]{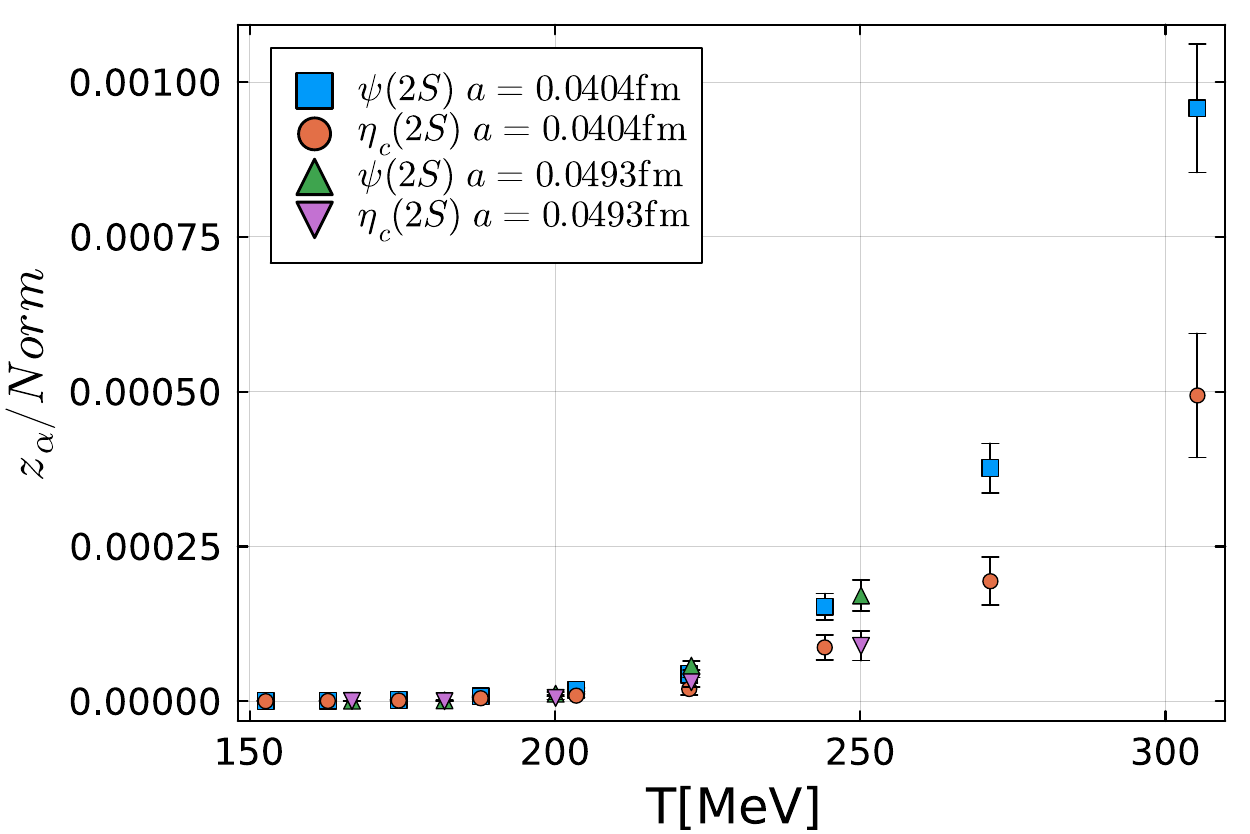}
\caption{The amplitude of the zero modes for the Lorentzian fit cut off at 4 times the width normalized by the integral over the cut Lorentzian. Errors are a combination of statistical and systematic effects from 2 different subtraction parameters. The left panel corresponds to 1S states, while the right panel corresponds to 2S states.}
\label{fig:zero_mode_ratio}
\end{figure}

In the vector channel we use Eq. \ref{eq:sigma_med_Ansatz} to fit the subtracted correlation function and treat $z_{\alpha}$
as a fit parameter. In the pseudo-scalar channel one can obtain good fits by setting $z_{\alpha}=0$. However, we also performed fits of the pseudo-scalar channel by treating $z_{\alpha}$ as a fit parameter. The relative amplitude of
the zero mode contribution is shown in Fig. \ref{fig:zero_mode_ratio}. 
We see that the relative contribution of the zero mode rapidly increases with increasing temperature.
For 1S
states the amplitude of the zero mode in the pseudo-scalar channel is much smaller
than for the vector channel and its numerical value is only marginally different
from zero. For 2S state the amplitude of the zero mode in the vector and pseudo-scalar channels is comparable. 

\subsection{Source dependence of S-wave charmonia}
\begin{figure}[H]
  \centering
 \includegraphics[scale=0.35]{pictures_new/local_mass_b7596_vector_Nt64_source_comparison_new.pdf}
  \includegraphics[scale=0.35]{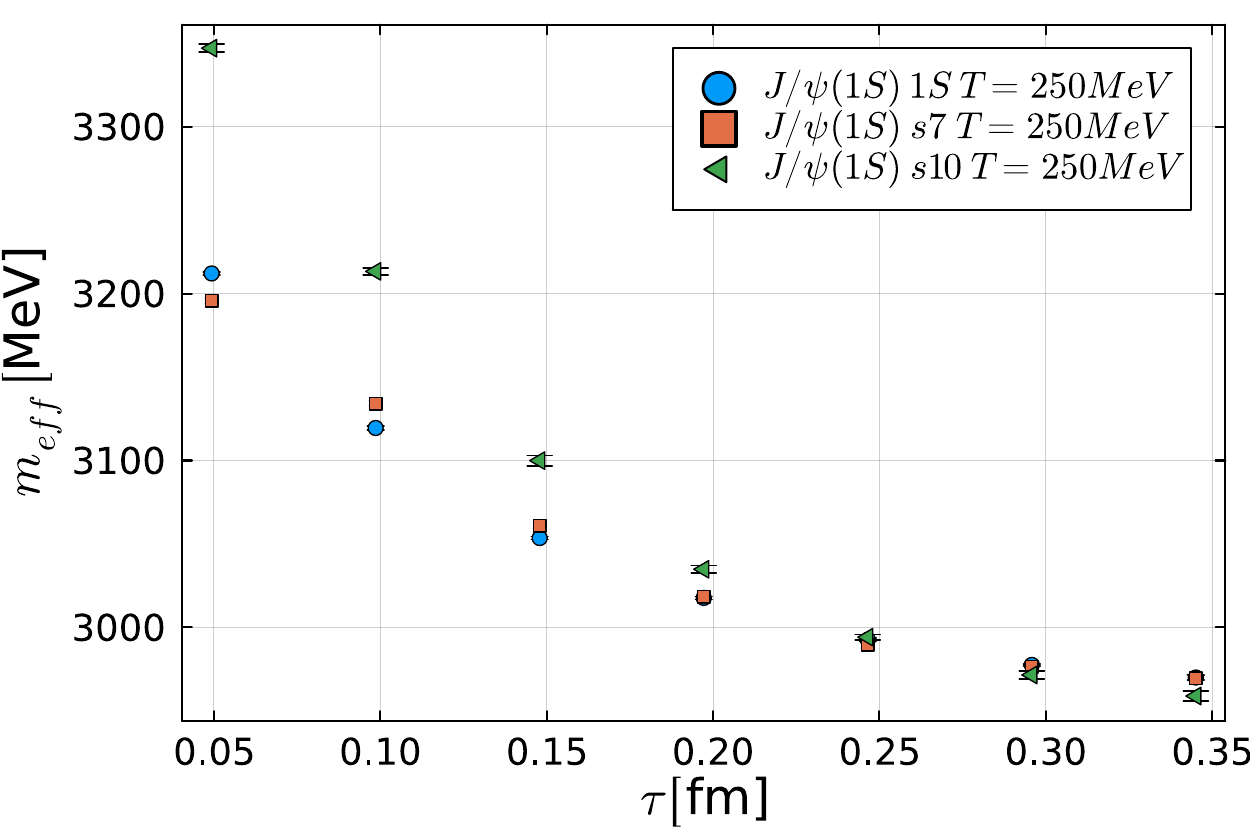}
\caption{Effective mass for different sources at zero (left) and finite (right) temperature. The first source was a 1S wave function, while the 2 other  sources are Gaussian sources  the size of 7 or 10 times the lattice spacing a, with $a=0.0493$fm.}
\label{fig:meff_comp_b7595}
\end{figure}

In this section of the Appendix we will explore the dependence of the effective masses 
and the extracted in-medium properties of 1S charmonia 
on the choice of meson operators. We use the cut Lorentzian Ansatz for the spectral function 
with $cut_{\alpha}=4 \Gamma_{\alpha}^0$ though we expect that the shape of the spectral peak may depend on
the choice of the meson operator.
For 1S state we can use either the wave function optimized operators or extended meson operators with Gaussian
smearing of different size. In Fig. \ref{fig:meff_comp_b7595} we show the effective masses of $\eta_c(1S)$
obtained using different meson operators for $a=0.0493$ fm at $T=0$ and $T=250$ MeV. For small $\tau$ the difference
in the effective masses corresponding to different operators is about the same at zero and finite temperature. At larger
$\tau$ the differences are small.

\begin{figure}[H]
  \centering
 \includegraphics[scale=0.35]{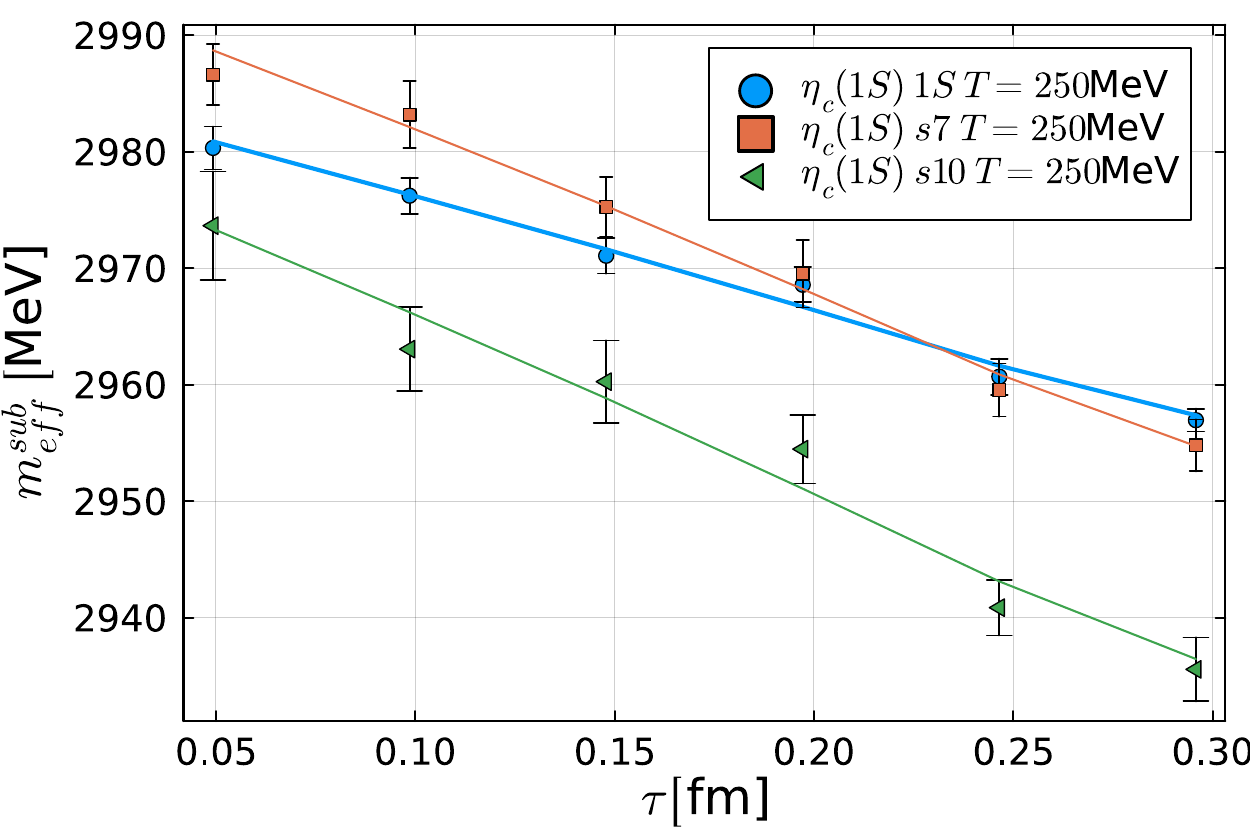}
  \includegraphics[scale=0.35]{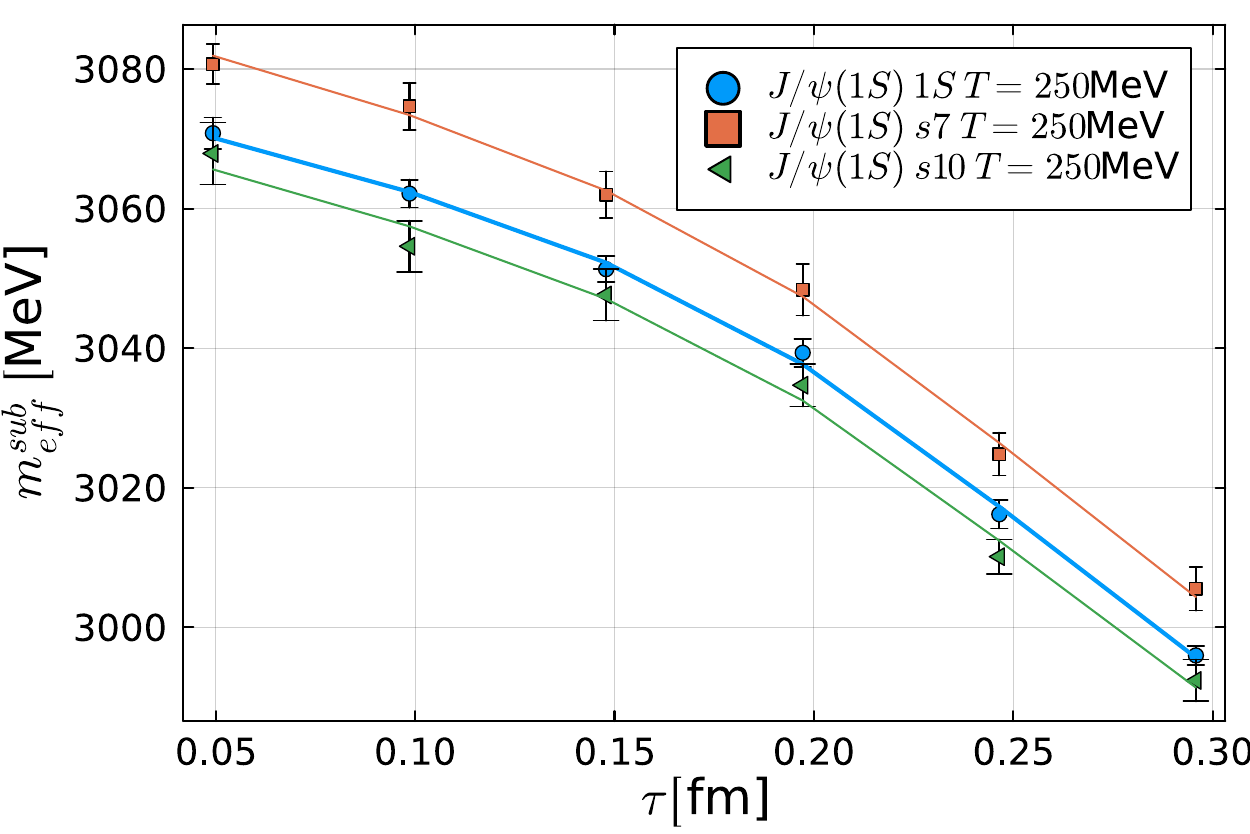}
\caption{Subtracted effective mass for different sources for pseudo-scalar (left) and vector (right) channel. The first source was a 1S wave function, while the 2 other sources are the Gaussian sources the size of 7 or 10 times the lattice spacing a, with $a=0.0493$fm.}
\label{fig:smear_comp_b7596}
\end{figure}

The subtracted effective masses for $\eta_c(1S)$ and $J/\psi$ for 
$a=0.0493$ and $T=250$ MeV are shown in Fig. \ref{fig:smear_comp_b7596} for various meson operators. We see small but visible differences in the subtracted effective masses. To quantify these differences in terms of the in-medium charmonium masses
in Tab. \ref{tab:Smear_comp_b7596} we show the extracted medium masses and width of 1S charmonia for this parameter set
obtained using different meson operators. The differences in the in-medium masses at this temperature obtained from using
different meson operators are at most 10 MeV. For the thermal width of $\eta_c(1S)$ the choice of meson operator can
lead to 25\% difference, while for $J/\psi$ the values of the thermal width corresponding to different choices of the meson operators
agree within errors. This is partly due to the fact that the errors on the subtracted $J/\psi$ correlator are larger.
Interestingly, we also see from Tab. \ref{tab:Smear_comp_b7596} that the zero mode contribution
does not depend on the choice of meson operators. 

\begin{figure}[H]
  \centering
 \includegraphics[scale=0.35]{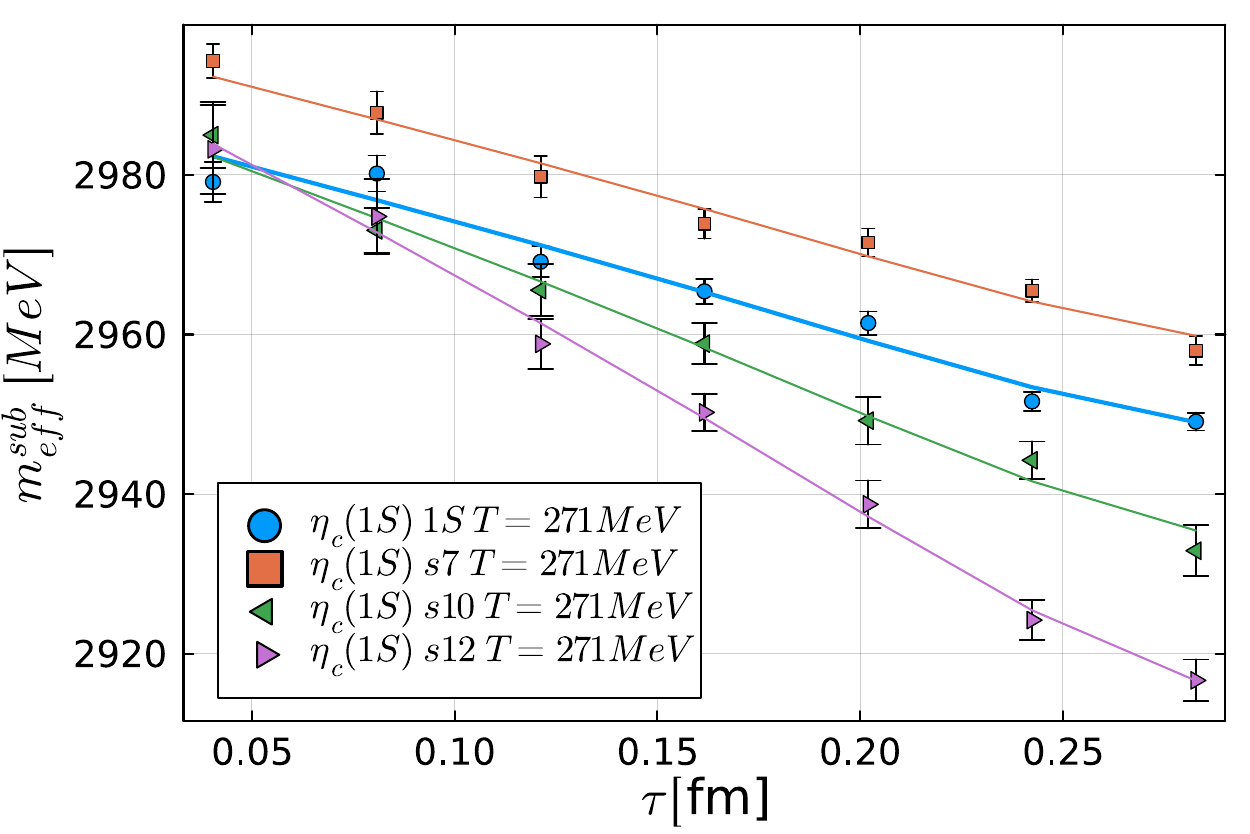}
  \includegraphics[scale=0.35]{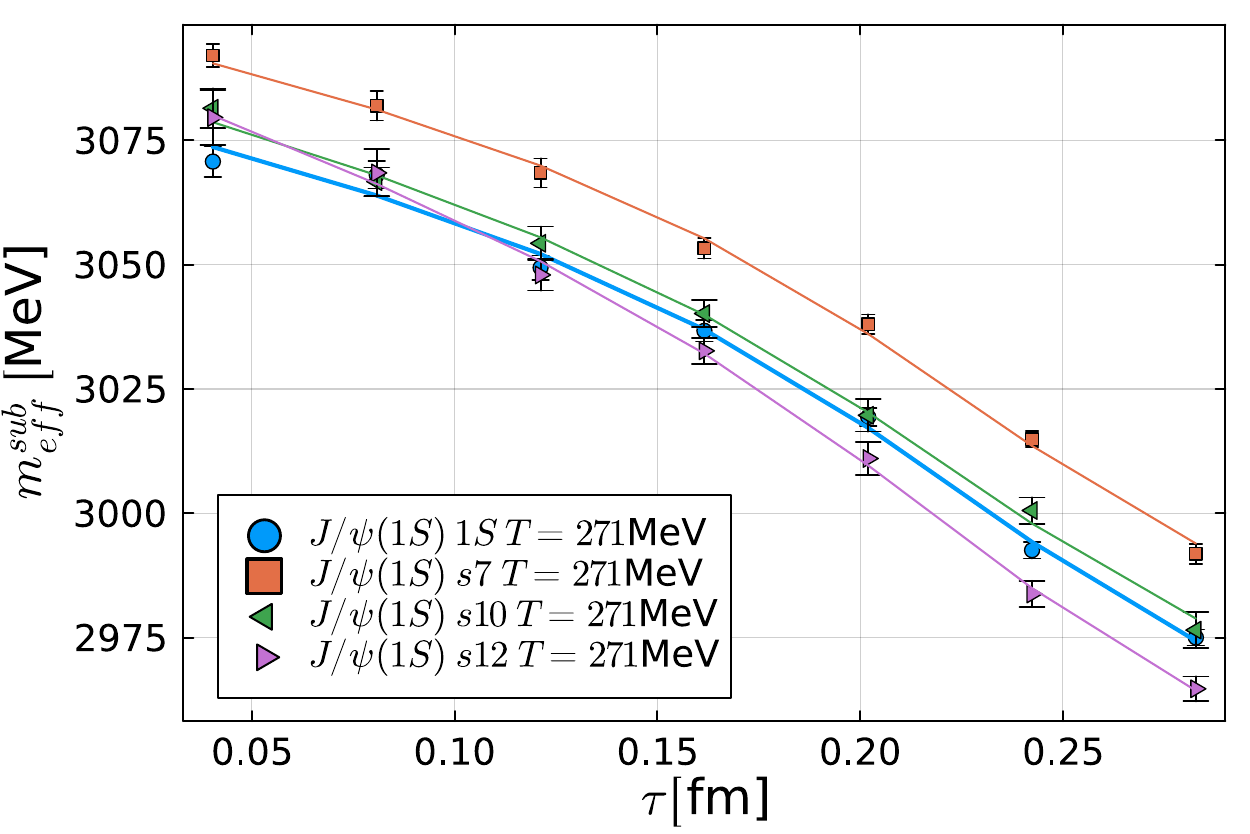}
\caption{Subtracted effective mass for different sources for pseudo-scalar (left) and vector (right) channel. The first source was a 1S wave function, while the 3 other sources are Gaussian sources of the size of 7, 10 and 12 times the lattice spacing a, with $a=0.0404$fm.   }
\label{fig:smear_comp_b7825}
\end{figure}

Similar analysis was performed for $a=0.0404$ fm and temperature $T=271$ MeV. Here we used wave function optimized
meson operator as well as meson operators with Gaussian smearing with sizes $\lambda=7,~10$ and 12. The results are shown
in Fig.  \ref{fig:smear_comp_b7825} and Tab. \ref{tab:Smear_comp_b7825}. Here the largest difference in the in-medium mass
is 16 MeV (the difference between the wave function, 1S operator and s7 Gaussian meson operator), while the largest difference in the extracted
width reaches 41\% in the case of $\eta_c$ with $s12$ meson operator. 

\begin{figure}[H]
  \centering
 \includegraphics[scale=0.35]{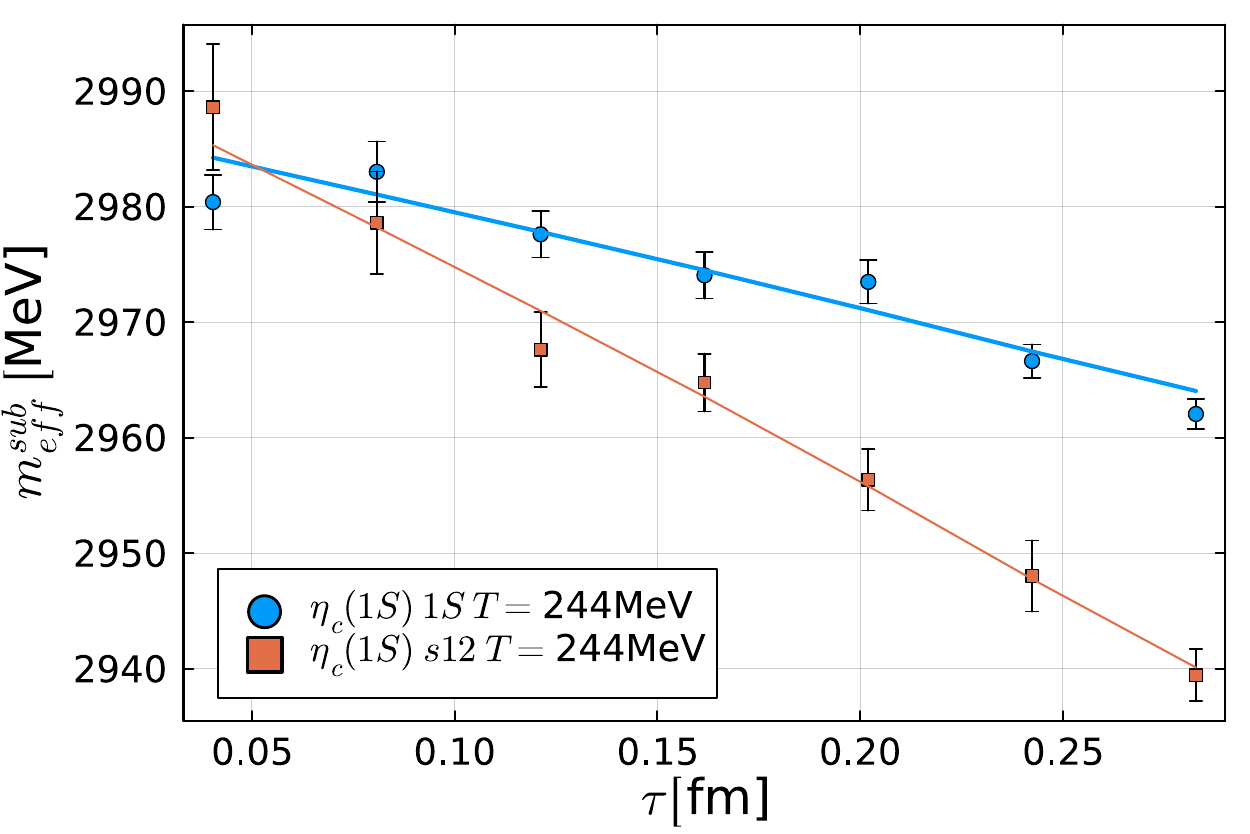}
  \includegraphics[scale=0.35]{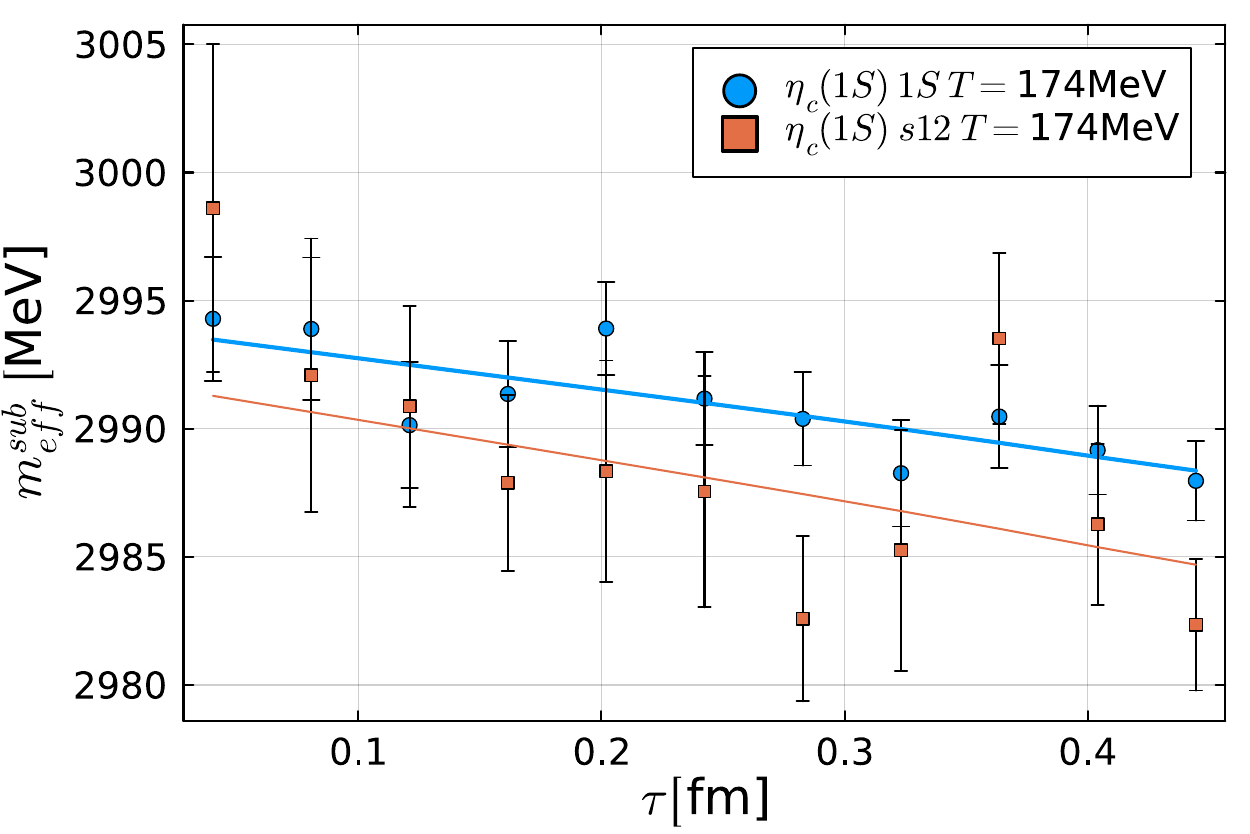}
\caption{Subtracted effective mass for different sources for pseudo-scalar for 2 lower temperatures. The first source was a 1S wave function, while the  other source is a Gaussian source of the size of 12 times the lattice spacing a, with $a=0.0404$fm.   }
\label{fig:smear_comp_b7825_Nt20_28}
\end{figure}

We expect that the differences in the obtained in-medium charmonia properties from using different meson operators
should reduce as the temperature reduces. For $a=0.0404$ fm we performed calculations using wave function optimized
meson operatior and Gaussian smeared meson operator, $s12$ for $T=174,~200$ and 244 MeV. The results are shown
in Fig. \ref{fig:smear_comp_b7825_Nt20_28} and Tab. \ref{tab:Smear_comp_b7825_temp}. We clearly see that the dependence
of in-medium meson properties is largely reduced. We do see significant dependence of the $\eta_c(1S)$ mass on the choice
of the meson operators and at the lowest temperature the thermal width is also independent of the choice of the meson
operator, cf.  Tab. \ref{tab:Smear_comp_b7596}.

We observe that the 2 smeared sources have a larger contribution of excited states, which is quite consistent between zero and finite temperature, while there is a slight difference in the middle also at finite temperature.

\begin{table}[H]
\begin{tabular}{ |c|c|c|c|c|c| } 
 \hline
State & Source & $\beta$ & Peak Position $M_\alpha$[MeV] & $\Gamma_{\alpha} ^0$[MeV] & zero mode \\ 
 \hline
$\eta _c$ & 1S & 7.596 & $2988.0 \pm 1.3$ & $95.3 \pm 2.7 $ & \\
$\eta _c$ & s7 & 7.596 & $2998.6 \pm 3.2$ & $113.6 \pm 5.5 $ & \\
$\eta _c$ & s10 & 7.596 & $2984.8 \pm 4.8$ & $120.0 \pm 8.5 $ & \\
\hline
$J/\psi$ & 1S & 7.596 & $3080.5 \pm 3.5$ & $106.3 \pm 11.9 $ & $(17.6 \pm 2.2) \cdot 10^{-5}$ \\
$J/\psi$ & s7 & 7.596 & $3093.4 \pm 5.7$ & $113.4 \pm 18.5 $ & $(16.8 \pm 3.3) \cdot 10^{-5}$ \\
$J/\psi$ & s10 & 7.596 & $3076.6 \pm 8.0$ & $110.8 \pm 23.6 $ & $(16.7 \pm 3.9) \cdot 10^{-5}$ \\
\hline
\end{tabular}
\caption{Results for fits shown in Fig. \ref{fig:smear_comp_b7596} using the fit ansatz shown in eq. (\ref{eq:sigma_med_Ansatz}) with a cut of size $4\Gamma_{\alpha} ^0$ for $\beta=7.596$ and $T=250$ MeV.}
\label{tab:Smear_comp_b7596}
\end{table}

\begin{table}[H]
\begin{tabular}{ |c|c|c|c|c|c| } 
 \hline
State & Source & $\beta$ & Peak Position $M_\alpha$[MeV] & $\Gamma_{\alpha} ^0$[MeV] & zero mode\\ 
 \hline
$\eta _c$ & 1S & 7.825 & $2990.5 \pm 1.7$ & $114.2 \pm 3.2 $ & \\
$\eta _c$ & s7 & 7.825 & $3000.3 \pm 3.0$ & $112.7 \pm 5.1 $ & \\
$\eta _c$ & s10 & 7.825 & $2993.8 \pm 3.4$ & $135.2 \pm 6.3 $ & \\
$\eta _c$ & s12 & 7.825 & $3000.3 \pm 5.2$ & $161.8 \pm 6.3 $ & \\
\hline
$J/\psi$ & 1S & 7.825 & $3087.4 \pm 3.4$ &  $133.6 \pm 10.7 $ & $(3.2 \pm 0.4) \cdot 10^{-4}$ \\
$J/\psi$ & s7 & 7.825 & $3103.6 \pm 4.2$ & $130.1 \pm 14.3 $ & $(3.0 \pm 0.5) \cdot 10^{-4}$ \\
$J/\psi$ & s10 & 7.825 & $3093.9 \pm 6.1$ & $144.0 \pm 18.4 $ & $(2.8 \pm 0.6) \cdot 10^{-4}$ \\
$J/\psi$ & s12 & 7.825 & $3099.8 \pm 8.3$ & $168.4 \pm 18.9 $ & $(2.6 \pm 0.8) \cdot 10^{-4}$ \\
\hline
\end{tabular}
\caption{The mass and width of 1S charmonia obtained using correlation function of optimized meson correlator and Gaussian smeared meson operators of various 
size for $\beta=7.825$ and $T=271$ MeV using cut Lorentzian form with $cut_{\alpha}=4 \Gamma_{\alpha}^0$.}
\label{tab:Smear_comp_b7825}
\end{table}

\begin{table}[H]
\begin{tabular}{ |c|c|c|c|c|c| } 
\hline
State  & Source & T[MeV] & $\beta$ & Peak Position $M_\alpha$[MeV]  & $\Gamma_{\alpha} ^0$[MeV] \\ 
\hline
$\eta _c$ & 1S & 244 & 7.825 & $2989.0 \pm 1.8$ & $87.5 \pm 4.4 $  \\
$\eta _c$ & s12 & 244 & 7.825 & $2995.9 \pm 4.6$ & $130.6 \pm 7.7 $  \\
\hline
$\eta _c$ & 1S & 203 & 7.825 & $2993.3 \pm 1.3$ & $60.9 \pm 4.3 $  \\
$\eta _c$ & s12 & 203 & 7.825 & $2991.8 \pm 3.3$ & $76.3 \pm 7.9 $  \\
\hline
$\eta _c$ & 1S & 174 & 7.825 & $2994.2 \pm 1.0$ & $34.6 \pm 4.7 $  \\
$\eta _c$ & s12 & 174 & 7.825 & $2992.2 \pm 3.9$ & $39.2 \pm 16.6 $  \\
\hline
\end{tabular}
\caption{Results for fits shown in Fig. \ref{fig:smear_comp_b7825_Nt20_28}  using the fit ansatz shown in eq. (\ref{eq:sigma_med_Ansatz}) with a cut of size $4\Gamma_{\alpha} ^0$.}
\label{tab:Smear_comp_b7825_temp}
\end{table}


There are a couple of different explanations for this source dependence. As observed in the effective masses before the subtraction, the larger source has a larger contribution from excited states. When the excited states get very wide, the subtraction in toy models shows that the tails still affect the effective mass of the ground state, and this is stronger the larger the amplitude of the excited states. Based on estimates on the excited states' contribution, this effect
, however, seems to not be large enough to explain the full difference we see. 

Another possibility is that the large source smears the spectral function of the point source.  On the other hand, testing on the free case indicates that the found hierarchy of states in terms of $1S$ $2S$ etc. does not appear, indicating that there are indeed specific sources to project into, and not a spectrum of smeared sources. We therefore believe that the states we see, even the excited states, are present in the medium and not a result of our chosen operator.

In the current state, we think the wave%
functions are the best to proceed with due to their small overlap with excited states, despite their larger fluctuations. 


Lastly, for the $\eta _c$ states, no zero mode should be present. We have however performed the fits with or without a zero-mode for the $\eta _c$ states, to check if the results 
depend on the inclusion of the zero mode. For temperatures at $250$ MeV and above, the fit without a zero-mode for the $2S$ state was however not good enough (too large $\chi ^2$), and was excluded.

\subsection{P-States source dependence}
\begin{figure}[H]
  \centering
 \includegraphics[scale=0.35]{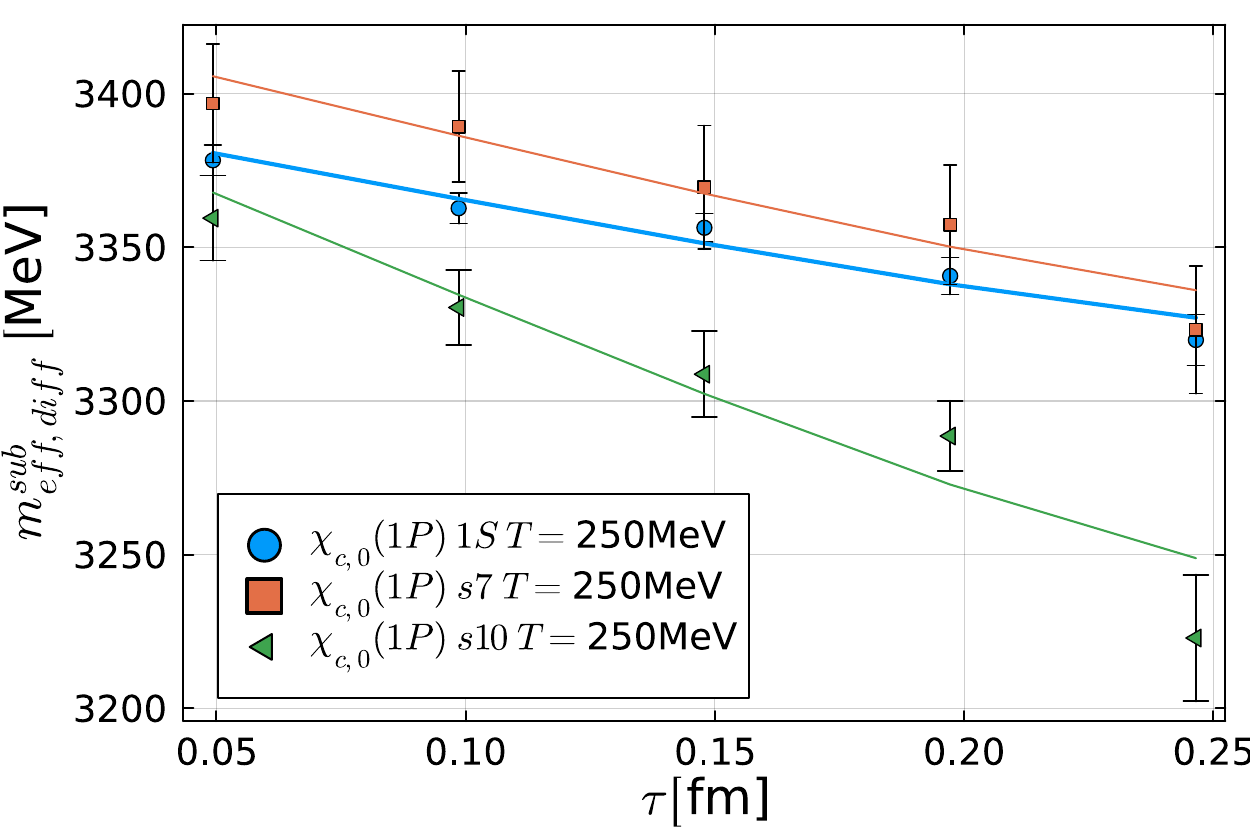}
 \includegraphics[scale=0.35]{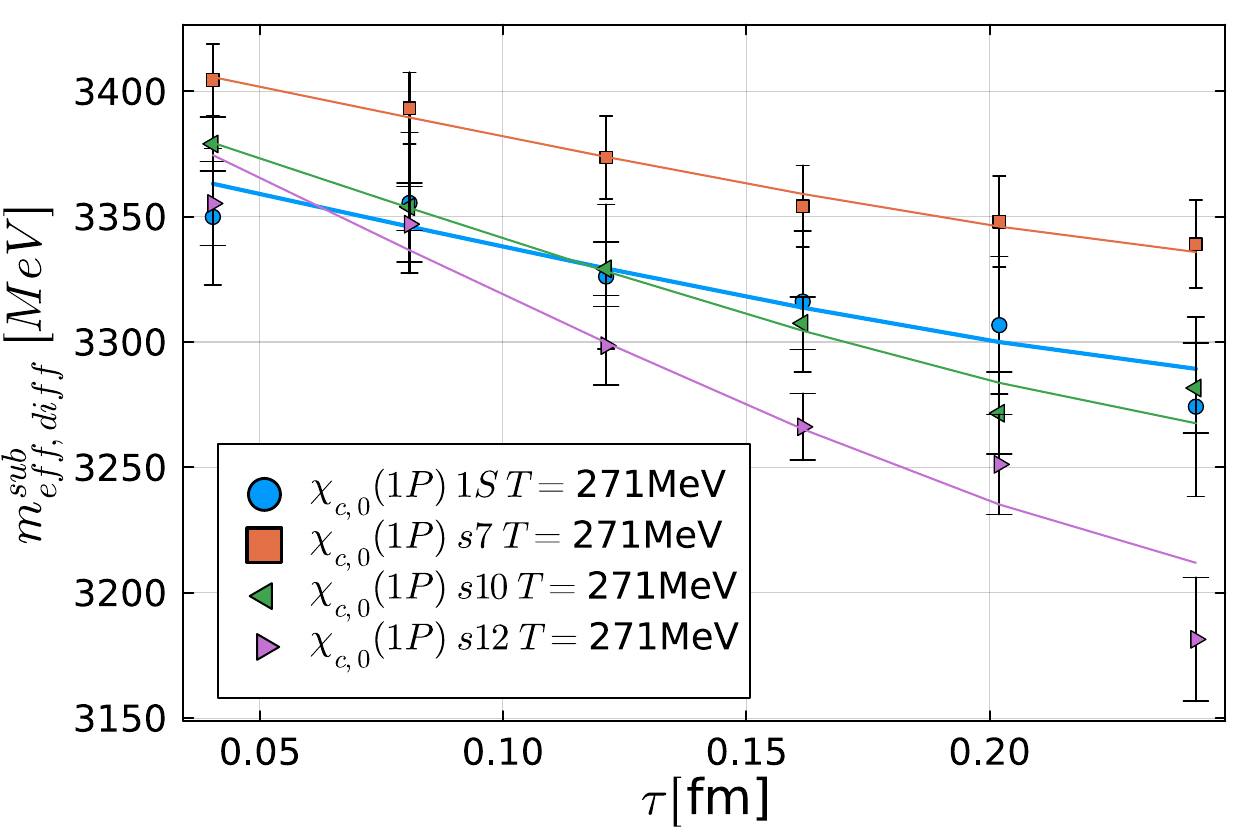}
\caption{Subtracted effective mass for the scalar channel from the difference in the correlator for different sources. The first source was a 1S wavefunction, while the other sources are Gaussian sources the size of 7, 10 and 12 times the lattice spacing a, with $a=0.0493$ fm (left) and $a=0.0404$ fm (right).   }
\label{fig:smear_comp_b7825_Nt18_scalar}
\end{figure}

\begin{figure}[H]
  \centering
 \includegraphics[scale=0.35]{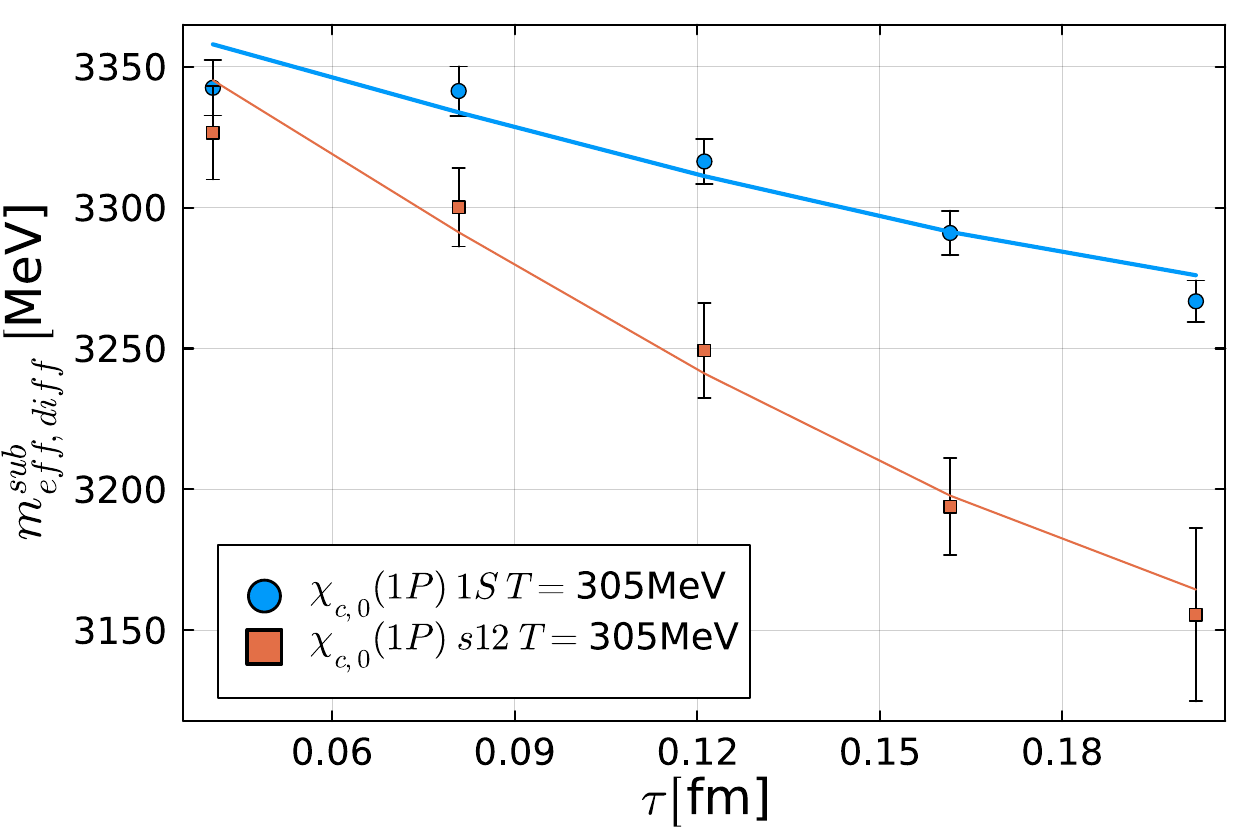}
  \includegraphics[scale=0.35]{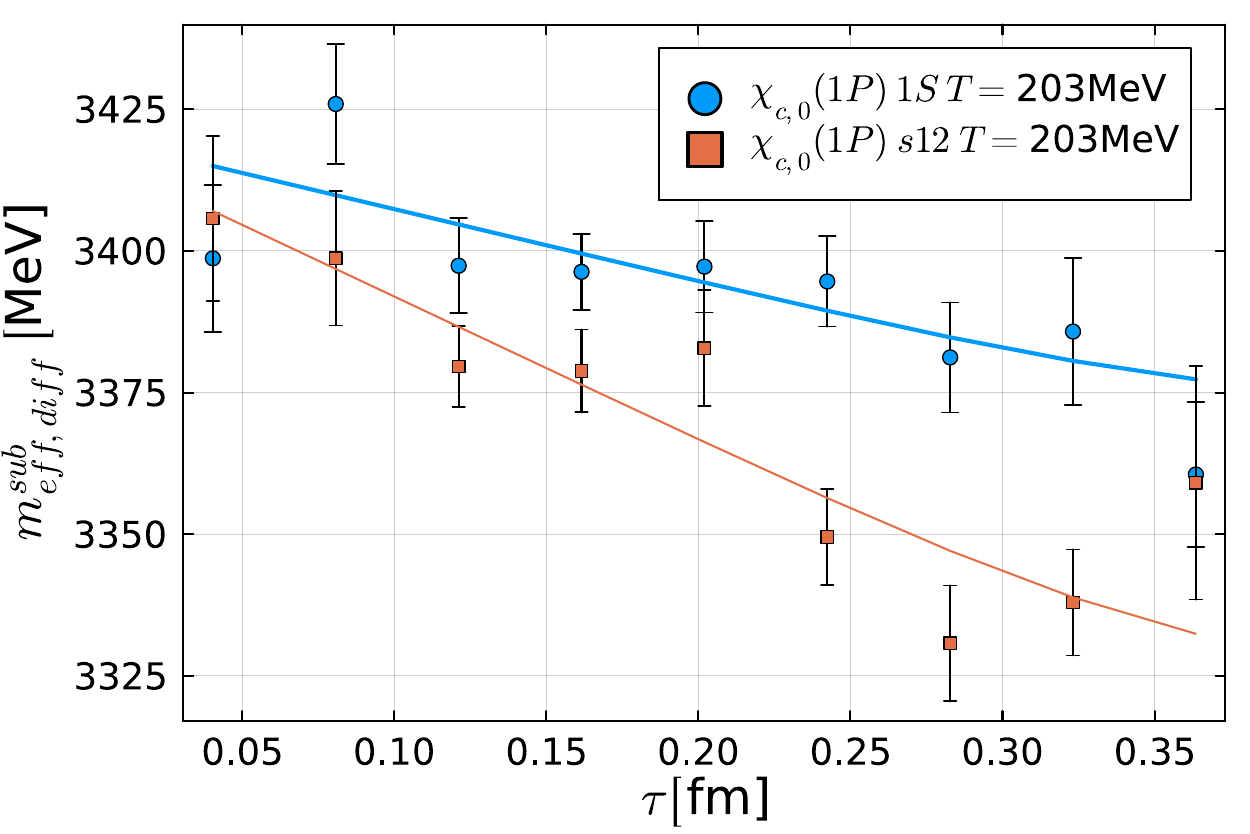}
\caption{Subtracted effective mass for the scalar channel from the difference in the correlator for different sources for 2 temperatures. The first source was a 1S wave function, while the  other source is a Gaussian source of size of 12 times the lattice spacing a, with $a=0.0404$ fm.   }
\label{fig:smear_comp_b7825_Nt16_24_scalar}
\end{figure}

We proceed with the same analysis for the P-states to study the dependence on the meson operator (meson source). Here we only show $\chi_{c,0}$ as the results for $\chi_{c,1}$ are very similar. As shown in Fig. \ref{fig:smear_comp_b7825_Nt18_scalar} and \ref{fig:smear_comp_b7825_Nt16_24_scalar} and Tab. \ref{tab:Smear_comp_b7825_temp_scalar} we see a similar behavior to the S-states, that is, that the larger sources create a larger width, while the smallest source has a small upward shift in the mass. 
This upward shift, however, is smaller than the estimated errors as one can see from Tab. \ref{tab:Smear_comp_b7825_temp_scalar}.
The source dependence for the in-medium width, however, does not reduce as much as it did for the S-states when the temperature is decreased, though it does slightly from $51\%$ to $44\%$.

\begin{table}[H]
\begin{tabular}{ |c|c|c|c|c|c| } 
\hline
State  & Source & T[MeV] & $\beta$ & Peak Position $M_\alpha$[MeV]  & $\Gamma_{\alpha} ^0$[MeV] \\ 
\hline
$\chi_{c,0} $ & 1S & 305 & 7.825 & $3369.4 \pm 5.0$ & $249.4 \pm 8.8 $  \\
$\chi_{c,0}$ & s12 & 305 & 7.825 & $3368.7 \pm 14.3$ & $378.5 \pm 14.3 $  \\
\hline
$\chi_{c,0}$ & 1S & 271 & 7.825 & $3371.4 \pm 25.1$ & $206.4 \pm 15.7 $  \\
$\chi_{c,0}$ & s7 & 271 & 7.825 & $3413.7 \pm 13.3$ & $199.9 \pm 11.6 $  \\
$\chi_{c,0}$ & s10 & 271 & 7.825 & $3392.0 \pm 9.4$ & $254.8 \pm 15.4 $  \\
$\chi_{c,0}$ & s12 & 271 & 7.825 & $3392.4 \pm 14.9$ & $309.4 \pm 15.1 $  \\
\hline
$\chi_{c,0}$ & 1S & 250 & 7.596 & $3391.8 \pm 4.8$ & $173.7 \pm 10.2 $  \\
$\chi_{c,0}$ & s7 & 250 & 7.596 & $3420.2 \pm 19.4$ & $198.0 \pm 13.9 $  \\
$\chi_{c,0}$ & s10 & 250 & 7.596 & $3392.4 \pm 16.2$ & $260.6 \pm 17.1 $  \\
\hline
$\chi_{c,0}$ & 1S & 244 & 7.825 & $3397.3 \pm 6.3$ & $153.9 \pm 14.2 $  \\
$\chi_{c,0}$ & s12 & 244 & 7.825 & $3403.3 \pm 10.8$ & $247.5 \pm 15.2 $  \\
\hline
$\chi_{c,0}$ & 1S & 203 & 7.825 & $3417.6 \pm 4.6$ & $111.9 \pm 11.5 $  \\
$\chi_{c,0}$ & s12 & 203 & 7.825 & $3412.2 \pm 8.8$ & $157.7 \pm 11.9 $  \\
\hline
$\chi_{c,0}$ & 1S & 174 & 7.825 & $3424.4 \pm 2.7$ & $67.3 \pm 7.8 $  \\
$\chi_{c,0}$ & s12 & 174 & 7.825 & $3419.7 \pm 7.3$ & $97.0 \pm 12.1 $  \\
\hline
\end{tabular}
\caption{Results for fits shown in Fig. \ref{fig:smear_comp_b7825_Nt18_scalar} and  \ref{fig:smear_comp_b7825_Nt16_24_scalar}  using fit ansatz shown in eq. (\ref{eq:sigma_med_Ansatz}) with a cut of size $4\Gamma_{\alpha} ^0$ but with no zero mode and on the difference of the correlator.}
\label{tab:Smear_comp_b7825_temp_scalar}
\end{table}

\subsection{Results for a larger value of $cut_{\alpha}$}

\begin{figure}[H]
  \centering
 \includegraphics[scale=0.5]{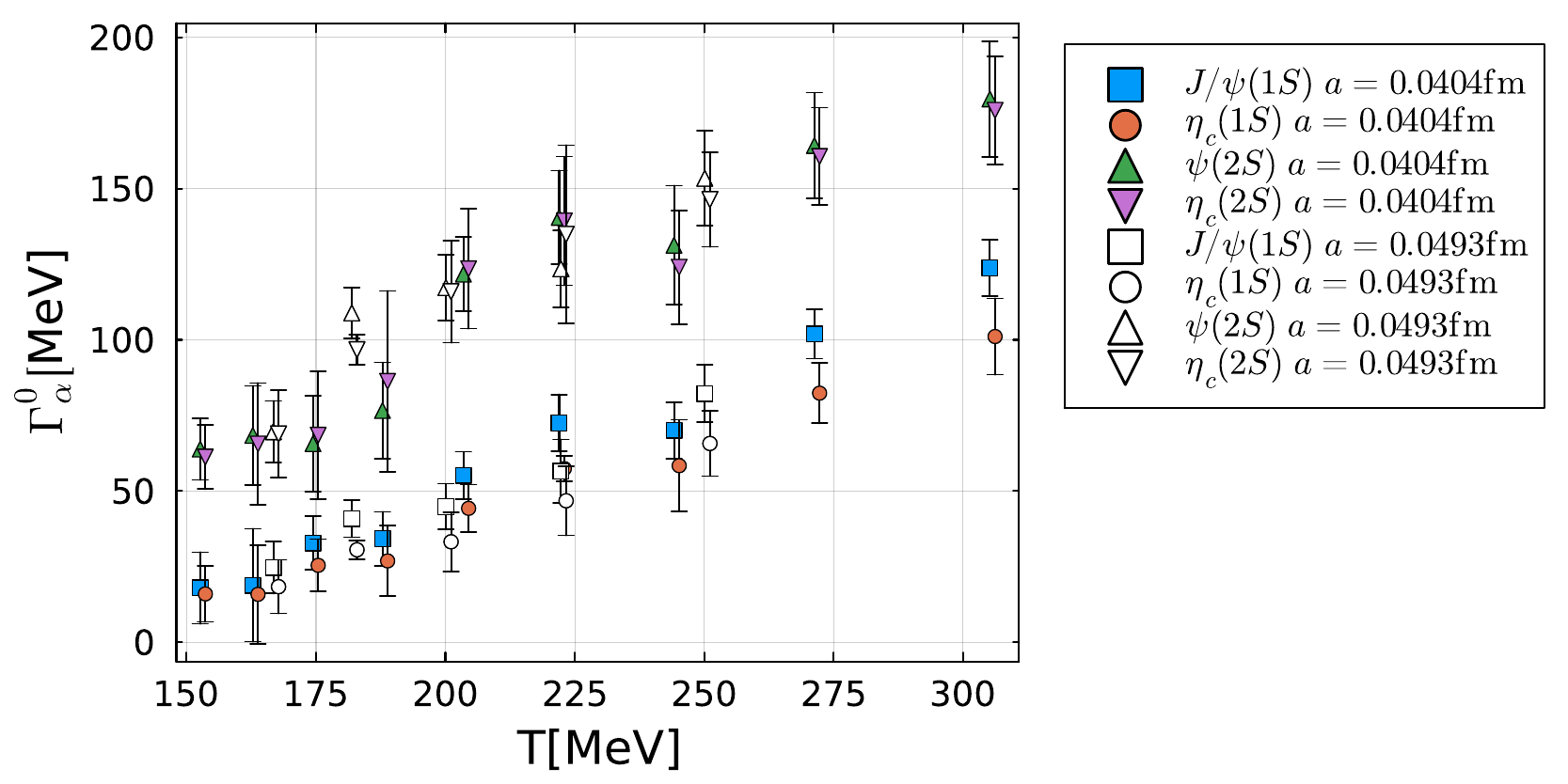}
  \label{fig:sub1}
\caption{The extracted width $\Gamma_{\alpha}^0$ from Lorentzian fits cut off at 6 times the width on the subtracted correlator for 1S and 2S wave function sources. Errors are a combination of statistical and systematics from 2 different fits for 2 different subtraction parameters, and 4 fits for pseudoscalar, with or without including a zero mode.}
\label{fig:results_width_cut6}
\end{figure}

\begin{figure}[H]
  \centering
 \includegraphics[scale=0.5]{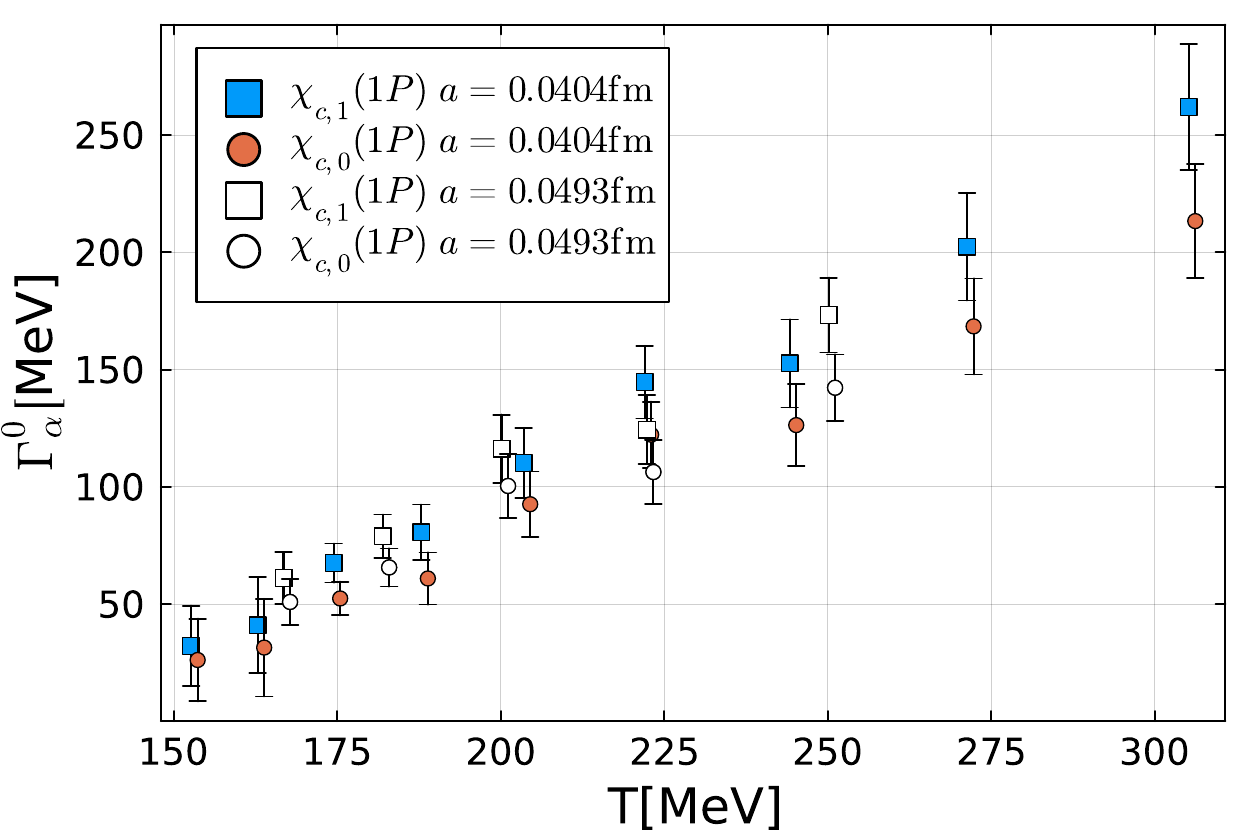}
  \label{fig:sub1}
\caption{The extracted width $\Gamma_{\alpha}^0$ from Lorentzian fits cut off at 6 times the width on the difference of the subtracted correlator for 1S wave function source for the P-states. Errors are a combination of statistical and systematics from 2 different fits for 2 different subtraction parameters. }
\label{fig:results_width_P_cut6}
\end{figure}

\begin{figure}[H]
  \centering
 \includegraphics[scale=0.5]{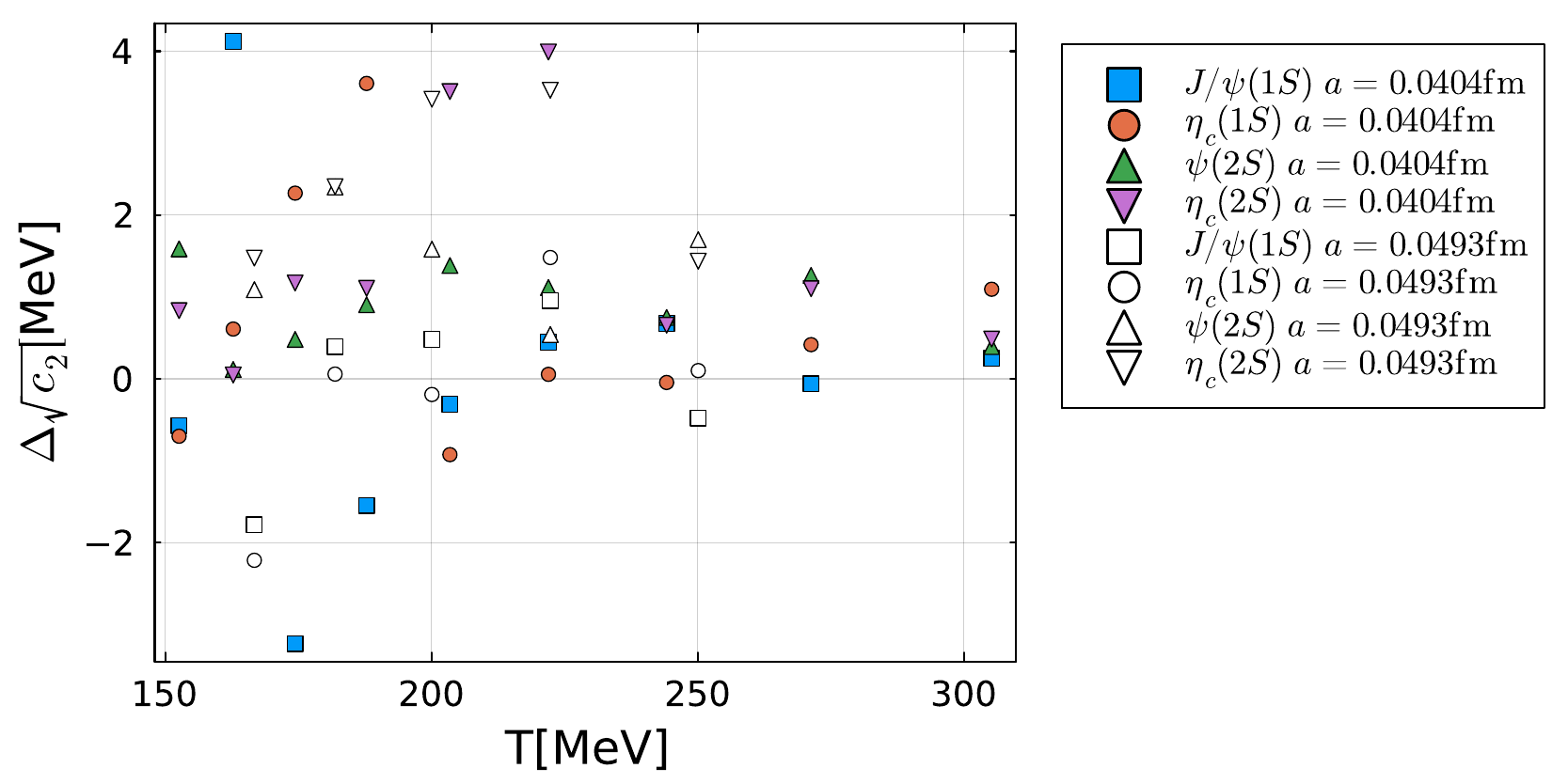}
\caption{Difference in second cumulant between fits with a cut at $4\Gamma_{\alpha}^0$ and $6\Gamma_{\alpha}^0$. The statistical errors
are not shown, since they are much larger than the difference.}
\label{fig:diff_width}
\end{figure}

Our fit Ansatz for the spectral function also depends on the parameter $cut_{\alpha}$ in addition to $M_{\alpha}$, $\Gamma_{\alpha}^0$
and $z_{\alpha}$. In our analysis so far we  set $cut_{\alpha}=4 \Gamma_{\alpha}^0$ based on calculations of the bottomonium spectral function in the T-matrix approach. We would like to check how sensitives are our finding to this choice of $cut_{\alpha}$.
Therefore, we also perform fits setting $cut_{\alpha}=4 \Gamma_{\alpha}^0$. 
\begin{figure}[H]
  \centering
 \includegraphics[scale=0.35]{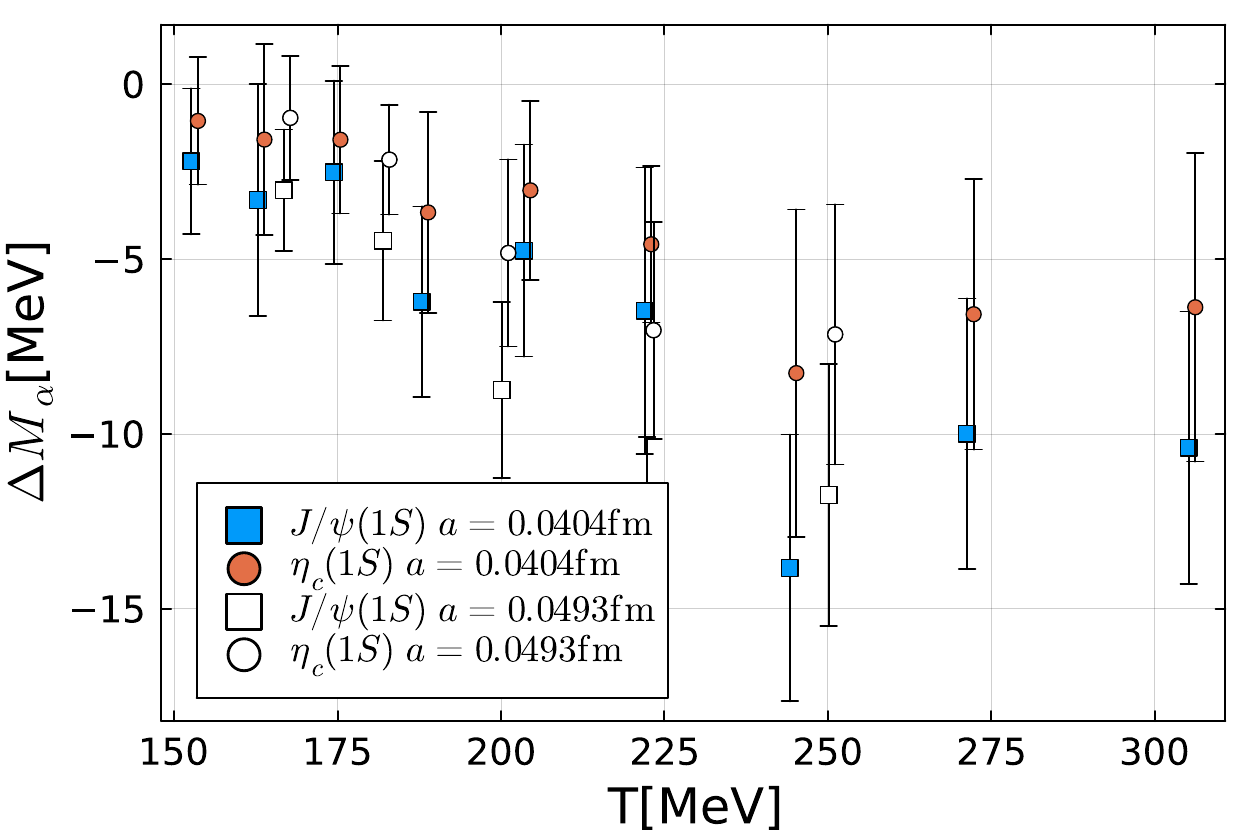}
  \includegraphics[scale=0.35]{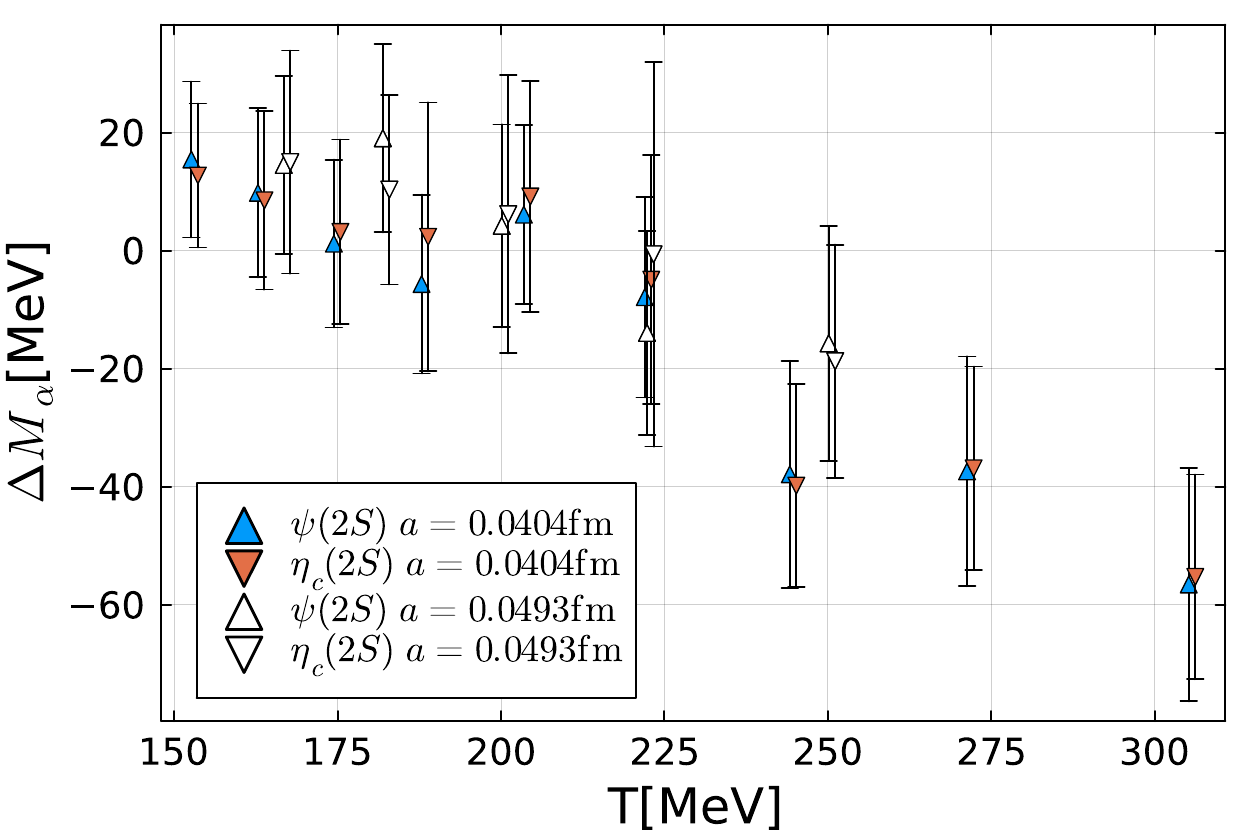}
  \includegraphics[scale=0.35]{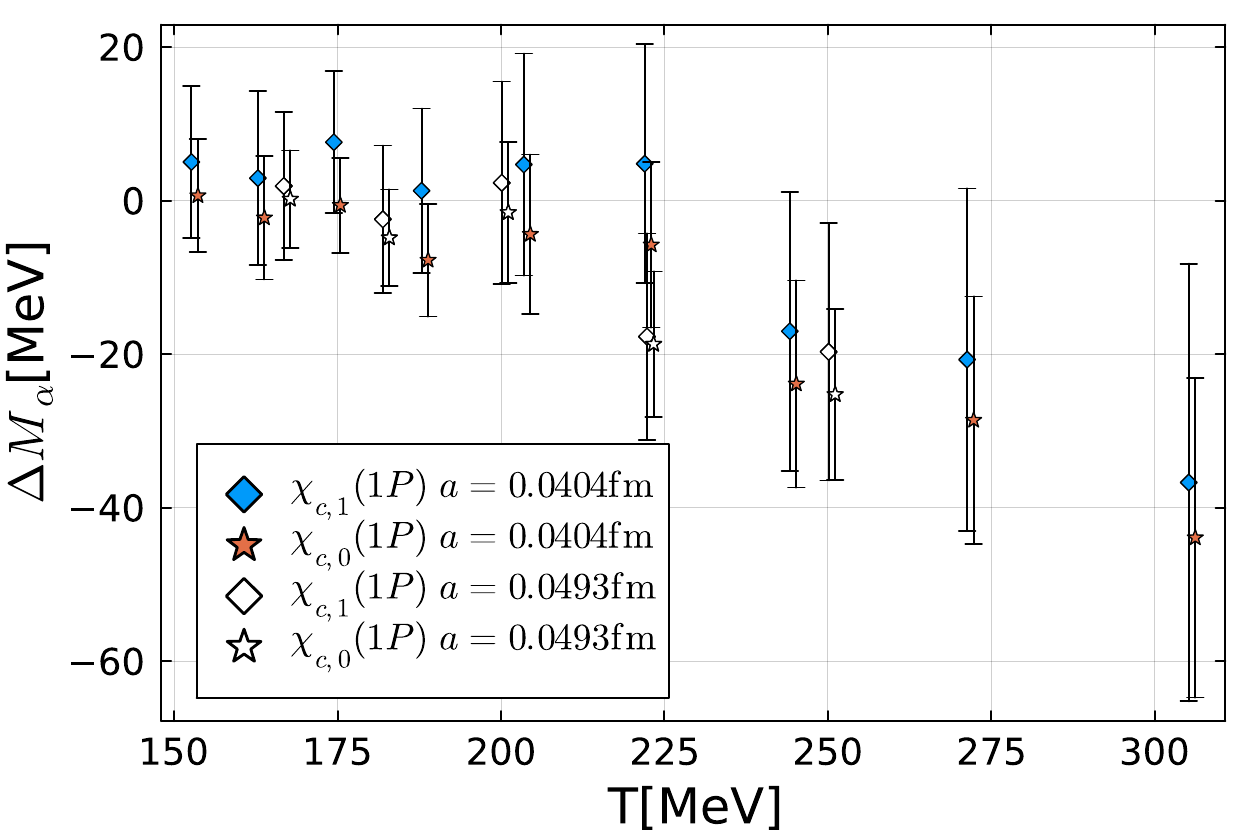}
  \label{fig:sub1}
\caption{The in-medium mass shift for 1S (top left), 2S (top right) and 1P (bottom) charmonium states obtained from cut Lorentzian fits
with $cut_{\alpha}=6 \Gamma_{\alpha}^0$.}
\label{fig:relative_peak_position_cut6}
\end{figure}
In Fig. \ref{fig:relative_peak_position_cut6} we show the in-medium mass obtained from fits $cut_{\alpha}=6 \Gamma_{\alpha}^0$
for 1S, 2S and 1P charmonia.
As one can see from the figures the in-medium mass shift for these charmonia states is about the same as the ones
obtained from fits with $cut_{\alpha}=4 \Gamma_{\alpha}^0$. In Figs. \ref{fig:results_width_cut6} and \ref{fig:results_width_P_cut6}
we show the thermal width of S-wave and P-wave charmonium states. We see that the temperature dependence of the charmonium width
is very similar to the one obtained with $cut_{\alpha}=4 \Gamma_{\alpha}^0$. However, the width overall values of the width is about 22\% smaller. 
To understand this we calculate the second cumulant of the spectral function $c_{\alpha,2}$ introduced in the main text. 
In Fig.  \ref{fig:diff_width} we show the difference in $c_{\alpha,2}$ obtained from fits with $cut_{\alpha}=4 \Gamma_{\alpha}^0$
and $cut_{\alpha}=6 \Gamma_{\alpha}^0$. One can see that this difference is tiny, much smaller than the statistical error.
This implies that the lattice data mostly constrain $c_{\alpha,2}$ are not very sensitive to the detailed shape of the
spectral function,

\section{Charmonium correlators on grown Lattices}\label{sec:grow}

\begin{figure}
    \centering
    \includegraphics[width=0.45\textwidth]{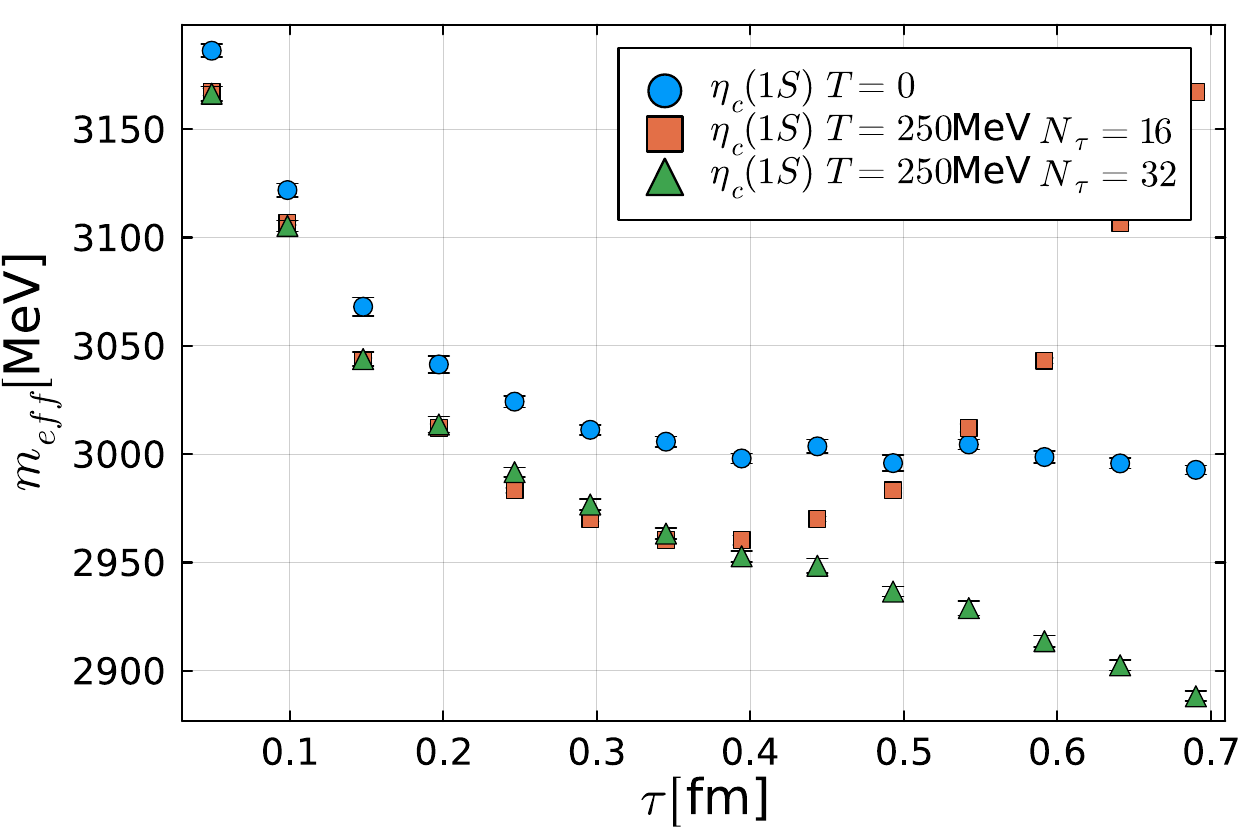}
    \includegraphics[width=0.45\textwidth]{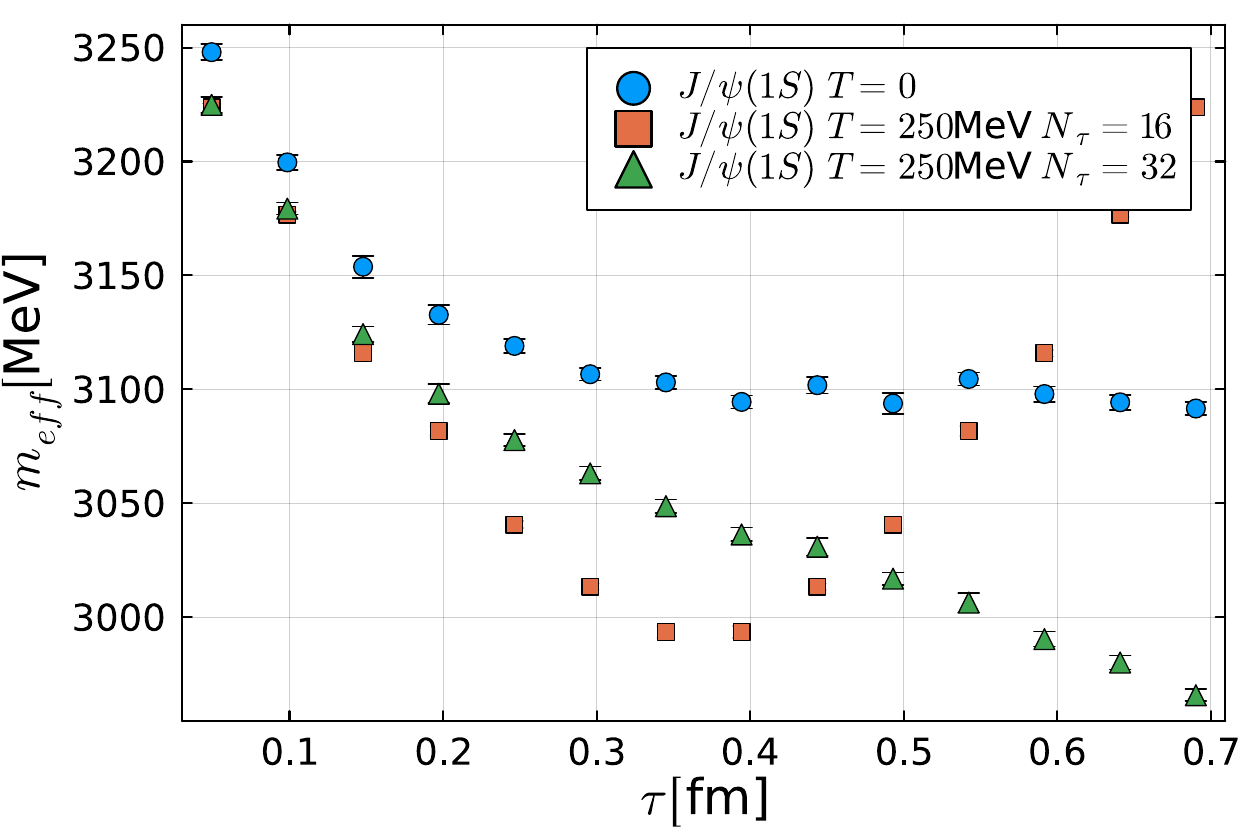}
    \caption{Comparison between the effective mass at $T=250$ MeV from the original correlator and the correlator from a lattice grown by 2 times in the temporal direction and the zero temperature results.}
    \label{fig:meff_grown}
\end{figure}

\begin{figure}
    \centering
    \includegraphics[width=0.45\textwidth]{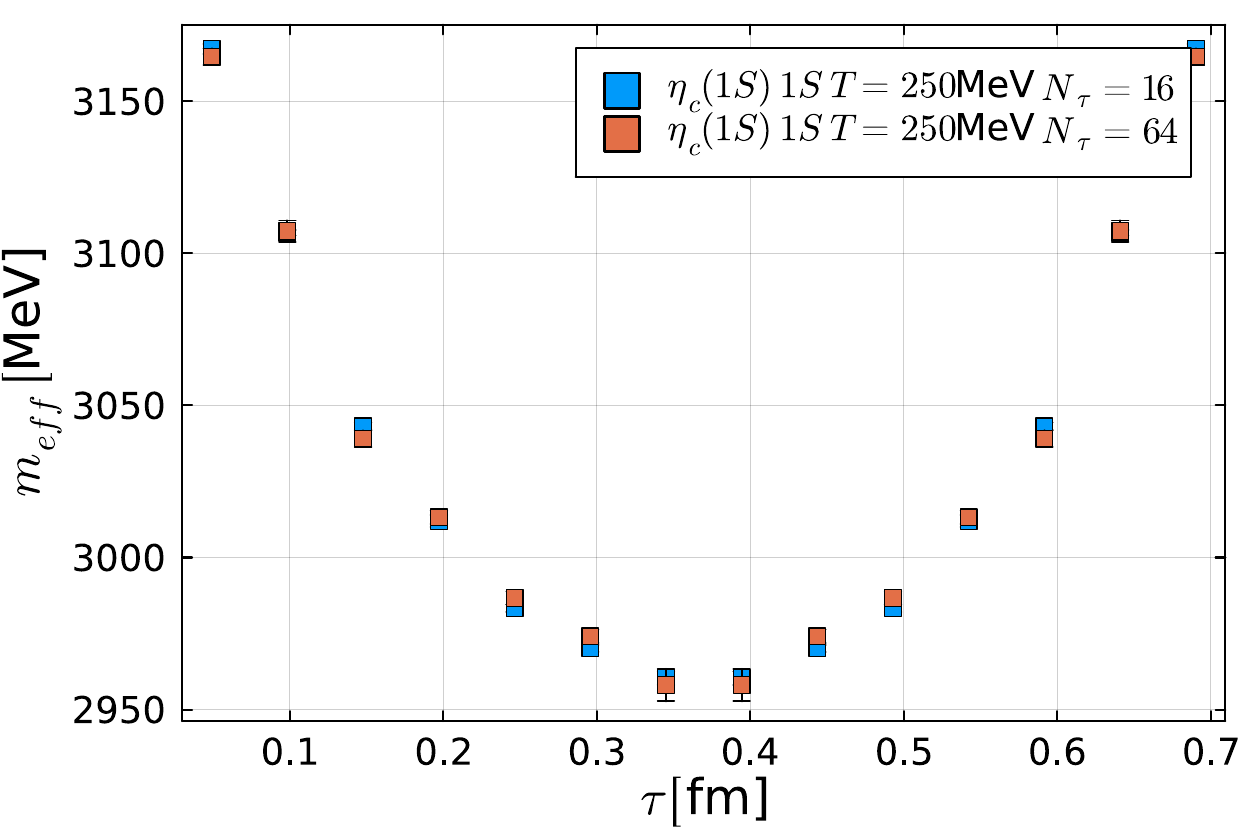}
    \includegraphics[width=0.45\textwidth]{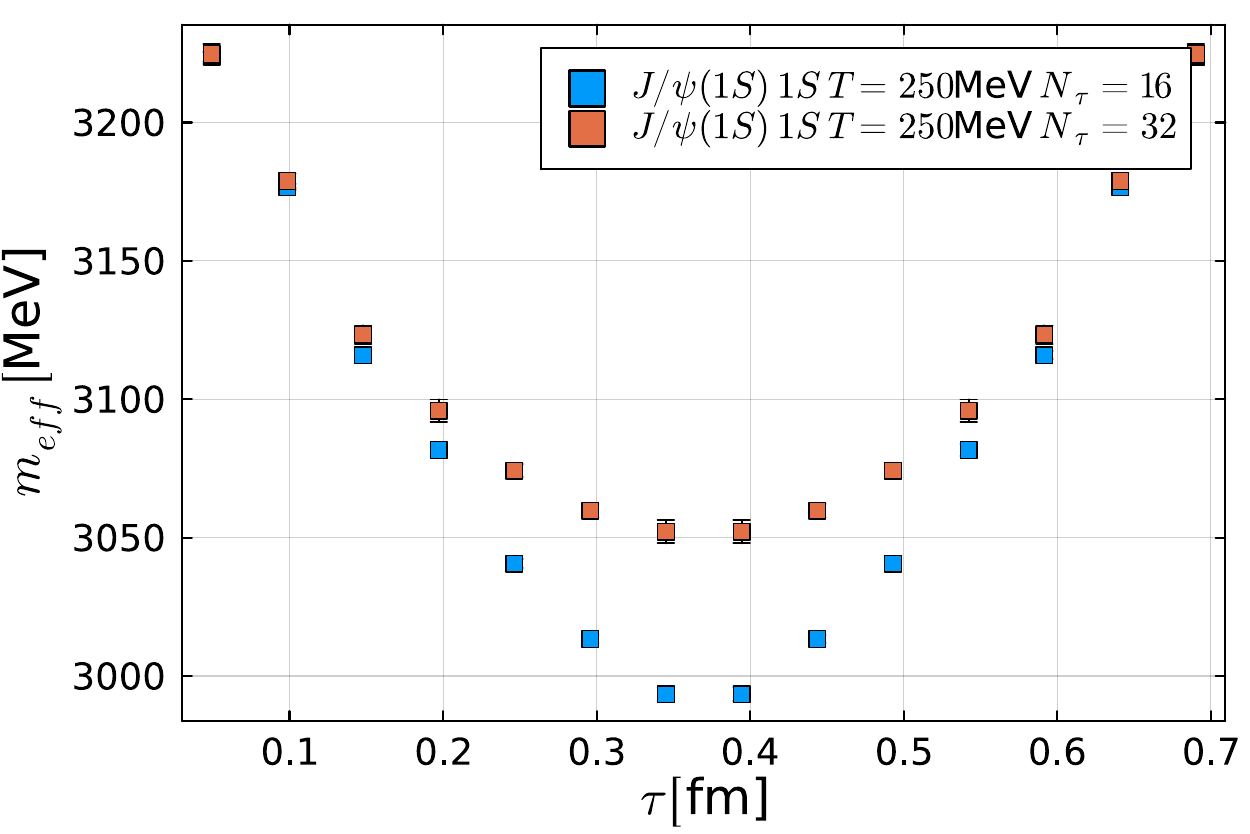}
    \caption{Comparison between the effective mass at $T=250$ MeV from the original correlator and the correlator from a lattice grown by 2 to 4 times in the temporal direction, which afterwards has been made periodic. It is seen, that $J/\psi$ is different in the middle, which is where the zero mode is important, while $\eta _c$, within error bars, stays the same.}
    \label{fig:meff_grown_periodic}
\end{figure}

\begin{figure}
    \centering
    \includegraphics[width=0.45\textwidth]{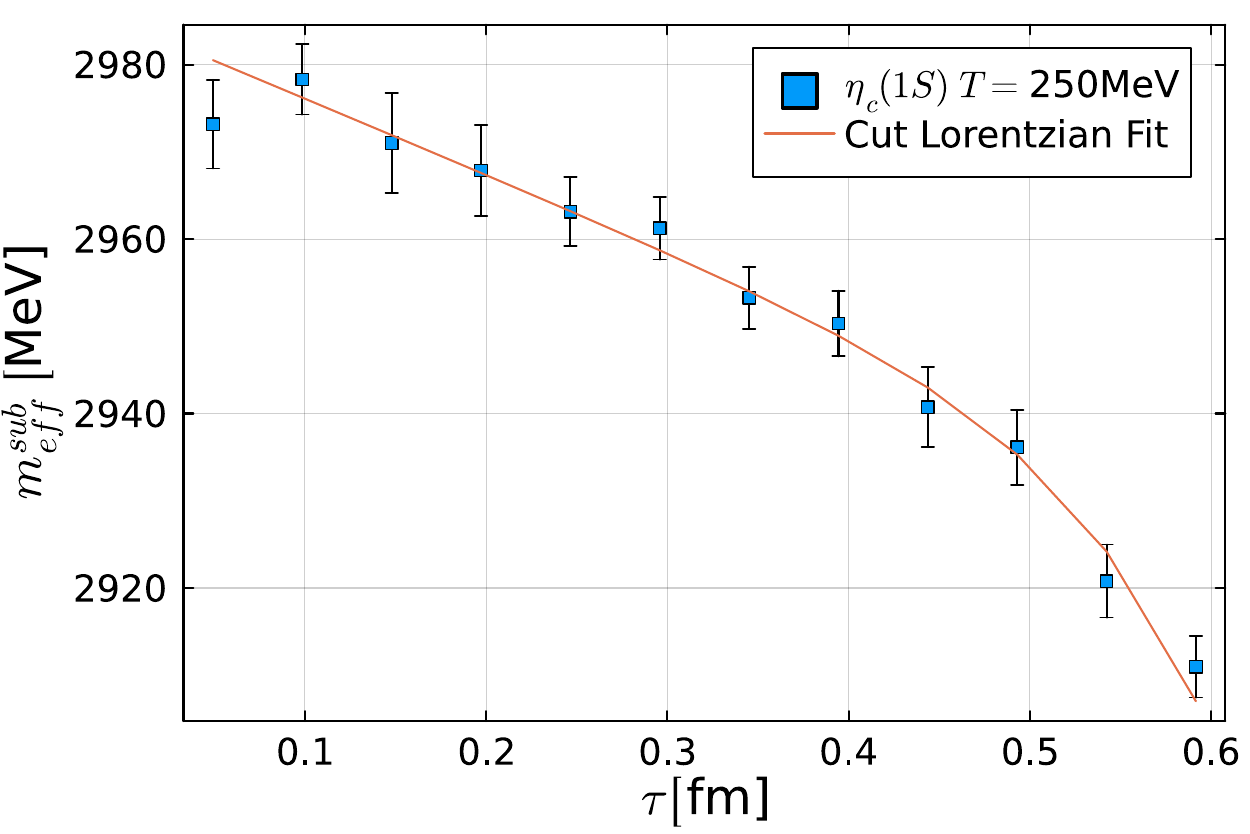}
    \includegraphics[width=0.45\textwidth]{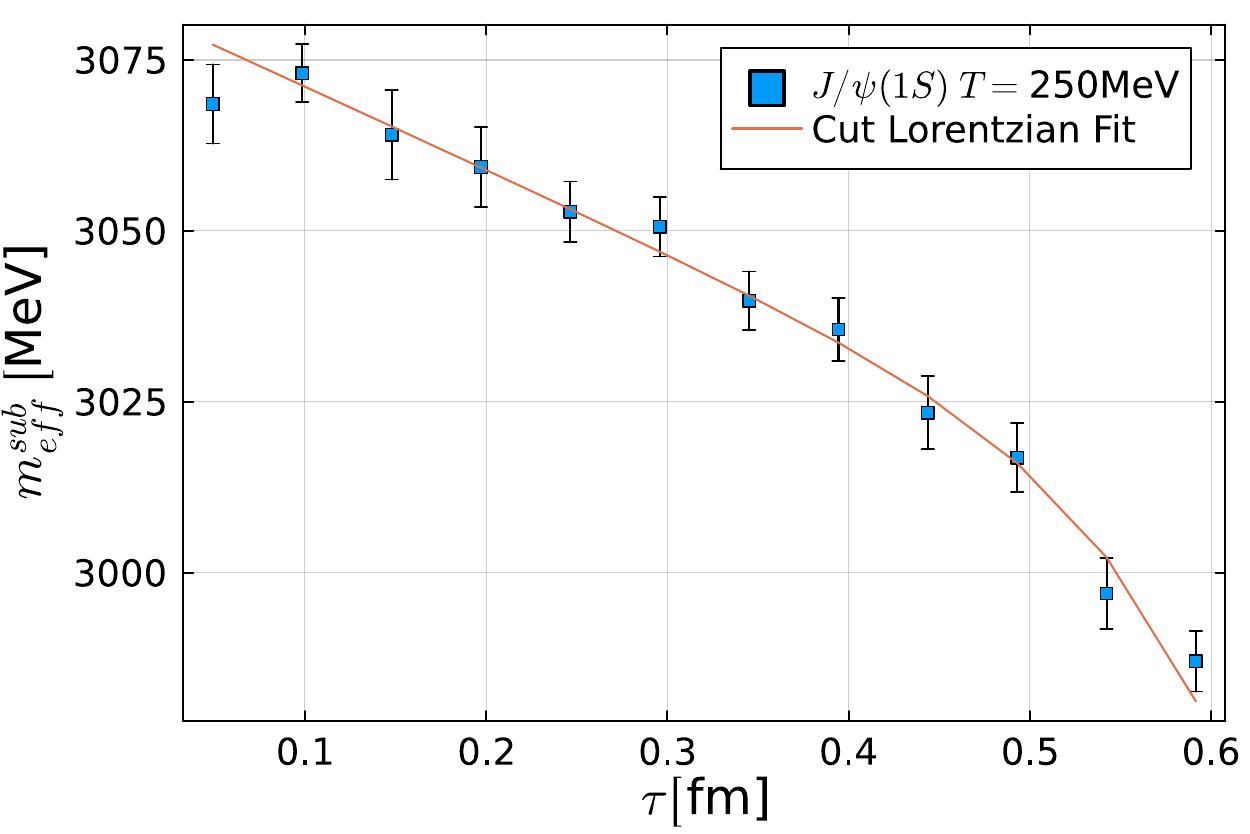}
    \caption{Subtracted effective mass from the correlator on lattices grown from $N_\tau = 16$ to $N_\tau = 32$ for $\eta _c(1S)$ on the left and $J/\psi (1S)$ on the right. The fit gives $M_\alpha=2986.9\pm 2.7$ MeV and $\Gamma_{\alpha}^0= 92.1 \pm 4.1 MeV$, while the fit from the non-grown lattice was $M_\alpha=2986.0\pm 2.8 $ MeV and $\Gamma_{\alpha}^0= 82.9 \pm 12.7 MeV$, while the $J/\psi$ fit gives $M_\alpha=3086.2\pm 3.4 $ MeV and $\Gamma_{\alpha} ^0= 108.7 \pm 4.0 MeV$, while the fit from the non-grown lattice was $M_\alpha=3080.9\pm 3.3 $ MeV and $\Gamma_{\alpha} ^0= 105.3 \pm 11.7 MeV$.}
    \label{fig:meff_grown_sub}
\end{figure}

The periodic boundary condition of the relativistic heavy quark formulation make the analysis of
the spectral function more difficult. First, the maximal temporal extent is $1/(2T)$ instead of $1/T$ of
the NRQCD formulation of quarkonia. Second the periodic boundary give rise to the zero modes in the
charmonium correlators. One possibility to overcome this difficulties is to construct grown lattices
of size $2 N_\tau$ or $4 N_\tau$ from the original lattices with temporal size $N_\tau$ by replicating them in
the time direction. We can then calculate the charmonium correlation functions on these grown lattices.
The medium effect are still contained in the gauge background but the charm quark propagator is now
non-thermal. This approach is similar to NRQCD approach, except no non-relativistic approximation is used at the level
of quark propagator. The zero mode contribution is expected to be much reduced on the grown lattices.
In Fig. \ref{fig:meff_grown} we show the effective masses for $\eta_c$ and $J/\psi$ obtained on the grown
lattices for $a=0.0493$fm and $T=250$ MeV with our previous results. We see from the figures that for $\eta_c$ there are only small differences
in the effective masses obtained on the original lattice and the grown lattice around $\tau \simeq 1/(2 T)$.
However, for $J/\psi$ the effective masses on the grown lattices are significantly larger for $\tau \simeq 1/(2 T)$
signaling that much of the zero mode contribution is eliminated on the grown lattice. To test this we can take the 
charmonium correlation
function calculated  on the grown lattice with temporal  size $2/T$ (or $4/T$) and make it into a periodic correlation function
corresponding to temporal size $1/T$ using the equation $C(\tau,1/T)=C(\tau,2/T)+C(\tau+1/T,2/T)$. There could be
additional terms in this equation corresponding to going around the periodic boundary more than once, but because
of the large charmonium mass such terms are very small and can be neglected. The effective masses corresponding
to this analysis are shown in Fig. \ref{fig:meff_grown_periodic} for $\eta_c$ and $J/\psi$ effective masses.
In the case of $\eta_c$ the procedure of making the charmonium correlator on the grown lattices periodic
with time extent $1/T$ more or less reproduces the correlator obtained on the original lattice with temperature $T$.
In the case of $J/\psi$, however, this procedure completely removes the zero mode contribution and makes the $J/\psi$
effective mass look very similar to that of $\eta_c$. This confirms that the differences in the effective masses of
$\eta_c$ and $J/\psi$ are indeed due to the zero mode contribution in the vector channel. 
We also performed the subtraction of the high energy part of the correlator and calculated the corresponding effective masses. These are shown in Fig. \ref{fig:meff_grown_sub} and look
similar to the effective masses  obtained from Wilson line correlators \cite{Bazavov:2023dci,Bala:2021fkm} and NRQCD bottomonium correlators  \cite{Larsen:2019zqv,Ding:2025fvo}.
Compared to our results in the main text we see the linear decrease of the effective masses in a larger region in $\tau$.
We can also fit the corresponding effective masses to determine the in-medium masses and width of $\eta_c$ and $J/\psi$. These fits are also shown in Fig. \ref{fig:meff_grown_sub}.
The determined masses and width from these fits are in good agreement with the ones obtained in the main text. The main difference is that the error on the width of $J/\psi$ is significantly
reduced due to the fact that no zero mode contribution has to be fitted. This analysis provides a valuable cross-check of the mass and width determinations presented in the main text.

\bibliographystyle{apsrev4-1}
\bibliography{ref}
\end{document}